\documentclass[11pt,twoside]{article}
\usepackage{fancyhdr}
\pdfoutput=1
\usepackage[colorlinks,citecolor=blue,urlcolor=blue,linkcolor=blue,bookmarks=false]{hyperref}
\usepackage{amsfonts,epsfig,graphicx}
\usepackage{afterpage}
\usepackage{pgfplots}
\usepackage{booktabs}
\usepackage{mathrsfs}
\usepackage{tikz}
\usetikzlibrary{arrows.meta,calc,angles}
\usepackage{nicefrac}
\usepackage{comment}
\usepackage{bbm}
\usepackage{amsmath,amssymb,amsthm} 
\usepackage{fullpage}
\usepackage[T1]{fontenc} 
\usepackage{epsf} 
\usepackage{cancel}

\usepackage{graphics} 
\usepackage{amsfonts,amsmath}
\usepackage[round]{natbib}
\usepackage{psfrag,xspace}
\usepackage{color,etoolbox}
\usepackage{subcaption} 
\usepackage{listings}
\usepackage{amsmath}
\usepackage{needspace}
\usepackage{threeparttable}
\usepgfplotslibrary{groupplots}
\usepackage[noabbrev,capitalize, nameinlink]{cleveref}
\newcommand{\Pstar}[1]{P^{*}_{#1}}

\usepgfplotslibrary{fillbetween}

\newcommand{\Pp}{\mathbb P}

\usepackage{pgfplots}
\pgfmathdeclarefunction{poiss}{1}{%
  \pgfmathparse{(#1^x)*exp(-#1)/(x!)}%
}

\DeclareMathOperator*{\argmin}{arg\,min}

\DeclareSymbolFont{bbold}{U}{bbold}{m}{n}
\DeclareSymbolFontAlphabet{\mathbbold}{bbold}

\newtheorem{theorem}{Theorem}
\newtheorem{lemma}{Lemma}
\newtheorem{corollary}{Corollary}
\newtheorem{proposition}{Proposition}

\theoremstyle{definition}
\newtheorem{definition}{Definition}

\theoremstyle{remark}

\newtheorem{remark}{Remark}

\let\hat\widehat
\definecolor{dkgreen}{rgb}{0,0.6,0}
\definecolor{gray}{rgb}{0.5,0.5,0.5}
\definecolor{mauve}{rgb}{0.58,0,0.82}
\definecolor{brightblue}{HTML}{00BFC4}

\definecolor{oiBlue}{HTML}{0072B2}
\definecolor{oiVermillion}{HTML}{D55E00}
\definecolor{oiGreen}{HTML}{009E73}
\definecolor{oiPurple}{HTML}{CC79A7}

\definecolor{emailpink}{RGB}{230,0,130}
\definecolor{emailblue}{HTML}{365F7D}

\newcommand\blfootnote[1]{%
  \begingroup
  \renewcommand\thefootnote{}%
  \footnote{#1}%
  \addtocounter{footnote}{-1}%
  \endgroup
}

\pgfplotsset{compat=1.18}

\begin{document}

\def\spacingset#1{\renewcommand{\baselinestretch}%
{#1}\small\normalsize} \spacingset{1}

\raggedbottom
\allowdisplaybreaks[1]

\begingroup
\hypersetup{colorlinks=true, urlcolor=blue} 

\begin{center}

{\LARGE Parallel Trends as a Spectrum\par}

\vspace{0.85cm}

{\normalsize
\begin{tabular*}{\linewidth}{@{\extracolsep{\fill}}cccc@{}}
Kyle Schindl$^\S$
& Joseph Fry$^\P$
& Florian Gunsilius$^\star$
& Larry Wasserman$^{\dagger\ddagger}$
\\[0.25em]
{\footnotesize
  \href{mailto:kschindl@iastate.edu}
       {kschindl@iastate.edu}}
&
{\footnotesize
  \href{mailto:joseph.fry@rutgers.edu}
       {joseph.fry@rutgers.edu}}
&
{\footnotesize
  \href{mailto:fgunsil@emory.edu}
       {fgunsil@emory.edu}}
&
{\footnotesize
  \href{mailto:larry@stat.cmu.edu}
       {larry@stat.cmu.edu}}
\end{tabular*}\par
}

\vspace{0.60cm}

{\small
\begin{tabular}{@{}r@{\hspace{0.15em}}l@{}}
$^\S$ &
Department of Statistics, Iowa State University\\
$^\P$ &
Department of Economics, Rutgers University\\
$^\star$ &
Department of Economics, Emory University\\
$^\dagger$ &
Department of Statistics \& Data Science, Carnegie Mellon University\\
$^\ddagger$ &
Department of Machine Learning, Carnegie Mellon University
\end{tabular}\par
}

\end{center}
\endgroup

  \blfootnote{Accompanying \texttt{R} code is available via \href{https://github.com/JosephPatrickFry/L-moments-DiD}{https://github.com/JosephPatrickFry/L-moments-DiD}.}
  \thispagestyle{empty}

\begin{abstract}
Difference-in-differences (DiD) designs are typically used to identify and estimate average causal effects, even though treatment can affect the entire outcome distribution. This has led to a growing interest in distributional DiD methods. However, identifying the full counterfactual distribution requires a parallel trends assumption on the entire distributional dynamics, which can be too strong in practice. We address this problem by introducing a spectrum of identifying restrictions that lie \textit{between} the average and fully distributional DiD by imposing parallel trends on $L$-moments. This corresponds to a basis expansion of the distributional dynamics. Importantly, each $L$-moment is interpretable as familiar distributional components: location, dispersion, skewness, kurtosis, etc. Researchers can pick and choose which features they want to target. For example, the second $L$-moment identifies the Gini mean difference, but does not require imposing mean-level or higher-order parallel trends. Taking a collection of $L$-moments can approximate distributional dynamics under weaker conditions since it leaves untargeted components unrestricted. We develop conditions under which higher-order $L$-moments decay, efficient estimators, and simultaneous inference across growing collections of moments. We demonstrate the benefits of our framework by exploring distributional changes to healthcare spending after the rollout of Medicare Part D.

\end{abstract}


\bigskip 

\spacingset{1.05}
\section{Introduction}

The main obstacle to causal inference with observational data is unmeasured confounding. When unobserved variables simultaneously influence treatment and outcomes, standard identification strategies based on conditioning or weighting give biased results of the actual causal effect \citep{Imbens_Rubin_2015}. Among the myriad of proposed methods to address this, the difference-in-differences framework has become a central tool for practitioners \citep{ROTH20232218}. In this framework, average treatment effects on the treated are identified by comparing changes in outcomes for treated and untreated groups under a parallel trends assumption \citep{ashenfelter1985using,meyer95,abadie2005semiparametric}. This mean-based formulation is useful, but many scientific and policy questions are inherently distributional, since a treatment may leave the mean nearly unchanged while altering other components of the distribution, such as its variance, skewness, or other higher-order effects \citep{bitler2006, chernozhukov2013}. In such settings, the average effect is not necessarily misleading, but it is an incomplete description of the full causal effect.

A growing literature has extended difference-in-differences beyond average treatment effects under a range of identifying assumptions. \cite{athey_imbens_2006} identify the full counterfactual distribution of the treated group under a nonparametric structural model, whereas \citet{BonhommeSauder2011} pursue identification through deconvolution in an additive model with unobserved heterogeneity. \citet{CALLAWAY2018395} and \citet{callaway_qte2019} combine restrictions on the distribution of untreated outcome changes by assuming copula invariance across groups and copula stability over time, respectively. Recently, \citet{KimWooldridge2025} identify quantile treatment effects under parallel changes in the distribution functions of untreated potential outcomes, while \citet{FernandezValEtAl2026} develop a distribution-regression approach based on parallel trends in transformed distribution functions. These approaches illustrate that ``distributional parallel trends'' can be formulated in several nonequivalent ways. Related work extends the scope of distributional difference-in-differences to more general treatment timing and outcome spaces. For example, \citet{LI2024111792} and \citet{ciaccio_did2024} extend copula-based identification of quantile treatment effects to settings with staggered adoption. There have been geometric generalizations of difference-in-differences, such as \cite{torous2024} who consider optimal transport maps between distributions and \cite{zhou2025geodesicdifferenceindifferences} who study outcomes in general metric spaces.

Each of these distributional difference-in-differences frameworks requires strong identification assumptions about the counterfactual distribution that may be unrealistic \citep{roth_functional_form2023}. Furthermore, these strong assumptions may not even be required to identify the causal effects that practitioners are actually interested in. Rather than trying to identify the entire counterfactual distribution, we identify a finite-dimensional projection of the quantile function onto the shifted Legendre polynomials, i.e., onto the $L$-moments of a distribution \citep{hosking1990}. Importantly, $L$-moments have interpretable connections to familiar distributional features. The first $L$-moment is the mean, the second captures dispersion, the third captures asymmetry, and higher-order moments capture increasingly refined tail and shape features. Combinations of $L$-moments also identify measures such as the Gini coefficient and can be used to construct lower bounds on the variance of individual treatment effects, without requiring rank invariance. Thus, our framework naturally complements existing work on partial identification \citep{athey_imbens_2006, FanYu2012} as well as works that treat functionals of the outcome distribution as causal targets \citep{Rothe2010, FirpoPinto2016, AtheyEtAl2023}. Imposing parallel trends on the first $L$-moment recovers conventional mean parallel trends and imposing restrictions on every $L$-moment is equivalent to parallel trends at the distribution level. However, these restrictions do not need to be imposed on consecutive moments; researchers may identify effects on selected features of the distribution even when parallel trends fails for others.

The remainder of this paper is organized as follows: in \cref{missing_middle} we introduce the parallel trends spectrum. Specifically, in \cref{mean_quantile_pt_section} we review mean and distributional parallel trends restrictions, in \cref{pt_spectrum_section} we introduce $L$-moment parallel trends under additive quantile parallel trends and the changes-in-changes framework, in \cref{higher_order_effects_section} we explore the new higher-order treatment effects that can be identified under our framework, and in \cref{decay_section} we establish the conditions under which higher-order $L$-moments decay; we show shared distributional features can cancel in difference-in-differences contrasts, allowing faster approximation than would be possible for the individual quantile functions. In \cref{estimation_inference} we establish estimation and inference for $L$-moments: in \cref{efficiency_theory_section} we derive the efficient influence function and nonparametric efficiency bound, in \cref{estimator_section} we explore tradeoffs between using the unbiased $U$-statistic estimator of \cite{hosking1990} versus the one-step estimator, in \cref{jackknife-bootstrap} we establish simultaneous inference across growing collections of $L$-moments, and in \cref{did_estimation} we extend our estimator to several difference-in-differences settings, including repeated cross sections, panel data, and staggered adoption. In \cref{choosing_k_section} we discuss decision rules for selecting $K$ in practice, along with a simulation study in \cref{risk_appendix}. In \cref{simulations_and_application} we include an applied data example, where we study higher-order changes to healthcare spending after the rollout of Medicare Part D. Finally, in \cref{discussion_and_conclusion} we conclude our analysis and discuss directions for future work.

\section{The Missing Middle} \label{missing_middle}

Mean-level parallel trends identify the untreated counterfactual mean for the treated group, and thus the average treatment effect on the treated. However, it cannot describe how treatment changes other features of the outcome distribution. On the other hand, parallel trends assumptions at the distribution level, such as for the cumulative distribution function or quantile function, identify the full marginal counterfactual distribution but are often too restrictive to be used in practice \citep{roth_functional_form2023}. In this section, we construct an intermediate regime of parallel trends assumptions indexed by $L$-moment coordinates \citep{hosking1990}. This hierarchy begins with mean parallel trends and progressively identifies richer features of the counterfactual distribution until, at the full-spectrum limit, it recovers distributional parallel trends.

\subsection{Mean and Quantile Parallel Trends} \label{mean_quantile_pt_section}

To begin, we consider the canonical two-group, two-period difference-in-differences design \citep{ashenfelter1985using, abadie2005semiparametric}. Let $G\in\{0,1\}$ indicate group membership, where group $1$ receives treatment in period $t = 1$, group $0$ is never treated, and $\mathbb{P}(G = g) > 0$ for $g \in \{0, 1\}$. For
$d\in\{0,1\}$, let $Y_t(d)$ denote the potential outcome under treatment regime $d$, where $d=0$ denotes remaining untreated and $d=1$ denotes treatment beginning in period $1$. We assume consistency holds, such that the observed outcomes satisfy $Y_t=Y_t(G)$, and that treatment has no anticipatory effect, so that $Y_0(1)=Y_0(0)$ almost surely. Furthermore, unless otherwise stated, we assume all relevant potential outcomes have finite absolute first moments. Let $F^d_{gt}(y) = \mathbb{P}(Y_t(d)\leq y\mid G=g)$ denote the counterfactual cumulative distribution function. Then, for $u \in (0, 1)$ we define
\begin{align*}
    Q_{gt}^d(u) =\inf\{y: F^d_{gt}(y) \geq u\}
\end{align*}
to be the quantile function of the distribution of the potential outcomes. Under consistency and no anticipation, $Q_{g0}^0=Q_{g0}^1=Q_{g0}$ for $g\in\{0,1\}$, $Q_{01}^0=Q_{01}$, and $Q_{11}^1=Q_{11}$. Among the distributions needed to identify treatment effects for the treated group, only $Q_{11}^0$ remains unobserved. For each group, we define the untreated quantile displacement to be $\Delta_g^0(u)=Q_{g1}^0(u)-Q_{g0}^0(u)$. This records the change over time in the $u$th quantile of the untreated marginal distribution. Because the control group remains untreated, $\Delta_0^0(u)$ is observed. By contrast, $\Delta_1^0(u)$ is the counterfactual trend that the treated group would have experienced at rank $u$, absent treatment. The difference between these two trends is the untreated quantile-trend defect,
\begin{align*}
     \mathcal{D}^0(u) &=\Delta_1^0(u)-\Delta_0^0(u) =\left\{Q_{11}^0(u)-Q_{10}^0(u)\right\} -\left\{Q_{01}^0(u)-Q_{00}^0(u)\right\}.
\end{align*}
For each $u \in (0, 1)$, $\mathcal{D}^0(u)$ represents the failure of the control group to match the evolution the treated group would have experienced without treatment. Notably, ordinary parallel trends at the mean-level (M-PT) requires
\begin{align} \label{mean_pt}
    \mathbb{E}[Y_1(0)-Y_0(0)\mid G=1] = \mathbb{E}[Y_1(0)-Y_0(0)\mid G=0].
\end{align}
Equivalently, \cref{mean_pt} is the same as requiring $\int_0^1 \mathcal{D}^0(u) du=0$. This identifies the average treatment effect on the treated, $ \tau_{\mathrm{ATT}} = \mathbb{E}[Y_1(1)-Y_1(0)\mid G=1]$. On the other hand, to identify the marginal quantile treatment effect on the treated, one can impose an \textit{additive} quantile parallel trends (Q-PT) assumption by requiring $\mathcal{D}^0(u)=0$ for Lebesgue-almost every $u\in(0,1)$. Note that this additive quantile difference-in-differences restriction is distinct from the changes-in-changes (CiC) model of \cite{athey_imbens_2006}. We explore CiC further in \cref{cic_section}. Under Q-PT, the untreated quantile at almost every rank $u$ must undergo the same displacement across groups and thus the missing counterfactual quantile function is identified as
\begin{align} \label{quantile_counterfactual}
    Q^0_{11}(u)=Q_{10}(u)+Q_{01}(u)-Q_{00}(u).
\end{align}
Because the right-hand side of \cref{quantile_counterfactual} must itself be a valid quantile, Q-PT imposes a nontrivial shape restriction that can often be unrealistic. At first, it appears we are at an impasse: M-PT, while weaker, doesn't adequately describe the treatment effect. Meanwhile, Q-PT, which fully describes the distributional effect, requires an implausibly strong restriction on the counterfactual distribution. However, it is possible to describe the world \textit{between} these restrictions by projecting $\mathcal{D}^0(u)$ onto an orthogonal basis of polynomials, as we show in the next section.

\subsection{The Parallel Trends Spectrum} \label{pt_spectrum_section}

Our goal in this section is to define a nested spectrum of parallel trends assumptions that interpolate between M-PT and Q-PT. To do so, we show how $\mathcal{D}^0(u)$ can be decomposed into differences in $L$-moments. First introduced by \cite{hosking1990}, $L$-moments represent a projection of the quantile function onto the basis formed by the shifted Legendre polynomials. Specifically, the shifted Legendre polynomials are defined by $P_k^*(u)=P_k(2u-1)$, where $P_k$ is the ordinary Legendre polynomial of degree $k$. They have the closed-form expression
\begin{align*}
   P^*_{k}(x) = (-1)^k\sum_{j=0}^{k}\binom{k}{j}\binom{k+j}{j}(-x)^{j}, 
\end{align*}
and in particular, $P^*_{0}(x)=1$. The shifted Legendre polynomials form an orthogonal basis for $L^2(0,1)$, where for $1 \leq p < \infty$, we let $L^p(0,1)$ denote the space of Lebesgue-measurable functions $f:(0,1)\to\mathbb{R}$ satisfying
\begin{align*}
    \|f\|_{L^p} = \left\{ \int_0^1|f(u)|^p du \right\}^{1/p} <\infty.
\end{align*}
Following \cite{hosking1990}, for any random variable $X$ with a finite first moment and quantile function $Q$, the $k$th $L$-moment is defined as $\lambda_k(Q) = \int_0^1 Q(u) P^*_{k-1}(u)du$. Thus, $L$-moments provide a robust analogue of conventional moments that are less sensitive to heavy tails and are always well-defined as long as the first moment of $X$ exists. In particular, note that
\begin{align*}
    \int_0^1|Q(u)|^p du = \mathbb{E}[|X|^p]
\end{align*}
whenever either side is finite; thus, we have the equivalences $Q\in L^1(0,1) \Longleftrightarrow \mathbb{E}[|X|] <\infty$ and $Q\in L^2(0,1) \Longleftrightarrow \mathbb{E}[X^2]<\infty$. Throughout most of the text we only require finite first moments; however, for some results (such as reconstruction of the quantile function) we require finite second moments. We discuss regularity conditions of $L$-moments and quantile functions in more detail in \cref{regularity_conditions}.

\begin{figure}[h]
    \centering
    \begin{tikzpicture}
    \begin{axis}[
        width=9.4cm,
        height=6.5cm,
        xmin=0, xmax=1,
        ymin=-1.1, ymax=1.1,
        domain=0:1,
        samples=400,
        axis lines=box,
        axis line style={black, line width=0.5pt},
        xlabel={$u$},
        ylabel={$P_{k-1}^*(u)$},
        xtick={0,0.2,0.4,0.6,0.8,1},
        ytick={-1,-0.5,0,0.5,1},
        ticklabel style={font=\small},
        label style={font=\small},
            xtick pos=bottom,
    ytick pos=left,
            grid=both,
    grid style={line width=0.1pt, draw=gray!20, opacity=0.5},
        minor tick num=1,
        axis background/.style={fill=white},
        legend cell align=left,
        legend style={
            at={(1.03,1)},
            anchor=north west,
            draw=black,
            fill=white,
            font=\small,
            inner xsep=5pt,
            inner ysep=5pt
        },
        legend image post style={line width=1.2pt},
        clip=false
    ]

    \addplot[thick, oiBlue] {2*x - 1};
    \addlegendentry{$\lambda_2$: Scale}

    \addplot[thick, oiVermillion, dashed] {6*x^2 - 6*x + 1};
    \addlegendentry{$\lambda_3$: Skewness}

    \addplot[thick, oiGreen, dotted] {20*x^3 - 30*x^2 + 12*x - 1};
    \addlegendentry{$\lambda_4$: Kurtosis}

    \addplot[thick, oiPurple, dashdotted] {70*x^4 - 140*x^3 + 90*x^2 - 20*x + 1};
    \addlegendentry{$\lambda_5$: Higher-order asymmetry}

    \end{axis}
    \end{tikzpicture}
    \caption{Shifted Legendre weight functions.}
    \label{fig:lmoment_weights}
\end{figure}
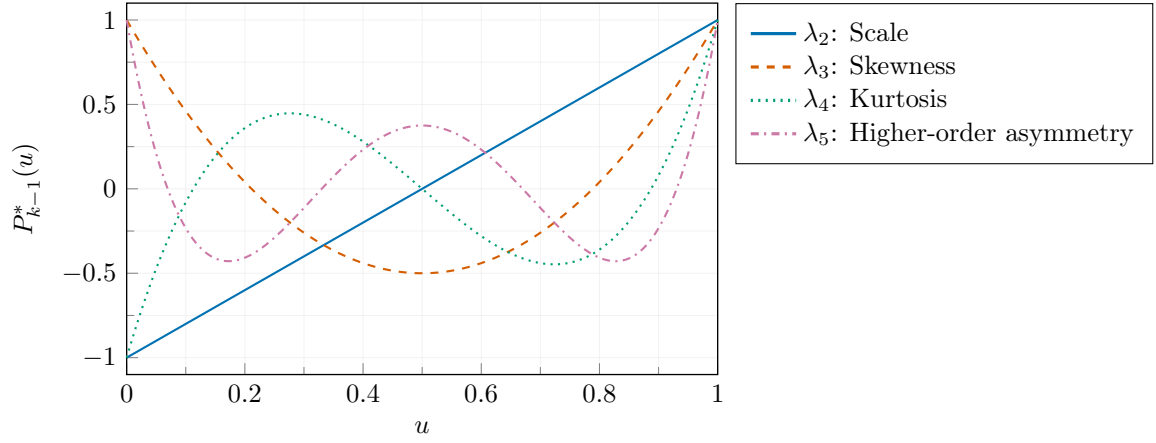

A major benefit $L$-moments provide is their simple and clean interpretation. This becomes clear when using the $L$-moment representation as an expectation of a linear combination of order statistics. If $X_{1:n}\leq\cdots\leq X_{n:n}$ are the order statistics from $n$ independent and identically distributed draws, then the first three $L$-moments are given by
\begin{align*}
    \lambda_1(Q) &=\mathbb{E}[X],\\
    \lambda_2(Q) &=\frac{1}{2}\mathbb{E}[X_{2:2}-X_{1:2}],\\
    \lambda_3(Q) &=\frac{1}{3}\mathbb{E}[( X_{3:3}-X_{2:3}) +(X_{1:3} -X_{2:3})].
\end{align*}
Thus, the first $L$-moment $\lambda_1$ is the mean.  The second $L$-moment $\lambda_2$ provides an alternative measure of dispersion to the standard deviation, the third $L$-moment $\lambda_3$ gives an alternative measure of asymmetry to the conventional skewness, and higher-order $L$-moments similarly capture finer shape features such as tail weight (analogous to kurtosis) and beyond; the increasing frequency of these moments is visualized in \cref{fig:lmoment_weights}. This interpretable spectrum is critically important for developing a hierarchy of parallel trends assumptions: there are many bases or approximation systems we could consider for decomposing the quantile function (spline bases, wavelet systems, etc.), but each component wouldn't be scientifically \textit{meaningful}. By contrast, each $L$-moment describes an intuitive shape of the overall distribution.

Because $\|\Pstar{k-1}\|_{L^2}^2 = (2k-1)^{-1}$, the shifted Legendre polynomials are orthogonal on $[0,1]$ but not orthonormal.  As a consequence, $L$-moments have a systematic scale that shrinks with $k$ purely because of the normalization of the basis. It will be convenient throughout the rest of the paper to work with the \emph{normalized} shifted Legendre polynomials $\phi_{k}(u) = \sqrt{2k-1}\Pstar{k-1}(u)$ so that $\|\phi_{k}\|_{L^2} = 1$. We define the normalized $L$-moments to be
\begin{align*}
    \lambda^*_k(Q) = \int_0^1 Q(u) \phi_{k}(u)du = \sqrt{2k - 1} \lambda_k(Q).
\end{align*}
The normalized $L$-moments are the Fourier coefficients of the quantile function in the orthonormal basis $\{\phi_{k}\}^{\infty}_{k = 1}$ of $L^2(0,1)$. With these definitions in place, we can now decompose the defect $ \mathcal{D}^0(u)$ and define a corresponding parallel trends assumption.

\begin{definition}[Additive $L$-Moment Parallel Trends] \label{lpt_definition} Let
\begin{align*}
    d^0_k = \int_0^1\mathcal{D}^0(u)\phi_k(u)du = \left\{\lambda^*_k(Q^0_{11}) -\lambda^*_k(Q^0_{10})\right\} -\left\{\lambda^*_k(Q^0_{01}) -\lambda^*_k(Q^0_{00})\right\}
\end{align*}
denote the quantile defect of the $k$th $L$-moment. Then, for a set $\mathcal{K} \subseteq \mathbb{N}$, we say $L$-moment parallel trends hold over $\mathcal{K}$ if $d^0_k = 0$ for every $k \in \mathcal{K}$. We denote this $\mathcal{L}(\mathcal{K})$. 
\end{definition}

In the particular case that $\mathcal{K} = \{1, 2, \ldots, K\}$ for some integer $K$, we denote the collective parallel trends assumption as $\mathcal{L}(1\mathord{:}K)$. Notably, $\mathcal{L}(1\mathord{:}K)$ defines a nested hierarchy of parallel trends assumptions, which we formalize in the following lemma.
\begin{lemma}[The Parallel Trends Spectrum]
\label{parallel_trends_spectrum_corollary}
Suppose $\mathcal{D}^0\in L^1(0,1)$. Then:
\begin{enumerate}
    \item $\mathcal{L}(\{1\})$ is equivalent to M-PT.
    \item $\mathcal{L}(1\mathord{:}K+1)$ implies $\mathcal{L}(1\mathord{:}K)$ for every $K\geq1$.
    \item $\mathcal{L}(\mathbb{N})$ is equivalent to Q-PT.
\end{enumerate}
\end{lemma}
By \cref{parallel_trends_spectrum_corollary} we formally recover the ``missing middle'' between M-PT and Q-PT. For some $K \geq 1$ we obtain a nested hierarchy of assumptions, such that $\text{Q-PT} \implies \mathcal{L}(1\mathord{:}K) \implies  \text{M-PT}$. From here, we establish causal identification. To begin, define the marginal quantile treatment effect on the treated by
\begin{align*}
    \tau_{\mathrm{QTT}}(u)=Q^1_{11}(u)-Q^0_{11}(u).
\end{align*}
Note that this is simply a difference between marginal potential-outcome quantiles; we do not assume any kind of rank-invariance assumptions now or elsewhere, and do not interpret these contrasts as individual-level treatment effects. Then, the $k$th $L$-moment coefficient is defined by
\begin{align*}
      \theta_k =\int_0^1\tau_{\mathrm{QTT}}(u)\phi_{k}(u)du = \lambda^*_k(Q^1_{11}) - \lambda^*_k(Q^0_{11}).
\end{align*}
Again, $\theta_k$ should be interpreted as a contrast between marginal outcome distributions, not as an individual-level effect. In the following theorem we establish formal identification.

\begin{theorem}[Identification Under $L$-Moment Parallel Trends] \label{identification_theorem} Suppose all potential outcome distributions have finite absolute first moments and that $\mathcal{L}(\mathcal{K})$ holds for some $\mathcal{K} \subseteq \mathbb{N}$. Suppose further that consistency holds, such that $Y_t = Y_t(G)$ for $t \in \{0, 1\}$, and that treatment has no anticipatory effect, such that $Y_0(1) = Y_0(0)$ almost surely. Then, for every $k \in \mathcal{K}$,
\begin{align} \label{additive_counterfactual}
     \theta_k = \lambda^*_k(Q_{11}) - \lambda^*_k(Q_{10}) -\lambda^*_k(Q_{01}) + \lambda^*_k(Q_{00}).
\end{align}
In particular, if $1 \in \mathcal{K}$, then $\theta_1 = \mathbb{E}\!\left[Y_1(1)-Y_1(0) \mid G=1 \right]$ is the average treatment effect on the treated. If $\mathcal{L}(\mathbb{N})$ holds, then the identified sequence $\{\theta_k:k\geq1\}$ uniquely determines $\tau_{\mathrm{QTT}}$ up to Lebesgue-almost-everywhere equality among functions in $L^1(0,1)$.
\end{theorem}

One important takeaway of \cref{identification_theorem} is that identification is component-wise. That is, to identify $\theta_k$ we only require parallel trends on its corresponding $L$-moments; we make no restriction on any of the remaining distributional components. By \cref{identification_theorem} we also obtain a very clean progression: $\mathcal{L}(\{k\}) \Rightarrow \theta_k$ is identified, $\mathcal{L}(1\mathord{:}K) \Rightarrow (\theta_1, \ldots, \theta_K)$ is identified, and $\mathcal{L}(\mathbb{N}) \Rightarrow \tau_{\mathrm{QTT}}$ is identified. Moreover, each of these results only requires finite absolute first moments. Importantly, these identification results hold only for the additive quantile parallel trends assumption; in the next section, we formalize an analogous decomposition for the changes-in-changes framework.

\subsubsection{Changes-in-Changes in $L$-Moment Coordinates} \label{cic_section}

The $L$-moment parallel trends assumptions of \cref{lpt_definition} should be seen as component-wise relaxations of \textit{additive} quantile difference-in-differences and, in general, are not the same as the changes-in-changes model of \cite{athey_imbens_2006}. Under CiC, untreated outcomes must satisfy $Y_t(0) = h_t(U)$ where $h_t$ is strictly increasing and $U\perp T\mid G$. Under this model, the counterfactual quantile for the treated group is constructed as
\begin{align*}
    Q_{11}^{0}(u) = Q_{01}\!\left(F_{00}\!\left(Q_{10}(u)\right)\right).
\end{align*}
Thus, instead of defining the defect as $\mathcal{D}^0(u) =\{Q_{11}^0(u)-Q_{10}^0(u)\} -\{Q_{01}^0(u)-Q_{00}^0(u)\}$, as in the additive case, the CiC-based defect is defined as
\begin{align*}
    \mathcal{D}^0_{\mathrm{CiC}}(u) =   Q_{11}^{0}(u) - Q_{01}\!\left(F_{00}\!\left(Q_{10}(u)\right)\right).
\end{align*}
Then, the CiC equivalent of parallel trends is to assume that $\mathcal{D}^0_{\mathrm{CiC}}(u) = 0$ for Lebesgue-almost every $u\in(0,1)$. With this definition in place, we can establish an analogous decomposition of $\mathcal{D}^0_{\mathrm{CiC}}(u)$ onto the shifted Legendre polynomials.

\begin{definition}[Changes-in-Changes $L$-Moment Restrictions] \label{cic_pt} 
     Let
\begin{align*}
    d^0_{k, \mathrm{CiC}} = \int_0^1 \mathcal{D}^0_{\mathrm{CiC}}(u)\phi_k(u)du = \lambda^*_k(Q^0_{11}) - \int_0^1 Q_{01}\!\left(F_{00}\!\left(Q_{10}(u)\right)\right) \phi_k(u)du
\end{align*}
denote the CiC-quantile defect of the $k$th $L$-moment. Then, for a set $\mathcal{K} \subseteq \mathbb{N}$, we say the CiC $L$-moment restriction holds over $\mathcal{K}$ if $d^0_{k, \mathrm{CiC}} = 0$ for every $k \in \mathcal{K}$. We denote this $\mathcal{L}_{\mathrm{CiC}}(\mathcal{K})$. 
\end{definition}

Note that in \cref{cic_pt} we refer to $d^0_{k, \mathrm{CiC}} = 0$ as a \textit{restriction} (as opposed to a parallel trends assumption) since CiC uses a nonlinear transport and not an additive trend. Notably, $\mathcal{L}_{\mathrm{CiC}}(\mathcal{K})$ does \textit{not} require the same latent-variable representation of CiC, and it does not impose a restriction on coordinates of $Q_{11}^{0}(u)$ outside of $\mathcal{K}$. For example, $\mathcal{L}_{\mathrm{CiC}}(\{1\})$ imposes only that
\begin{align*}
    \mathbb{E}[Y_1(0) \mid G = 1] = \int^1_0  Q_{01}\!\left(F_{00}\!\left(Q_{10}(u)\right)\right) du
\end{align*}
since $\phi_1(u) = 1$. Thus, we are able to identify the ATT using the mean of the CiC-imputed counterfactual without specifying that the rest of the distribution be correctly specified. We formalize this idea in the following theorem.

\begin{theorem}[Identification Under the Changes-in-Changes $L$-Moment Spectrum] \label{identification_cic}

Suppose all potential outcome distributions have finite absolute first moments and that the consistency and no anticipation assumptions outlined in \cref{identification_theorem} hold. Next, for $u \in (0, 1)$ define the CiC imputation as
\begin{align*}
    Q_{\mathrm{CiC}}(u) = Q_{01}\!\left( F_{00}\!\left(Q_{10}(u)\right) \right).
\end{align*}
Finally, assume $Q_{\mathrm{CiC}} \in L^1(0, 1)$, $ 0 < F_{00}(Q_{10}(u)) < 1$ for Lebesgue-almost every $u \in (0, 1)$, and $\operatorname{supp}(Y_0\mid G=1) \subseteq \operatorname{supp}(Y_0\mid G=0)$. Then, the following statements hold:
\begin{enumerate}
    \item[1.] If $\mathcal{L}_{\mathrm{CiC}}(\mathcal{K})$ holds for some $\mathcal{K} \subseteq \mathbb{N}$, then, for every $k\in\mathcal{K}$,
    \begin{align} \label{cic_counterfactual}
         \theta_k = \lambda_k^*(Q_{11}) - \int^1_0 Q_{\mathrm{CiC}}(u) \phi_k(u) du. 
    \end{align}
    In particular, if $1\in\mathcal{K}$, then
    \begin{align*}
        \theta_1 = \mathbb{E}[Y_1\mid G=1] - \int_0^1 Q_{\mathrm{CiC}}(u)du = \mathbb{E}\!\left[ Y_1(1)-Y_1(0) \mid G=1 \right].
    \end{align*}
    \item[2.] For every $K \geq 1$, $\mathcal{L}_{\mathrm{CiC}}(1\mathord{:}K+1) \Rightarrow \mathcal{L}_{\mathrm{CiC}}(1\mathord{:}K)$.
    \item[3.] $\mathcal{L}_{\mathrm{CiC}}(\mathbb{N})$ holds if and only if $\mathcal{D}^0_{\mathrm{CiC}}(u) = 0$ for Lebesgue-almost every $u\in(0,1)$. In particular, under these conditions, $Q_{11}^0(u) = Q_{\mathrm{CiC}}(u)$ and $\tau_{\mathrm{QTT}}(u) = Q_{11}(u) - Q_{\mathrm{CiC}}(u)$ for Lebesgue-almost every $u\in(0,1)$.
\end{enumerate}
\end{theorem}

The structural model imposed by the changes-in-changes framework requires a few additional regularity assumptions for identification not present in \cref{identification_theorem}. Specifically, we require the support condition $\operatorname{supp}(Y_0\mid G=1) \subseteq \operatorname{supp}(Y_0\mid G=0)$; this ensures that the transformation $Q_{01}\circ F_{00}$ is applied only at outcome levels
represented in the control group. Conceptually, it is an analogue of the latent-support restriction of \citet[Assumption~3.4]{athey_imbens_2006}, which we impose directly without actually requiring $Y_t(0) = h_t(U)$. Separately, we require $0 < F_{00}(Q_{10}(u)) < 1$ to ensure that the outer quantile function $Q_{01}$ is evaluated within its domain $(0, 1)$ in order to avoid potentially undefined endpoint values.

There are several important implications of \cref{identification_cic}. First, analogous to \cref{identification_theorem}, under the CiC $L$-moment spectrum we obtain a clear progression of identified components: $\mathcal{L}_{\mathrm{CiC}}(\{k\}) \Rightarrow \theta_k$ is identified, $ \mathcal{L}_{\mathrm{CiC}}(1\mathord{:}K) \Rightarrow (\theta_1,\ldots,\theta_K)$ is identified, and $\mathcal{L}_{\mathrm{CiC}}(\mathbb{N}) \Rightarrow \tau_{\mathrm{QTT}}$ is identified up to almost-everywhere equality. Second, as noted earlier, the finite-dimensional restrictions imposed by $\mathcal{L}_{\mathrm{CiC}}(\mathcal{K})$ are strictly weaker than the structural model outlined in \cite{athey_imbens_2006}. For finite $\mathcal{K}$, we impose
\begin{align*}
    \int_0^1 Q_{11}^0(u)\phi_k(u)du = \int_0^1 Q_{\mathrm{CiC}}(u)\phi_k(u)du
\end{align*}
only for $k \in \mathcal{K}$. It neither requires the latent-variable representation $Y_t(0)=h_t(U)$ nor implies that $Q_{11}^0=Q_{\mathrm{CiC}}$ almost everywhere. Finally, we can see that the additive quantile parallel trends and changes-in-changes spectra impose substantively different restrictions. Under the additive parallel trends assumption of \cref{lpt_definition}, the condition $\mathcal{L}(\{k\})$ requires the $k$th normalized $L$-moment of the missing counterfactual distribution to satisfy
\begin{align*}
    \lambda_k^*(Q_{11}^0) = \lambda_k^*(Q_{10}) +\lambda_k^*(Q_{01}) -\lambda_k^*(Q_{00}).
\end{align*}
Notably, this restriction is coordinate-separable; identification of the $k$th effect depends only on the $k$th $L$-moment of each observed distribution. By contrast, under the CiC restriction of \cref{cic_pt}, $\mathcal{L}_{\mathrm{CiC}}(\{k\})$ requires
\begin{align*}
    \lambda_k^*(Q_{11}^0) = \int_0^1 Q_{01}\!\left(F_{00}\!\left(Q_{10}(u)\right)\right) \phi_k(u)du.
\end{align*}
This imputation is not coordinate-separable: its value depends on the full distributions $F_{00}$, $F_{01}$, and $F_{10}$ through the nonlinear composition $Q_{01}\circ F_{00}\circ Q_{10}$. Thus, the first $K$ restrictions in the two spectra do not produce two parameterizations of the same identifying assumption; they represent two different rules for extrapolation. We provide a longer discussion on the differences between these models in \cref{additive_vs_cic}.

Importantly, the difference between the extrapolation rules defined in \cref{additive_counterfactual} and \cref{cic_counterfactual} is an \textit{observed} quantity. Consequently, an interesting direction for future work would be to develop a testing framework for choosing, or combining, the predictions given by each of these models. However, throughout the rest of this text we assume that the practitioner has already chosen an extrapolation method. Next, we explore and define the causal effects that can be identified and estimated under $L$-moment-based parallel trends.

\subsection{Higher-Order Treatment Effects} \label{higher_order_effects_section}
In this section, we explore the suite of higher-order causal effects identified by the parallel trends assumption based on $L$-moments. There are two primary causal applications of this idea: component-wise treatment effects and distributional reconstruction. We explore both of these further in this section. Both applications connect the framework to the literature on distributional policy effects \citep[e.g.][]{Rothe2010} and inequality treatment effects \citep[e.g.][]{FirpoPinto2016}. The identifying restrictions can be chosen to match the specific features of interest in the distribution (such as the dispersion, mean, etc.), without requiring identification of the entire untreated distribution. Throughout this section, the derivations apply to both the additive restrictions of \cref{identification_theorem} and the CiC restrictions of \cref{identification_cic}, with their respective identifying conditions.

\subsubsection{Component-Wise Effects}
The first and most direct application of the $L$-moment parallel trends assumption is to focus on a single component, or targeted combinations of components. Of course, $\mathcal{K} = \{1\}$ simply identifies the ATT; however, other subsets of $\{1, 2, 3, 4\}$ can also be natural targets, which we now explore. In the following, for $X \sim F$ with quantile function $Q$, we write $\lambda_k(F)=\lambda_k(Q)$ and similarly for $\lambda_k^*$.

\paragraph{The Second $L$-Moment.}
A noteworthy case is $\mathcal{K}=\{2\}$ because the second $L$-moment is proportional to the Gini mean difference \citep{Yitzhaki82}. To be more precise, for independent draws $X,X^\prime\sim F$, the Gini mean difference is defined as $\mathrm{GMD}(F)=\mathbb{E}[|X-X^\prime|]$. To derive its $L$-moment representation, write $X=Q(U)$ and $X^\prime=Q(V)$ for independent uniform random variables $U,V$ on $(0,1)$. Since $Q$ is nondecreasing, symmetry and Fubini's theorem directly yield
\begin{align*}
    \mathrm{GMD}(F) = 2\int_0^1\! \! \int_0^u( Q(u)-Q(v)) dv\,du  =2\int_0^1Q(u)(2u-1)du= 2\lambda_2(F).
\end{align*}
This derivation only requires finite first moments and holds for distributions with atoms.

In the same vein, the Gini coefficient is a standard measure of relative inequality \citep{Shorrocks1982}. For nonnegative outcomes with a strictly positive mean $\mu =\mathbb{E}[X]$, it is defined as $G(F)=\mathbb{E}[|X-X^\prime|]/(2\mu)$. Equivalently, it is twice the area between the Lorenz curve and the line of perfect equality \citep{Gastwirth1971}. To see this, let $L(p)=\mu^{-1}\int_0^p Q(u)du$ denote the Lorenz curve. Another application of Fubini's theorem then yields
\begin{align*}
    2\int_0^1( p-L(p)) dp =1-\frac{2}{\mu}\int_0^1(1-u)Q(u)du =\frac{\lambda_2(F)}{\mu}=G(F).
\end{align*}
Since $\mathrm{GMD}(F)=2\lambda_2(F)$ \citep{hosking1990} and $\lambda_2^*(F)=\sqrt{3}\lambda_2(F)$, taking the difference between the two potential-outcome distributions gives
\begin{align*}
    \theta_2 = \frac{\sqrt{3}}{2}\Big(\mathrm{GMD}(F_{11}^1)-\mathrm{GMD}(F_{11}^0)\Big).
\end{align*}
Therefore, assuming parallel trends over $\mathcal{K}=\{2\}$ identifies the causal effect of treatment on absolute pairwise dispersion in the treated population, \emph{even if the classic average parallel trends assumption does not hold}. This target has the same units as the outcome and requires only finite absolute first moments. 

Now, if instead one considers $\mathcal{K}=\{1,2\}$ together, then the counterfactual Gini coefficient and the corresponding treatment effect are identified, provided both potential-outcome distributions are nonnegative with positive means, since then 
\begin{align*}
    G(F)=\frac{\mathrm{GMD}(F)}{2\mu}
    =\frac{\lambda_2(F)}{\lambda_1(F)}
    =\frac{1}{\sqrt{3}}\frac{\lambda_2^*(F)}{\lambda_1^*(F)}.
\end{align*}
Since $\lambda_k^*(F_{11}^1)$ is observed, identification of $\theta_k$ also identifies $\lambda_k^*(F_{11}^0)=\lambda_k^*(F_{11}^1)-\theta_k$. Hence,
\begin{align*}
    \Delta_{\mathrm{Gini}} = G(F_{11}^1)-G(F_{11}^0)
    =\frac{1}{\sqrt{3}}\left(\frac{\lambda_2^*(F_{11}^1)}{\lambda_1^*(F_{11}^1)}-\frac{\lambda_2^*(F_{11}^0)}{\lambda_1^*(F_{11}^0)}\right).
\end{align*}
Unlike the Gini mean difference, the Gini coefficient normalizes the dispersion by the average. Absolute dispersion and relative inequality can therefore respond differently to treatment. For example, adding the same positive amount to every outcome leaves $\theta_2$ equal to zero but reduces a positive Gini coefficient by increasing the mean.

\begin{figure}[h]
\centering

\definecolor{hoOrange}{HTML}{2E86C1}
\definecolor{hoBlue}{HTML}{C0392B}

\begin{tikzpicture}[
    declare function={
        hoU(\z)=1/(1+exp(-\z));
        hoPtwo(\u)=6*\u^2-6*\u+1;
        hoPthree(\u)=20*\u^3-30*\u^2+12*\u-1;
        hoV(\z)=hoU(\z)*(1-hoU(\z));
    }
]
\begin{groupplot}[
    group style={
        group size=2 by 2,
        horizontal sep=1.95cm,
        vertical sep=2.1cm
    },
    width=0.455\linewidth,
    height=0.33\linewidth,
    axis lines*=left,
    axis line style={black!55},
    tick align=outside,
    tick style={black!55},
    tick label style={font=\small},
    label style={font=\small},
    title style={font=\small},
    every axis plot/.append style={
        line width=1pt,
        mark=none
    },
    scaled ticks=false,
    enlargelimits=false,
    clip=true
]

\nextgroupplot[
    title={$(a)$ $L$-skewness densities},
    xmin=-6.5,xmax=6.5,
    ymin=0,ymax=0.8,
    xtick={-6,-3,0,3,6},
    ytick={0,0.2,0.4,0.6,0.8},
    xlabel={Outcome},
    ylabel={Density},
    legend style={
        at={(0.98,0.98)},
        anchor=north east,
        draw=none,
        fill=none,
        font=\scriptsize,
        cells={anchor=west}
    }
]

\addplot[
    hoBlue,dashed,
    domain=-6.5:6.5,samples=301
] {hoV(x)};
\addlegendentry{Counterfactual}

\addplot[
    hoOrange,
    variable=\t,domain=-9:9,samples=501
]
    ({\t+1.25*hoPtwo(hoU(\t))},
     {hoV(\t)/(1+1.25*hoV(\t)*(12*hoU(\t)-6))});
\addlegendentry{Treated}

\nextgroupplot[
    title={$(b)$ $L$-kurtosis densities},
    xmin=-6.5,xmax=6.5,
    ymin=0,ymax=1.08,
    xtick={-6,-3,0,3,6},
    ytick={0,0.25,0.5,0.75,1},
    xlabel={Outcome},
    ylabel={Density}
]

\addplot[
    hoBlue,dashed,
    domain=-6.5:6.5,samples=301
] {hoV(x)};

\addplot[
    hoOrange,
    variable=\t,domain=-9:9,samples=501
]
    ({\t+hoPthree(hoU(\t))},
     {hoV(\t)/(1+hoV(\t)*
        (60*hoU(\t)^2-60*hoU(\t)+12))});

\nextgroupplot[
    title={$(c)$ $L$-skewness quantile effect},
    xmin=0,xmax=1,
    ymin=-0.8,ymax=1.4,
    xtick={0,0.25,0.5,0.75,1},
    ytick={-0.5,0,0.5,1},
    xlabel={Quantile rank $u$},
    ylabel={$Q^1(u)-Q^0(u)$},
    domain=0:1,samples=201
]

\addplot[draw=none,fill=hoOrange!15]
    {1.25*hoPtwo(x)} \closedcycle;

\addplot[black!50,line width=0.5pt,samples=2] {0};
\addplot[hoOrange] {1.25*hoPtwo(x)};

\nextgroupplot[
    title={$(d)$ $L$-kurtosis quantile effect},
    xmin=0,xmax=1,
    ymin=-1.15,ymax=1.15,
    xtick={0,0.25,0.5,0.75,1},
    ytick={-1,-0.5,0,0.5,1},
    xlabel={Quantile rank $u$},
    ylabel={$Q^1(u)-Q^0(u)$},
    domain=0:1,samples=201
]

\addplot[draw=none,fill=hoOrange!15]
    {hoPthree(x)} \closedcycle;

\addplot[black!50,line width=0.5pt,samples=2] {0};
\addplot[hoOrange] {hoPthree(x)};

\end{groupplot}
\end{tikzpicture}

\caption{Higher-order treatment effects with $\theta_1 = \theta_2 = 0$. Here, $Q^0(u)=\log(u/(1-u))$ and $Q^1(u)=Q^0(u)+(5/4)P_2(2u-1)$ in panels $(a,c)$ and $Q^1(u)=Q^0(u)+P_3(2u-1)$ in panels $(b,d)$. The first example increases $L$-skewness from $0$ to $1/4$; the second increases $L$-kurtosis from $1/6$ to $13/42$. The lower panels display the corresponding marginal quantile effects.}
\label{higher_order_effects_figure}
\end{figure}
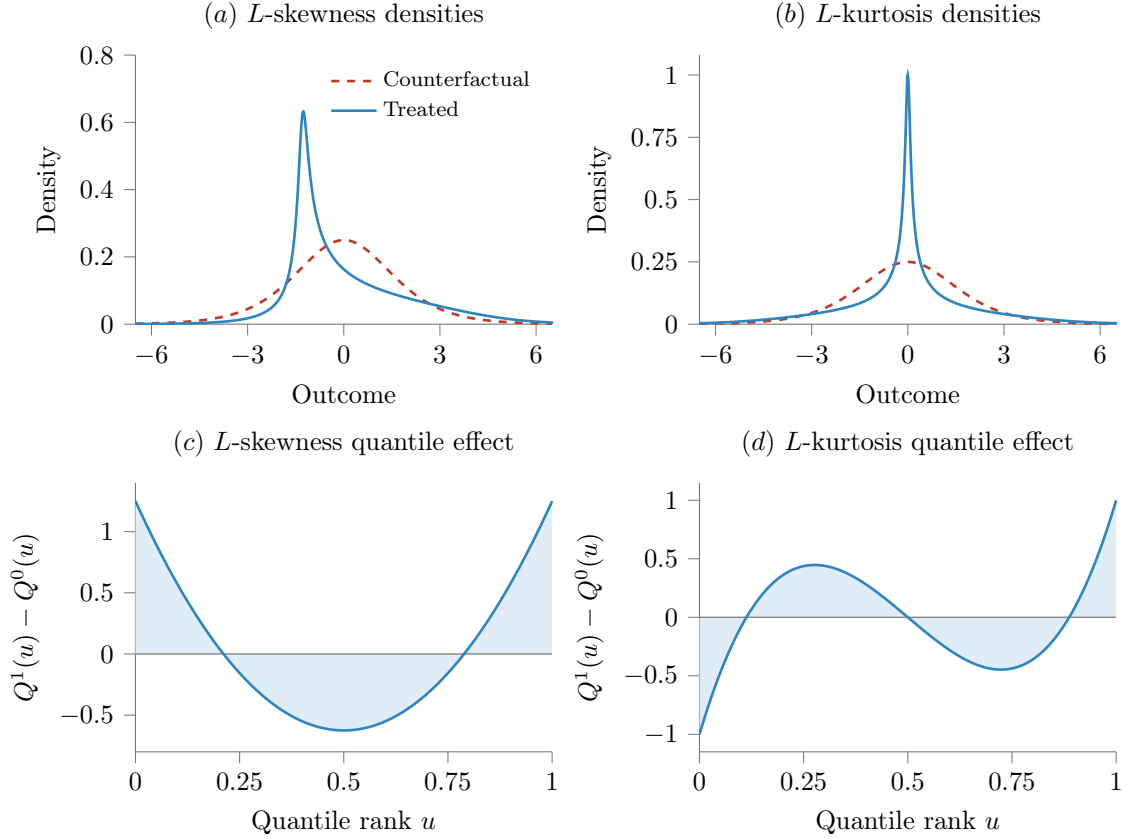

\paragraph{Combinations of Higher-Order $L$-Moments.}
As discussed in \cref{pt_spectrum_section}, $\lambda_3(F)$ measures the expected difference in upper and lower spacings for a randomly drawn triplet, and therefore captures the asymmetry within the distribution; e.g., $\lambda_3(F)=0$ for any symmetric distribution. However, note that the converse need not hold. Beyond that, one could also consider the set $\mathcal{K}=\{2,3\}$, which identifies the $L$-skewness of the counterfactual distribution and its change due to treatment. For a nondegenerate distribution, it holds that
\begin{align*}
    \tau_3(F) = \frac{\lambda_3(F)}{\lambda_2(F)} \in (-1,1),
\end{align*}
and this ratio has the added benefit of being standardized \citep{hosking1990}. Finally, $\mathcal{K}=\{4\}$ or $\mathcal{K}=\{2,4\}$ can be used to identify measures of kurtosis. Most notably,
\begin{align*}
    \lambda_4(F)=\frac{1}{4}\mathbb{E}\!\left[(X_{4:4}-X_{3:4})-2(X_{3:4}-X_{2:4})+(X_{2:4}-X_{1:4})\right],
\end{align*}
which compares the outer spacings to the middle spacing and therefore captures a notion of ``peakedness'' or tail weight. Specifically, $L$-kurtosis is defined for nondegenerate distributions as
\begin{align*}
    \tau_4(F) = \frac{\lambda_4(F)}{\lambda_2(F)} \in \left[\frac{5\tau_3(F)^2-1}{4},1\right),
\end{align*}
and thus summarizes the relative size of outer and central spacings \citep{hosking1990}. Both of these types of higher-order treatment effects are visualized in \cref{higher_order_effects_figure}. More generally, it can also be of interest to consider even- or odd-order $L$-moments of the distribution, which can be used to explore treatment-induced changes in dispersion or symmetry.

\paragraph{Interpretation.}
The distinction between a coordinate and a standardized ratio is important  for interpreting its treatment effect. The third and fourth coordinates are invariant to location shifts, but they do scale with the outcome: for $a\in\mathbb{R}$ and $b>0$, $\lambda_k^*(a+bX)=b\lambda_k^*(X)$ for $k\geq2$. A nonzero $\theta_3$ or $\theta_4$ could hence reflect a change in scale as well as in standardized shape. By contrast, the ratios are invariant to positive changes in scale and to location shifts. If both potential-outcome distributions are nondegenerate, restrictions on $\mathcal{K}=\{2,r\}$ identify the standardized treatment effects
\begin{align*}
    \Delta_{\tau_r} = \tau_r(F_{11}^1)-\tau_r(F_{11}^0)
    =\sqrt{\frac{3}{2r-1}}\left(\frac{\lambda_r^*(F_{11}^1)}{\lambda_2^*(F_{11}^1)}-\frac{\lambda_r^*(F_{11}^0)}{\lambda_2^*(F_{11}^0)}\right),\qquad r\in\{3,4\}.
\end{align*}
\cref{tab:component_targets} provides a summary of the coordinate restrictions and additional conditions for these effects, where we assume finite absolute first moments, consistency, no anticipation, and the indicated additive or CiC coordinate restrictions of \cref{identification_theorem} and \cref{identification_cic}.

\Needspace{16\baselineskip}
\begin{table}[htbp]
    \centering
    \small
    \begin{tabular}{lll}
        \toprule
        Treatment-effect target & Coordinate set $\mathcal{K}$ & Additional conditions \\
        \midrule
        Mean (ATT) & $\{1\}$ & None \\
        Gini mean difference & $\{2\}$ & None \\
        Gini coefficient & $\{1,2\}$ & Nonnegative outcomes, positive means \\
        $L$-skewness & $\{2,3\}$ & Nondegenerate distributions \\
        $L$-kurtosis & $\{2,4\}$ & Nondegenerate distributions \\
        \bottomrule
    \end{tabular}
    \caption{Sufficient coordinate restrictions for selected marginal treatment effects. }
    \label{tab:component_targets}
\end{table}

\subsubsection{Distributional Reconstruction} \label{reconstruction_subsection}
We now turn to the second application: using the identified coordinates to approximate the counterfactual quantile function and the marginal QTT curve. To start, note that the collective $L$-moment parallel trends assumption $\mathcal{L}(1\mathord{:}K)$, or $\mathcal{L}_{\mathrm{CiC}}(1 \mathord{:}K)$, identifies the projection
\begin{align*}
    \Pi_K\tau_{\mathrm{QTT}}(u)=\sum_{k=1}^K\theta_k\phi_k(u)
\end{align*}
of the marginal quantile treatment effect onto the polynomials of degree at most $K-1$. More precisely, if $\tau_{\mathrm{QTT}} \in L^2(0,1)$, this is the unique best $L^2$ polynomial approximation of degree at most $K-1$, such that
\begin{align*} 
    \Pi_K\tau_{\mathrm{QTT}}
    &=\operatorname*{arg\,min}_{v\in\mathcal{V}_K}\|\tau_{\mathrm{QTT}}-v\|_{2}, 
\end{align*}
where $\mathcal{V}_K=\operatorname{span}\{\phi_1,\ldots,\phi_K\}$, i.e.\ the set of polynomials of degree at most $K-1$. The projection remainder is given by $ \|\tau_{\mathrm{QTT}}-\Pi_K\tau_{\mathrm{QTT}}\|_{2}^2 =\sum_{k>K}\theta_k^2$. As discussed in \cref{regularity_conditions}, finite second moments of both post-treatment potential outcomes are sufficient for this $L^2$ condition.

Letting $K$ increase provides a finer and finer approximation to the full quantile function. In fact, by \citet{sillitto69}, when $X$ has a finite variance,
\begin{align*}
    Q(u)=\sum_{k=1}^{\infty}\lambda_k^*(Q)\phi_k(u),
\end{align*}
with convergence in $L^2(0,1)$, so as $K\to\infty$ we recover the quantile function in this sense. In our setting, the identified counterfactual coefficients give the truncated expansion
\begin{align*}
    Q_{11,K}^0(u)=\sum_{k=1}^K\left(\lambda_k^*(F_{11}^1)-\theta_k\right)\phi_k(u).
\end{align*}
For finite $K$, this truncation need not be nondecreasing and hence need not be a valid quantile function itself; one can impose monotonicity by monotone rearrangement \citep[e.g.][]{chernozhukov_crossing}, which preserves the first $L$-moment but can change higher order coefficients. Increasing $K$ also adds identifying restrictions, so convergence of the expansion must be distinguished from identification under a fixed, finite spectrum.

\paragraph{Distributional Distance, Location, and Shape.}
Another way to leverage higher-order treatment effects is through its close relationship to the Wasserstein distance. When both post-treatment potential-outcome distributions have finite
second moments, the Wasserstein distance between counterfactual distributions can be written as \citep{vallender1974, villani2009optimal,schindl2026distributionaldiscontinuitydesign}
\begin{align} \label{wasserstein_treatment_decomposition}
    W_2^2(F_{11}^1,F_{11}^0) =\int_0^1\tau_{\mathrm{QTT}}(u)^2du =\sum_{k=1}^{\infty}\theta_k^2.
\end{align}
Under either $\mathcal{L}(1\mathord{:}K)$ or $\mathcal{L}_{\mathrm{CiC}}(1\mathord{:}K)$, the quantity
\begin{align*}
    W_{2,K}^2 =\sum_{k=1}^K\theta_k^2 =\|\Pi_K\tau_{\mathrm{QTT}}\|_{L^2}^2 \leq W_2^2(F_{11}^1,F_{11}^0)
\end{align*}
is an identified lower bound on the total squared distributional distance. Its gap from the full distance is exactly the projection remainder $\|\tau_{\mathrm{QTT}}-\Pi_K\tau_{\mathrm{QTT}}\|_2^2$. Provided the total distance is positive, the share attributable to these coordinates is $W_{2,K}^2/W_2^2(F_{11}^1,F_{11}^0)$. Again, this share is generally unidentified under finite-coordinate restrictions alone, because its denominator contains the unidentified tail. The Wasserstein decomposition also distinguishes a change in location from a change in marginal shape:
\begin{align*}
    W_2^2(F_{11}^1,F_{11}^0)-\theta_1^2
    =\int_0^1\{\tau_{\mathrm{QTT}}(u)-\theta_1\}^2du
    =\sum_{k=2}^{\infty}\theta_k^2.
\end{align*}
Thus, any identified nonzero higher-order coordinate establishes that the two marginal distributions differ by more than a constant location shift. For example, if $Q_{11}^0(u)=2u-1$ and $Q_{11}^1(u)=2(2u-1)$, the mean effect is zero, but $\theta_2=1/\sqrt{3}$ and $W_2^2=1/3$. This means that  the treatment expands dispersion without changing the mean.

\paragraph{Individual Treatment-Effect Heterogeneity.}
These summaries describe differences between marginal potential-outcome distributions. However, they do not identify the distribution of individual effects $Y_1(1)-Y_1(0)\mid G=1$ in general, since those also depend on the joint distribution of the two potential outcomes \citep{FanPark2010}. With the established results, we can nevertheless give a lower bound on individual treatment-effect heterogeneity. The following result does this by combining the standard optimal-coupling bound \citep{villani2009optimal,DingFellerMiratrix2019} with identification of selected $L$-moment coordinates.

\begin{proposition}[Individual Heterogeneity Bounds]
\label{individual_heterogeneity_bound}
Suppose that $\mathbb{E}[Y^2_1(d) \mid G=1]<\infty$ for $d\in\{0,1\}$, let $\mathcal{K}\subseteq\{2,3,\ldots\}$ be finite, and suppose either $\mathcal{L}(\mathcal{K})$ or $\mathcal{L}_{\mathrm{CiC}}(\mathcal{K})$ holds. Then, the quantity $H_{\mathcal{K}}^2 =\sum_{k\in\mathcal{K}}\theta_k^2$ is identified and satisfies
\begin{align*}
    H_{\mathcal{K}}^2 \leq W_2^2(F_{11}^1,F_{11}^0)-\theta_1^2 \leq \mathbb{V}\!\left(Y_1(1)-Y_1(0)\mid G=1\right).
\end{align*}
The second inequality holds for every joint distribution with these potential-outcome marginals. For fixed marginals, its lower bound is attained by a common-uniform coupling.
\end{proposition}

\cref{individual_heterogeneity_bound} is interesting in that it turns selected marginal distributional effects into a quantitative statement about variation in \emph{individual treatment effects}. To see the intuition, suppose every treated unit experienced the same individual effect $a$. The two potential-outcome distributions would then differ only by a location shift, with $Q_{11}^1(u)=Q_{11}^0(u)+a$, and every higher-order coordinate would be zero. An identified nonzero higher-order effect therefore establishes that treatment affects units differently. The quantity $H_{\mathcal{K}}^2$ provides a lower bound on the variance needed to produce the identified marginal changes. Note in particular that this conclusion does not require rank invariance or identification of the ATT.

To combine this with the previously established relationships to measures of inequality, note that a restriction on the second coordinate alone yields
\begin{align*}
    \mathbb{V}\!\left(Y_1(1)-Y_1(0)\mid G=1\right)
    \geq \frac{3}{4}\left\{\mathrm{GMD}(F_{11}^1)-\mathrm{GMD}(F_{11}^0)\right\}^2.
\end{align*}
This bound therefore provides information about treatment heterogeneity \emph{even when the average effect is unidentified}. Note, however, that while a positive bound provides evidence of heterogeneity, a zero bound does not establish that effects are homogeneous. For instance, $Y_1(0)=U$ and $Y_1(1)=1-U$ with $U\sim\operatorname{Uniform}(0,1)$ have identical marginals and hence zero coordinate effects, but the individual effect has variance $1/3$. Note that in general there is no upper bound on individual treatment effect variation. In the next section, we establish bounds on the projection remainder under moment and smoothness conditions and examine their implications for the choice of $K$.

\subsection{Approximation, Decay, and the Choice of $K$} \label{decay_section}

In \cref{pt_spectrum_section} and \cref{reconstruction_subsection} we established the parallel trends spectrum and showed how a collection of $L$-moments can be used to reconstruct the counterfactual distribution. The natural next question to consider is, ``how many $L$-moments do we need for reconstruction?'' The answer to this question loosely depends on how many conventional moments we assume exist, the smoothness and regularity of the quantile function across $u \in (0, 1)$, and, in difference-in-differences, if there are any shared features across quantile functions. Let
\begin{align*}
    R^2_K(f) = \| f - \Pi_K f\|^2_2 = \sum_{k > K} c^2_k(f)
\end{align*}
denote the projection remainder, for $f \in L^2(0, 1)$ and $c_k(f) = \int^1_0 f(u) \phi_k(u) du$. In what follows, we establish the conditions under which $R_K(f)$ decays.

\subsubsection{Moment-Based Bounds} \label{moment_subsection}

The most straightforward approach to establishing some kind of decay rate in $K$ is through some kind of moment-based restriction. Existing work has established moment bounds for individual $L$-moments \citep{JONES2025110249}; however, we are interested in the reconstruction error $R_K$. This requires dealing with distributional tails and thus our arguments and bounds differ. Let $X$ be some random variable with quantile function $Q$. Under the assumption that $Q \in L^2(0, 1)$ then it follows that $R_K(Q) \to 0$ as $K \to \infty$ \citep{sillitto69}, following the regularity discussion in \cref{regularity_conditions}. However, this simply guarantees convergence in $K$, but not anything about the rate at which it achieves convergence. Fortunately, if we assume more than finite second moments, it is possible to achieve a better rate, which we establish in the following theorem.

\begin{theorem}[Moment Bound] \label{moment_bound}
  Let $M_p(Q) = \inf_{a\in\mathbb{R}} \left(\mathbb{E}\! \left[ |X-a|^p \right] \right)^{1/p}<\infty$ for some $p > 4$. Then, there exists some constant $C_p < \infty$ depending only on $p$ such that
    \begin{align*}
        R_K(Q) \leq \frac{C_p M_p(Q)}{\sqrt{K}}.
    \end{align*}
\end{theorem}

One might expect assuming a greater number of moments exist automatically yields a faster rate in $K$, but \cref{moment_bound} shows that this is not the case. In fact, there is a limited gain that can be obtained as we move from $p = 2$ to $p > 4$, but that saturates at an $O(K^{-1/2})$ rate. The reason for this phenomenon is that $L$-moment decay is both tied to the regularity of the underlying quantile function and to how heavy its tails are. For example, one could consider the bounded quantile function $Q_B(u) = \mathbb{I}(u > 1/2)$. It can be shown that 
\begin{align*}
    KR^2_K(Q_B) \longrightarrow \frac{1}{2 \pi}.
\end{align*}
Thus, even though its distribution has finite moments at every order, $R_K(Q_B) \asymp (2\pi K)^{-1/2}$; the obstruction is the discontinuity of the quantile function. We explore this phenomenon further in the next section, where we show that more assumed regularity yields faster rates of convergence.

\subsubsection{Regularity Conditions for $L$-Moment Decay} \label{smoothness_section}

Establishing appropriate regularity conditions for $L$-moment decay is trickier than one might anticipate. At first glance, one might expect that imposing some kind of smoothness constraint on $f$ would be sufficient for controlling the remainder $R_K(f)$. For example, we may want to assume $f \in C^s(\mathcal{S})$, i.e.\ the space of continuous real-valued functions on some set $\mathcal{S}$ whose derivatives of order $1, \ldots, s$ exist. Then, we could reasonably hope to show faster rates of decay in $K$ for higher smoothness values $s$. However, choosing $\mathcal{S} = (0, 1)$ is not a sufficiently strong enough condition to control $R_K(f)$ because it doesn't control how derivatives of $f$ behave near zero and one. On the other hand, letting $\mathcal{S} = [0, 1]$ is typically \textit{too} strong of a condition, such that it rules out many common distributions with unbounded quantiles, such as a standard Gaussian.

To better illustrate this point, suppose $X \sim \mathcal{N}(\mu, \sigma^2)$. Then, its quantile function is $Q(u) = \mu + \sigma \Phi^{-1}(u)$, where $\Phi$ is the standard normal cumulative distribution function (and $\varphi$ is its density function). Even though $Q$ is infinitely differentiable on $(0, 1)$, its derivatives become unbounded at zero and one. Observe that differentiating $Q$ yields $ Q^\prime(u) = \sigma / \varphi(\Phi^{-1}(u))$. Then, as $u \to 0$ or $u \to 1$, the denominator tends to zero. Consequently,
\begin{align*}
   \sup_{0 < u < 1} |Q^\prime(u)| = \infty,
\end{align*}
and therefore, $Q \in C^s((0, 1))$ but not $C^s([0, 1])$. This distinction matters, since our goal is to control $R^2_K(f) = \int^1_0 |f(u) - \Pi_K f(u)|^2 du,$ so in order to establish a bound we need control over neighborhoods near the endpoints.

In order to balance these interior and endpoint concerns, we begin by assuming
\begin{align} \label{f_regularity}
    f\in L^2(0,1)\cap C^{s-1}((0,1))
\end{align}
such that $f^{(s-1)}$ is Lipschitz continuous on every compact subinterval of $(0,1)$, and where $f^{(0)} = f$. Note that membership of $L^2(0, 1)$ ensures that $R_K(f)$ is well-defined. For example, letting $s = 2$, our assumption is that $f \in L^2(0, 1) \cap C^1((0, 1))$ and that, for every $0 < \delta < 1/2$, there exists some $L_\delta$ such that
\begin{align*}
    |f^\prime(u) - f^\prime(v)| \leq L_\delta |u - v|
\end{align*}
for all $u, v \in [\delta, 1 - \delta]$. \cref{f_regularity} provides us with everything we need to control the interior ranks. To control the intervals near zero and one we require some kind of additional regularity conditions. However, the exact conditions we need are not immediately obvious. Thus, we first establish a technical lemma that provides us with the ability to control the endpoint errors. The general idea is that we want to bound $ R^2_K(f) = \| f - \Pi_K f\|^2_2$ by constructing some smooth auxiliary curve $g$ to replace $f$ with, that is controlled near zero and one. We can then establish endpoint regularity conditions by requiring $f$ to behave similarly to $g$. We formalize this result in what follows.

\begin{lemma}[Higher-Order Smoothness Bound]
\label{higher_order_energy_bound}
Let $s\geq1$. Suppose $g\in C^{s-1}([0,1])$,
$g^{(s-1)}$ is absolutely continuous on $[0,1]$, $w(u) = u(1-u)$, and
\begin{align*}
    \mathcal{E}_s(g) =\int_0^1 w(u)^s|g^{(s)}(u)|^2du<\infty.
\end{align*}
Then, for every integer $K\geq s$, $  R^2_K(g) \leq \frac{(K-s)!}{(K+s)!}\mathcal{E}_s(g).$ Consequently, for every $f\in L^2(0,1)$, 
\begin{align} \label{rk_bound}
     R_K(f)\leq\|f-g\|_2+ \sqrt{\frac{(K-s)!}{(K+s)!}}\mathcal{E}^{1/2}_s(g).
\end{align}
\end{lemma}

The first term in \cref{rk_bound} represents the cost of replacing $f$ by $g$ and the second term represents the cost of approximating $g$ with the first $K$ coefficients. Roughly speaking, a very smooth $g$ may be very far from $f$ and a $g$ that closely follows $f$ may have a large derivative energy. Thus, the bound on $R_K(f)$ balances both. Importantly, \cref{higher_order_energy_bound} allows us to control the neighborhood of $R_K(f)$ near zero and one. Specifically, for $0 < \delta < 1/2$ we define the Taylor polynomials
\begin{align*}
     p^-_{\delta}(u) &=\sum_{r=0}^{s-1} \frac{f^{(r)}(\delta)}{r!}(u-\delta)^r, \\
    p^+_{\delta}(u) &=\sum_{r=0}^{s-1} \frac{f^{(r)}(1-\delta)}{r!}(u-1+\delta)^r.
\end{align*}
Then, using these two polynomials, we can define the auxiliary function
\begin{align*}
     g_\delta(u)=
    \begin{cases}
        p^-_\delta(u),&0\leq u\leq\delta,\\
        f(u),&\delta<u<1-\delta,\\
        p^+_\delta(u),&1-\delta\leq u\leq1.
    \end{cases}
\end{align*}
Now, we can see that each of the endpoint polynomials matches $f$ and its derivatives, and thus $g_\delta \in C^{s-1}([0, 1])$. Furthermore, because the derivative of order $s-1$ is constant on the endpoint intervals near zero and one and Lipschitz on the interior, $g^{(s-1)}_\delta$ is globally Lipschitz. Hence, the conditions of \cref{higher_order_energy_bound} are met. Then, we may split apart the interior and endpoint approximation error as
\begin{align*}
    E^2_s(f;\delta) &= \int_0^\delta [f(u)-p^-_{\delta}(u)]^2du +\int_{1-\delta}^1 [f(u)-p^+_{\delta}(u)]^2du,\\
    I^2_s(f;\delta) &= \int_\delta^{1-\delta}[u(1-u)]^s [f^{(s)}(u)]^2du.
\end{align*}
Notably, $E^2_s(f;\delta) = \|f-g_\delta\|^2_2$ measures the error given by replacing the neighborhoods near zero and one with polynomials and $I^2_s(f;\delta) = \mathcal{E}_s(g_\delta)$ measures the remaining variation in the interior. Furthermore, we may directly apply \cref{higher_order_energy_bound} to see that
\begin{align*}
     R_K(f) \leq \inf_{\delta \in (0, 1/2)} \left\{ E_s(f;\delta) + \sqrt{\frac{(K-s)!}{(K+s)!}}  I_s(f;\delta) \right\}.
\end{align*}
With these regularity conditions in place, we now have a way to control the global approximation $R_K(f)$. Specifically, the rate in $K$ we are able to obtain depends on how much control we have over the derivatives of $f$ near zero and one. We formalize this in the following theorem.

\begin{theorem}[$L$-Moment Decay]
\label{rk_bound_general}
Let $s\geq2$ be a fixed integer, and suppose that $f$ satisfies \cref{f_regularity}, with $f^{(s-1)}$ Lipschitz on every compact subinterval of $(0,1)$. Write $d(u)=u\wedge(1-u)$. If, for some $C\in[0,\infty)$ and $\alpha\in[0,1/2)$,
\begin{align*}
     |f^{(s)}(u)| \leq C d(u)^{-s-\alpha}
\end{align*}
for almost every $u \in (0, 1)$, then there exists a finite constant $C_{s,\alpha}$, depending only on $s$, $\alpha$, and $C$, such that for $K \geq s$,
\begin{align*}
      R_K(f)\leq \frac{C_{s,\alpha}}{K^{1- 2\alpha}}.
\end{align*}
\end{theorem}

By \cref{rk_bound_general}, we establish a decay rate in $K$ that allows for the $s$th derivative to grow as it moves toward zero and one, as long as it doesn't grow too quickly; this derivative growth is measured by $\alpha$. Thus, we can now establish decay rates for distributions that have unbounded supports, such as the standard normal distribution. We can also see how adding some kind of smoothness condition increases the approximation rate relative to a moment condition. For example, when $\alpha = 0$ then we now have an $O(K^{-1})$, thereby improving upon the $O(K^{-1/2})$ rate we obtained under $p > 4$ moments. More generally, for $\alpha < 1/4$ then the rate in $K$ is determined by the minimum between the moment bound of \cref{moment_bound} and the bound established in \cref{rk_bound_general}. Finally, it is worth noting that if $f$ has a bounded support then the decay rate can dramatically improve with its assumed smoothness.

\begin{corollary}[Decay for Smooth Bounded Functions]
\label{decay_smoothness}
Let $s \geq 1$ be an integer. Suppose $f\in C^{s-1}([0,1])$ and, for some $L_s\in[0,\infty)$ and all $u,v\in[0,1]$, $|f^{(s-1)}(u)-f^{(s-1)}(v)| \leq L_s|u-v|.$ Then, for every integer $K\geq s$,
\begin{align*}
     R_K(f) \leq \frac{s!}{\sqrt{(2s+1)!}} \sqrt{\frac{(K-s)!}{(K+s)!}} L_s.
\end{align*}
In particular, for fixed $s$ and $L_s$, $R_K(f)=O(K^{-s})$ as
$K\to\infty$.
\end{corollary}

\cref{decay_smoothness} follows directly from \cref{higher_order_energy_bound}  with $g=f$: the Lipschitz condition implies $|f^{(s)}|\leq L_s$ almost everywhere, and $\int_0^1[u(1-u)]^sdu=(s!)^2/(2s+1)!$. Intuitively, this shows that well-behaved functions with bounded support require much fewer coefficients for a full distributional reconstruction. 
Examples of $R_K$ for some common distributions are given in \cref{table::R}.

\begin{table}[tbp]
\centering
\begingroup
\renewcommand{\arraystretch}{1.08}
\begin{tabular}{@{}l@{\hspace{1.5em}}l@{}}
\toprule
Distribution & $R_K$ as $K\to\infty$ \\
\midrule
$\operatorname{Uniform}(a,b)$
    & $0$ for $K\ge 2$ \\
$\operatorname{Exponential}(r)$
    & $1/(rK)$, exactly \\
$\operatorname{Logistic}(\mu,b)$
    & $\sim \sqrt{2}\,b/K$ \\
$\mathcal{N}(\mu,\sigma^2)$
    & $\sim \sigma/(K\sqrt{2\log K})$ \\
$\operatorname{Pareto}(x_{\min},\tau)$, $\tau>2$
    & $\asymp K^{-1+2/\tau}$ \\
$t_\nu$, $\nu>2$
    & $\asymp K^{-1+2/\nu}$ \\
$\operatorname{Cauchy}(\mu,b)$; $t_\nu$, $0<\nu\le 2$
    & $\infty$, since $Q\notin L^2(0,1)$ \\
Analytic quantile across $[0,1]$
    & $\le Ce^{-cK}$ \\
\bottomrule
\end{tabular}
\endgroup
\caption{The behavior of $R_K$ for some common distributions.}
\label{table::R}
\end{table}

\subsubsection{Decay in Difference-in-Differences} \label{did_decay}

In the previous sections we explored decay conditions for some general function $f$, under certain moment constraints or smoothness conditions. However, in a difference-in-differences framework we consider \textit{differences} and compositions of quantile functions. This is notable, because quantile functions may exhibit shared features that cancel out, thereby leading to \textit{faster} convergence than would be possible for individual distributions alone. In this section we focus on the additive quantile parallel trends assumption, although the same principles can be applied to changes-in-changes. Specifically, we consider cancellations in
\begin{align} \label{cancellations_eq}
    \Delta(u) = \Delta_1(u)-\Delta_0(u) =\left\{Q_{11}(u)-Q_{10}(u)\right\} -\left\{Q_{01}(u)-Q_{00}(u)\right\}
\end{align}
as this determines the number of causal coefficients required to reconstruct $\tau_{\mathrm{QTT}}(u)$. To see this, recall that $\theta_k =\int_0^1\tau_{\mathrm{QTT}}(u)\phi_{k}(u)du$, and by \cref{additive_counterfactual} of \cref{identification_theorem}, we have that
\begin{align*}
    \theta_k = \lambda^*_k(Q_{11}) - \lambda^*_k(Q_{10}) -\lambda^*_k(Q_{01}) + \lambda^*_k(Q_{00}).
\end{align*}
To establish cancellations in \cref{cancellations_eq} we can directly leverage the results established in \cref{smoothness_section}. However, since we are now focusing on differences in quantiles we slightly change our perspective from derivatives to comparisons across ranks.

In order to avoid differentiating marginal quantiles, we leverage the forward rank difference operator. For some function $v$ and increment $h > 0$, the rank difference is recursively defined as $D^{s+1}_h v(u) = D^s_h v(u + h) - D^s_h v(u)$ \citep{devore1993constructive}. The first two orders are given by
\begin{align*}
    D^1_h v(u) &= v(u + h) - v(u), \\
    D^2_h v(u) &= \{v(u+2h)-v(u+h)\}-\{v(u+h)-v(u)\},
\end{align*}
and in general, $D_h^s v(u)=\sum_{r=0}^s(-1)^{s-r}\binom{s}{r} v(u+rh)$. Thus, the rank difference operator quantifies regularity across quantiles not through derivatives but by how adjacent spacings change, particularly as $h$ becomes small. In fact, repeated application of the fundamental theorem of calculus gives the identity,
\begin{align*}
    \frac{D_h^sv(u)}{h^s} = \int_{[0,1]^s} v^{(s)}\!\left(u+h(t_1+\cdots+t_s)\right) dt_1\cdots dt_s,
\end{align*}
for $[u, u + sh] \subseteq (0, 1)$, so $h^{-s} D_h^sv(u)$ represents an average of the $s$th derivative over nearby ranks. Moreover, if $v$ has continuous derivatives through order $s$ then $h^{-s} D_h^sv(u) \to v^{(s)}(u)$ as $h \to 0$. For our difference-in-differences contrast in \cref{cancellations_eq}, by linearity of the difference operator, $D_h^s\Delta= D_h^s\Delta_1-D_h^s\Delta_0$, and so the first order difference is given by
\begin{align*}
     D_h\Delta(u) &=\{Q_{11}(u+h)-Q_{11}(u)\} -\{Q_{10}(u+h)-Q_{10}(u)\} \, - \\
    &\phantom{{}={}} \{Q_{01}(u+h)-Q_{01}(u)\} +\{Q_{00}(u+h)-Q_{00}(u)\}.
\end{align*}
Our goal now is to establish regularity of $D^s_h \Delta$. Fix $0\leq\delta<1/2$ and an integer $s\geq1$. Then, for $0<h<(1-2\delta)/s$ and $u\in(\delta,1-\delta-sh)$, note that every rank $u,u+h,\ldots,u+sh$ lies within the interval $(\delta,1-\delta)$. Now, define the cancellation seminorm,
\begin{align*}
     \mathcal{C}_s(\delta) = \sup_{0<h< (1-2\delta) / s} \left\{ \frac{1}{h^s} \|D_h^s\Delta_1-D_h^s\Delta_0\|_{L^\infty(\delta,1-\delta-sh)} \right\}.
\end{align*}
In \cref{smoothness_section} we required some form of control over $f^{(s)}$, whereas now we control $\mathcal{C}_s(\delta)$. For example, if $\mathcal{C}_s(\delta)\leq C\delta^{-s-\alpha}$ for every $0<\delta<1/2$, then \cref{rk_bound_general} implies $ R_K(\Delta)\leq C_{s,\alpha} K^{-1+2\alpha}$ where $C_{s,\alpha}$ depends only on $C$, $s$, and $\alpha$. If instead $\mathcal{C}_s <\infty$ at any order $s\geq1$ where $\mathcal{C}_s = \mathcal{C}_s(0)$, then by \cref{decay_smoothness},
\begin{align*}
    R_K(\Delta) \leq \frac{s!}{\sqrt{(2s+1)!}}
             \sqrt{\frac{(K-s)!}{(K+s)!}} \mathcal{C}_s.
\end{align*}
Notably, the decay mechanism is different here than in \cref{smoothness_section}, since now our results depend on shared cancellations between quantiles. In fact, even if the individual quantiles in \cref{cancellations_eq} decay at an $O(K^{-1/2})$ rate, the shared cancellation between them can be much faster. For example, suppose that $Q^1(u) = Q^0(u) + h(u)$ for some counterfactual quantile functions. Here, $h(\cdot)$ represents some kind of shared effect profile. For example, we could say
\begin{align*}
     h(u)=a+b(u-1/2)+\sum_{j=1}^{J}\gamma_j\psi_j(u)
\end{align*}
for some basis functions $\psi_j \in C^1([0, 1])$, which encodes a linear combination of shape features. Then, even if $Q^0$ and $Q^1$ decay at slower rates, the difference between them is determined by $h$. We formalize this in the following proposition.
\begin{proposition}[Cancellation-Based Decay]
\label{counterfactual_cancellation}
Let $Q^0\in L^2(0,1)$ be a quantile function satisfying, for some
$m>0$ and every $0<u<v<1$,
\begin{align*}
    Q^0(v)-Q^0(u)\geq m(v-u).
\end{align*}
For any $h\in C^1([0,1])$ with $\|h'\|_\infty < m$, the function
$Q^1=Q^0+h$ is a strictly increasing quantile in $L^2(0,1)$ and for all $K \geq 1$, $|R_K(Q^1)-R_K(Q^0)|\leq R_K(h)$. Consequently, if, for a positive sequence $r_K\to0$,
\begin{align*}
     R_K(Q^0)\asymp r_K,\qquad R_K(h)=o(r_K),
\end{align*}
then $R_K(Q^1)\sim R_K(Q^0)$ but $R_K(Q^1-Q^0)=o(r_K)$.
\end{proposition}

By \cref{counterfactual_cancellation} we can see that the true decay rate depends on the shared features between $Q^0$ and $Q^1$. In fact, it doesn't matter how slowly either individual quantile converges; only the decay rate on $R_K(h)$ matters. This could lead to substantially faster cancellations than one might initially anticipate. For example, leveraging \cref{higher_order_energy_bound}, if, for some fixed $L > 0$, $\| h^\prime\|_\infty \leq m / 2$ and $\|h^{(s)} \|_\infty \leq L^s$ for all $s \geq 1$ then
\begin{align*}
    |R_K(Q^1)-R_K(Q^0)| = O\! \left( \frac{(L/4)^K}{K!} \right).
\end{align*}

\begin{remark}
A subtle point to note is that $\Delta = \tau_{\mathrm{QTT}} + \mathcal{D}^0$, where $\mathcal{D}^0$ is the untreated quantile-trend defect discussed in \cref{pt_spectrum_section}. If we assume $\mathcal{L}(1\mathord{:}K)$, then $d^0_k = 0$ for every $k = 1, \ldots, K$ and so each causal component $\theta_k$ is identified. However, this does not place any kind of restriction on higher order coordinates. More generally,
\begin{align*}
    R_K(\tau_{\mathrm{QTT}}) \leq R_K(\Delta) + R_K(\mathcal{D}^0).
\end{align*}
Thus, if one wanted to formally establish some kind of decay rate specifically for $R_K(\tau_{\mathrm{QTT}})$ then similar regularity conditions would need to be imposed on the counterfactual $\mathcal{D}^0$ as well.
\end{remark}

\section{Estimation and Inference} \label{estimation_inference}

In this section we develop efficient estimators for $L$-moment-based effects. We begin by deriving the efficient influence function for general $L$-functionals and then apply these results to various difference-in-differences settings. We focus on estimating the consecutive vector of $L$-moments $\lambda^*_1(F), \ldots, \lambda^*_K(F)$ corresponding to the $\mathcal{L}(1\mathord{:}K)$ assumption; however, most results hold for arbitrary combinations of $k \in \mathcal{K}$. Note that our approach differs from \citet{ALVAREZ2025106101}: they develop inference for a fixed-dimensional model parameter using an increasing number of $L$-moments as estimating restrictions; we treat the $L$-moment effects themselves as estimands. In particular, we study joint inference for a collection whose dimension may grow with the sample size, such that $K_n \to \infty$. 

\subsection{Efficiency Theory for $L$-Moments} \label{efficiency_theory_section}
To begin, we derive the efficient influence function for an $L$-moment, in general. At first glance, it would appear that we can directly leverage \cite{firpo2007}, since $\lambda_k = \int^1_0 Q(u) P^*_{k-1}(u)du$ is simply a linear function of the quantile function. In fact, one \textit{could} technically leverage this approach, which would yield the influence function,
\begin{align} \label{bad_eif}
    \varphi_k(y; F) = \int^1_0 P^*_{k-1}(u) \left( \frac{u - \mathbb{I}(y \leq Q(u))}{f(Q(u))} \right) du.
\end{align}
However, this is problematic for two reasons. First and foremost, \cref{bad_eif} relies on the full quantile function, as opposed to the particular components we are interested in. Second, \cref{bad_eif} contains the term $f(Q(u))^{-1}$ which requires both density and quantile estimation, and is often numerically unstable. Fortunately, $L$-moments can be represented in terms of probability weighted moments \citep{greenwood_pwm}, which allows us to define a density-free estimator. Specifically, let
\begin{align*}
     \beta_j(F) =\int_0^1 Q_F(u)u^j du =\frac{1}{j+1} \mathbb{E}_F\!\left[\max\{Y_0,\ldots,Y_j\}\right]
\end{align*}
denote the probability weighted moments, where $Y_0,\ldots,Y_j$ are independent draws from $F$. Then, using the fact that $P^*_{k-1}(u) = \sum^{k-1}_{j=0} \xi_{kj} u^j$ where
\begin{align*}
    \xi_{kj} = (-1)^{k-1+j} \binom{k-1}{j}\binom{k-1+j}{j}
\end{align*}
we may write the $k$th normalized $L$-moment as
\begin{align} \label{normalized_lmoment_pwm}
    \lambda^*_k(F) = \sqrt{2k-1} \sum_{j=0}^{k-1} \xi_{kj} \beta_j(F).
\end{align}
In the following theorem, we formalize the influence function for each $L$-moment.
\begin{theorem}[Efficient Influence Function of an $L$-Moment] \label{lmoment_eif} Let $k \geq 1$ be some integer and suppose $F \in \mathcal{P}_2(\mathbb{R})$. Furthermore, for some integer $j \geq 0$ let $M_j(y;F) =\mathbb{E}_F\!\left[\max\{y,Y_1,\ldots,Y_j\}\right]$ where we say $M_0(y;F) = y$. Then, in the nonparametric model for $F$, the efficient influence function for $\lambda^*_k(F)$ is given by
\begin{align*}
    \ell_k(y;F) &= \sqrt{2k - 1}\sum_{j=0}^{k-1}\xi_{kj} \Big(M_j(y;F)-(j+1)\beta_j(F)\Big) \\
    &= \int_{\mathbb{R}} \phi_k(F(z)) \left(F(z)-\mathbb{I}(y\leq z)\right)dz.
\end{align*}
The influence function for the unnormalized moment $\lambda_k(F)$ is simply $\ell_k/\sqrt{2k-1}$.  
\end{theorem}

By \cref{lmoment_eif}, we can see that each influence function retains the order-statistic interpretation of their corresponding $L$-moments. Let $Y_{1:n}\leq\cdots\leq Y_{n:n}$ denote the order statistics of independent draws $Y_1,\ldots,Y_n \sim F$. We interpret conditioning on $Y_1=y$ as fixing one unsorted observation at $y$ and averaging over the remaining draws; so, for example, $M_{n-1}(y;F)=\mathbb{E}[Y_{n:n}\mid Y_1=y]$. Then, the first three influence functions can then be written as
\begin{align*}
 \ell_1(y;F) &=y-\lambda_1^*(F),\\
 \ell_2(y;F) &=\sqrt{3}\mathbb{E}\!\left[ Y_{2:2}-Y_{1:2}\mid Y_1=y \right]-2\lambda_2^*(F),\\
 \ell_3(y;F) &=\sqrt{5}\,\mathbb{E}\!\left[ (Y_{3:3}-Y_{2:3})-(Y_{2:3}-Y_{1:3}) \mid Y_1=y \right]-3\lambda_3^*(F).
\end{align*}
Thus, we retain the same shape-based interpretation as before. The cumulative distribution function representation provides a complementary perspective: it integrates the discrepancies $F(z)-\mathbb I(y\leq z)$ against their rank-dependent weights $\phi_k(F(z))$. This representation in particular will be useful for deriving joint covariance formulas and controlling higher-order remainders.

The coordinatewise influence functions also determine the efficient influence function for the first $K$ normalized $L$-moments jointly. For fixed $K<\infty$, define
\begin{align*}
     \boldsymbol{\lambda}_K^*(F) &= \bigl(\lambda_1^*(F),\ldots,\lambda_K^*(F)\bigr)^{\top},  \\
   \boldsymbol{\ell}_K(y;F) &= \bigl(\ell_1(y;F),\ldots,\ell_K(y;F)\bigr)^{\top}.
\end{align*}
By linearity of the pathwise derivative and \cref{lmoment_eif}, every regular parametric submodel $\{F_\varepsilon\}$ through $F$ with score $s\in L_0^2(F)$ satisfies
\begin{align*}
    \left.\frac{d}{d\varepsilon} \boldsymbol{\lambda}_K^*(F_\varepsilon)\right|_{\varepsilon=0} = \mathbb{E}\!\left[\boldsymbol{\ell}_K(Y;F)s(Y)\right].
\end{align*}
Since each component of $\boldsymbol{\ell}_K$ is the canonical gradient of its corresponding moment, $\boldsymbol{\ell}_K$ is the efficient influence function for the vector $\boldsymbol{\lambda}_K^*$. The covariance matrix of $\boldsymbol{\ell}_K$ therefore determines the nonparametric efficiency bound for joint estimation, as formalized in what follows.

\begin{corollary}[Nonparametric Efficiency Bound] \label{lmoment_efficiency_bound}
Let $F\in\mathcal{P}_2(\mathbb{R})$ and fix $K<\infty$. For $\boldsymbol{\lambda}_K^*(F) =(\lambda_1^*(F),\ldots,\lambda_K^*(F))^{\top}$, the
nonparametric efficiency bound is
\begin{align*}
    \Sigma_K(F) = \mathbb{E}\!\left[\boldsymbol{\ell}_K(Y;F)\boldsymbol{\ell}_K(Y;F)^{\top} \right].
\end{align*}
For $1 \leq j, k \leq K$ its entries admit the density-free representation
\begin{align*}
    \Sigma_{jk}(F) &= \sqrt{(2j-1)(2k-1)} \sum_{a=0}^{j-1}\sum_{b=0}^{k-1} \xi_{ja}\xi_{kb} \operatorname{Cov}\!\left(M_a(Y_1;F),M_b(Y_1;F) \right) \\
 &= \int_{\mathbb{R}}\int_{\mathbb{R}} \phi_j(F(z))\phi_k(F(t))  \bigl( F(z\wedge t)-F(z)F(t)\bigr) dz\,dt.
\end{align*}
In particular, we say $\sigma^2_k = \Sigma_{kk}(F)$. Furthermore, $|\Sigma_{jk}(F)| \leq \sqrt{(2j-1)(2k-1)}\mathbb{V}(Y)$. 
\end{corollary}

An important takeaway from \cref{lmoment_efficiency_bound} is that, even though the shifted Legendre polynomials are orthogonal in $L^2(0, 1)$, the off-diagonals of $\Sigma_K(F)$ need not vanish. The same observation contributes to several shape summaries, and its sampling covariance is determined by the covariance kernel $F(z\wedge t)-F(z)F(t)$. In the next section, we define an estimator.

\subsection{Constructing an Estimator} \label{estimator_section}

Let $Y_1,\ldots,Y_n \overset{iid}{\sim} F$ for some $F\in\mathcal{P}_2(\mathbb{R})$, and consider the first $K<n$ normalized $L$-moments. In this section, we establish an estimator for $\boldsymbol{\lambda}^*_K(F)$. There are two approaches one could take: first, one could construct an unbiased $U$-statistic through the probability-weighted-moment definition given in \cref{normalized_lmoment_pwm}, and second, one could leverage the efficient influence function in \cref{lmoment_eif} to define the one-step estimator. Both estimators share the same expansion for the $k$th normalized $L$-moment,
\begin{align} \label{lmoment_linearization}
     \widehat{\lambda}^*_{k, n} -\lambda_k^*(F) =\frac{1}{n}\sum_{i=1}^n\ell_k(Y_i;F)+o_p(n^{-1/2})
\end{align}
and thus, both estimators share the same asymptotic covariance as described in \cref{lmoment_efficiency_bound}. However, their finite-sample behavior may be different and inference can differ depending on the estimator. Thus, we explore both approaches.

\subsubsection{The Unbiased $U$-Statistic} \label{u_statistic_section}

The most direct estimator is to leverage the $U$-statistic representation from \cite{hosking1990}. For a symmetric kernel $h(\cdot)$ of $r$ arguments, write
\begin{align*}
    U_{n,r}(h) = \binom nr^{-1} \sum_{1\leq i_1 < \cdots < i_r\leq n} h(Y_{i_1},\ldots,Y_{i_r}).
\end{align*}
Taking $h_r(y_1,\ldots,y_r)=\max\{y_1,\ldots,y_r\}$ gives us the following estimator for the probability-weighted moment,
\begin{align*}
    \widehat{\beta}_{j, n}=\frac{U_{n,j+1}(h_{j+1})}{j+1} \tag{$j=0,\ldots,K-1$}
\end{align*}
and so we may define $\widehat{\lambda}_{k,n}^* =\sqrt{2k-1}\sum_{j=0}^{k-1}\xi_{kj}\widehat{\beta}_{j, n}$ to be our estimator of the $k$th normalized $L$-moment. Although at first it appears as though this estimator requires us to enumerate over all subsets of $Y_1, \ldots, Y_n$, an equivalent estimator is given by \cite{hosking1990},
\begin{align*}
     \widehat{\beta}_{j, n} = \frac{1}{n} \sum_{i=j+1}^n \frac{\binom{i-1}{j}}{\binom{n-1}{j}}Y_{(i)}.
\end{align*}
Consequently, our estimator is simply a weighted sum of the ordered observations.

A key benefit of the $U$-statistic estimator is its finite-sample unbiasedness. Because each subset contains independent draws from $F$, it follows that $\mathbb{E}[\widehat{\beta}_{j, n}] =\beta_j(F)$ and therefore $\mathbb{E}[\widehat{\lambda}_{k,n}^*] = \lambda_k^*(F)$ exactly. Furthermore, we can establish a direct connection to \cref{lmoment_eif} by noting that
\begin{align*}
    \mathbb{E}[\widehat\beta_{j,n}\mid Y_1=y]-\beta_j(F) = \frac{1}{n}\left\{M_j(y;F)-(j+1)\beta_j(F)\right\}
\end{align*}
which gives us that $n (\mathbb{E}[\widehat\lambda_{k,n}^*\mid Y_1=y] -\lambda_k^*(F)) = \ell_k(y;F)$. Consequently, the Hoeffding decomposition \citep{hoeffding1948} implies the linearization \cref{lmoment_linearization}, and furthermore, that
\begin{align*}
    \mathbb{V}(\widehat{\lambda}_{k,n}^*) =\frac{\Sigma_{kk}(F)}{n}+ \mathbb{E}[R_{k,n}^2]
\end{align*}
where $R_{k,n}$ is some remainder and $\mathbb{E}[R_{k,n}^2] = O(n^{-2})$. Notably, \cite{ELAMIR2004337} show there is an unbiased estimator for the variance and covariance of sample $L$-moments. One final benefit of the $U$-statistic estimator is that $\widehat{\lambda}_{k,n}^*$ is a minimum variance unbiased estimator \citep{halmos46, hosking1990}. In the next section, we contrast the $U$-statistic estimator with its one-step estimator counterpart.

\subsubsection{The Empirical Plug-In and One-Step Estimation}

The second approach to consider is the empirical plug-in estimator, which coincides with the one-step estimator \citep{bickel1998efficient}. Specifically, applying the efficient influence function from \cref{lmoment_eif} to the empirical distribution gives the one-step update
\begin{align*}
    \widetilde{\lambda}_{k,n}^* = \lambda_k^*(F_n) + \frac{1}{n}\sum_{i=1}^n \ell_k(Y_i;F_n),
\end{align*}
where $F_n$ is the empirical cumulative distribution function. The correction removes the first-order effect of preliminary estimation error. In this setting, however, the correction vanishes. Because the influence function is centered under the distribution at which it is evaluated, $\frac{1}{n}\sum_{i=1}^n\ell_k(Y_i;F_n) = \int\ell_k(y;F_n) dF_n(y)=0$ exactly, and so the one-step update leaves the empirical plug-in unchanged and the estimator simplifies to
\begin{align*}
    \widetilde{\lambda}_{k,n }^* =\lambda_k^*(F_n) =\int_0^1 Q_n(u)\phi_k(u)du =\sum_{i=1}^n Y_{(i)} \int_{(i-1)/n}^{i/n}\phi_k(u)du.
\end{align*}
For each fixed $k$, assuming finite second moments, this estimator satisfies \cref{lmoment_linearization}. Its first-order behavior is therefore governed by the efficient influence function $\ell_k$, even though no explicit correction is needed. Consequently, its limiting variance attains the nonparametric efficiency bound in \cref{lmoment_efficiency_bound}, so no regular estimator has a smaller asymptotic variance \citep{van2000asymptotic, kennedy2024semiparametric}.

Unlike the $U$-statistic approach, this is biased in finite samples. For example, observe that
\begin{align*}
    \widetilde{\lambda}_{2,n}^* =\frac{\sqrt{3}}{n^2}\sum_{i<j}|Y_i-Y_j| =\frac{n-1}{n}\widehat{\lambda}_{2,n}^*,
\end{align*}
and so its bias is $-\lambda_2^*(F)/n$. However, this exact same equality implies that $\mathbb{V}(\widetilde{\lambda}_{2,n}^*) < \mathbb{V}(\widehat{\lambda}_{2,n}^*)$, so the bias comes with a smaller variance for this particular coordinate. More generally, the mean squared error depends on the distribution and sample size, so there is no universal ordering of the mean squared error for each estimator. For each fixed $k$, $ \widetilde{\lambda}_{k,n}^* -\widehat{\lambda}_{k,n}^* =O_p(n^{-1})$. Notably, the empirical cumulative distribution function construction has other advantages. Namely, its coefficients are exactly those of $Q_n$, so the collective estimator is given by
\begin{align*}
\widetilde{Q}_{K,n} =  \sum_{k=1}^K\widetilde{\lambda}_{k,n}^*\phi_k =\Pi_K Q_n.
\end{align*}
Thus, $\widetilde{\lambda}_{k,n }^*$ is especially useful for distribution reconstruction, and it allows us to easily and directly apply classical quantile-process theory to derive functional Gaussian convergence. In fact, the empirical estimator has the very important property that
\begin{align*}
     \|\widetilde{Q}_{K,n}-\Pi_K Q\|_2 \leq \|Q_n-Q\|_2.
\end{align*}
This inequality is notable because it does not depend on $K$. Consequently, we can derive Gaussian process limits that hold in a growing-$K$ setting without restricting the rate at which $K$ grows.

\begin{theorem}[Functional Gaussian Limit for Empirical $L$-Moments] \label{gaussian_process_limit}
Let $Y_1,\ldots,Y_n\overset{iid} \sim F$ with $F \in \mathcal{P}_2(\mathbb{R})$. Suppose the quantile function $Q$ is Lipschitz continuous on $(0, 1)$. Then,
\begin{align*}
    \sqrt{n}(Q_n-Q) \rightsquigarrow \mathbb{G}_F
\end{align*}
in $L^2(0, 1)$ where $\mathbb{G}_F$ is a centered Gaussian random element with the same law as $u \mapsto Q^\prime(u)\mathbb{B}(u)$ for a standard Brownian bridge $\mathbb{B}$. Its covariance kernel is
\begin{align*}
    C(u,v)=Q^\prime(u)Q^\prime(v)( u\wedge v-uv)
\end{align*}
for almost every $(u,v)\in(0,1)^2$. Furthermore, for every integer sequence $K_n\to\infty$,
\begin{align*}
    \sqrt{n}\left( \widetilde{Q}_{K_n,n}-\Pi_{K_n}Q \right) \rightsquigarrow \mathbb{G}_F
\end{align*}
in $L^2(0, 1)$. No upper growth restriction on $K_n$ is required. If, in addition, $\sqrt{n}\|Q-\Pi_{K_n}Q\|_2 \longrightarrow 0$ then the same limit holds for $\sqrt{n}(\widetilde{Q}_{K_n,n}-Q)$.
\end{theorem}

The Lipschitz condition on the quantile function in \cref{gaussian_process_limit} allows for flat quantile portions and does not require $Q^\prime$ to be continuous or bounded away from zero. However, it does exclude jumps in $Q$. Furthermore, there is no restriction on the rate at which $K_n$ grows; the empirical coefficients remain defined for every $k$, including $k > n$. On the other hand, for the unbiased $U$-statistic estimator, in order to obtain a Gaussian process limit the higher-order terms must be explicitly controlled. Consequently, we must impose some kind of bounded rate in $K_n$, which we formalize in the following corollary.

\begin{corollary}[Functional Gaussian Limit for $U$-Statistic Estimator]
\label{u_statistic_functional_limit}
Suppose the assumptions of \cref{gaussian_process_limit} hold and let
\begin{align*}
    \widehat{Q}_{K,n}=\sum_{k=1}^K\widehat{\lambda}_{k,n}^*\phi_k
\end{align*}
denote the collective $U$-statistic estimator. Let $K_n \leq n$ be deterministic integers satisfying $K_n\to\infty$ and $K_n^3/n \to 0$. Then
\begin{align*}
    \sqrt{n} \|\widehat{Q}_{K_n,n}-\widetilde{Q}_{K_n,n}\|_2
 \overset{P}{\longrightarrow} 0.
\end{align*}
Consequently, $\sqrt{n}(\widehat{Q}_{K_n,n}-\Pi_{K_n}Q) \rightsquigarrow \mathbb{G}_F$ in $L^2(0, 1)$ where $\mathbb{G}_F \overset{d}{=} Q^\prime \mathbb{B}$.
\end{corollary}

The additional rate condition $K^3_n / n \to 0$ in \cref{u_statistic_functional_limit} is used to control the higher order sampling variation. Specifically, in the Hoeffding decomposition we have $\widehat{\lambda}_{k,n}^*-\lambda_k^*(F) =\frac1n\sum_{i=1}^n\ell_k(Y_i;F)+R_{k,n}$. In order to control the remainder $R_{k,n}$ we show that
\begin{align*}
    \mathbb{E}\left[ \left\| \sqrt{n}\sum_{k=1}^{K_n}R_{k,n}\phi_k \right\|_2^2 \right] =O\!\left(\frac{K_n^3}{n}\right).
\end{align*}
Notably, in the proof of \cref{u_statistic_functional_limit} we only assume a Lipschitz condition on $Q$; if one were to assume stronger regularity conditions, such as the smoothness conditions explored in \cref{decay_section}, it is possible to obtain a more relaxed dependence on $K_n$. In the next section, we establish simultaneous inference for a collection of $L$-moments.

\subsection{Standard Errors and Simultaneous Inference}
\label{jackknife-bootstrap}

In this section we describe a jackknife estimator of the asymptotic variance of our $L$-moment estimators and a Gaussian multiplier bootstrap for simultaneous inference over the first $K_n$ moments; we explicitly allow for $K_n \to \infty$, although $K_n$ could also be treated as fixed. Note that, as discussed in \cref{estimator_section}, both the unbiased and empirical estimators share the same asymptotic variance given in \cref{lmoment_efficiency_bound}. Thus, our confidence bands hold for either approach. We let $\lambda_{k,n}^{\circ}$ denote either $\widehat{\lambda}_{k,n}^*$ or $\widetilde{\lambda}_{k,n}^*$, and similarly, we let $Q_{K,n}^{\circ}=\sum_{k=1}^K\lambda_{k,n}^{\circ}\phi_k$. Now, for $k<n$, we define
\begin{align*}
    \widehat{\ell}_{i,k} =(n-1)\left(\widehat{\lambda}_{k,n}^* -\widehat{\lambda}_{k,n}^{*,(-i)}\right),
\end{align*}
the covariance $\widehat{\Sigma}_{jk,n} =\frac{1}{n}\sum_{i=1}^n\widehat{\ell}_{i,j}\widehat\ell_{i,k}$, and variance $\widehat{\sigma}_{k,n}^2=\widehat{\Sigma}_{kk,n}$, where $\widehat{\lambda}_{k,n}^{*,(-i)}$ is the unbiased estimator after deleting observation $i$. This construction estimates the influence contributions without a density estimate or sample splitting; see \citet{arvesen1969} for classical jackknife theory for $U$-statistics. Following the discussion  in \cref{u_statistic_section} we know
\begin{align*}
    \mathbb{E}[\widehat\ell_{i,k}\mid Y_i]
 =\frac{n-1}{n}\ell_k(Y_i;F).
\end{align*}
Moreover, averaging the deleted estimators returns the full-sample $U$-statistic, so $\sum^n_{i=1}\widehat{\ell}_{i,k}=0$ exactly. The matrix $\widehat{\Sigma}_{K,n}=(\widehat{\Sigma}_{jk,n})_{j,k\le K_n}$ is therefore an empirical covariance matrix and is positive semidefinite.

\begin{remark}
    Throughout this section and in the associated proofs we use the unbiased estimator $\widehat{\lambda}^*_{k, n}$ to construct the jackknife variance estimator. We do so out of convenience, in order to leverage previous results established elsewhere in the manuscript. However, all results still hold using the empirical estimator $\widetilde{\lambda}^*_{k, n}$. The construction under the empirical estimator differs slightly. Define
    \begin{align*}
        \widetilde{\ell}_{i,k} = (n-1)\left\{ \frac{1}{n}\sum_{j=1}^n\widetilde{\lambda}_{k,n}^{*,(-j)} -\widetilde{\lambda}_{k,n}^{*,(-i)} \right\}
    \end{align*}
    and the covariance $\widetilde{\Sigma}_{jk,n} = \frac{1}{n}\sum_{i=1}^n\widetilde{\ell}_{i,j}\widetilde{\ell}_{i,k}$ where $\widetilde{\lambda}_{k,n}^{*,(-i)}= \lambda_k^*(F_{n,-i})$ and $F_{n,-i}$ is the empirical distribution after deleting observation $i$. Here, the deleted estimates are centered at their average, which is not necessarily equal to the full-sample empirical estimate $\widetilde{\lambda}_{k,n}^*$. We center to ensure that $\sum^n_{i=1}\widetilde{\ell}_{i,k}=0$, which makes $\widetilde{\Sigma}_{K,n}$ positive semidefinite. 
\end{remark}
Conditionally on the data $\mathcal{D}_n$, generate independent draws $\xi_1, \ldots, \xi_n \sim \mathcal{N}(0,1)$ and define
\begin{align*}
    Z_{k,n}^{(\mathrm{b})}=\frac{1}{\sqrt{n}}\sum_{i=1}^n \xi_i\widehat{\ell}_{i,k} \qquad \text{and} \qquad T_n^{(\mathrm{b})}=\max_{k \leq K_n} \left\{ \frac{|Z_{k,n}^{(\mathrm{b})}|}{\widehat{\sigma}_{k,n}} \right\}.
\end{align*}
Conditional on the data, $(Z^{(\mathrm{b})}_{1, n},\ldots,Z^{(\mathrm{b})}_{K_n, n})^\top$ is exactly Gaussian.  Its correlation structure is determined by the empirical covariances of the jackknife influence contributions and therefore allows the different $L$-moment estimators to be arbitrarily correlated. Now, for some $\alpha \in (0, 1)$ let
\begin{align*}
     c_{n, 1-\alpha}^* = \inf \! \left\{ c\in\mathbb{R}: \mathbb{P}\!\left(T^{(\mathrm{b})}_n \leq c \mid \mathcal{D}_n\right) \geq 1-\alpha \right\}
\end{align*}
denote the critical value. In practice, we generate $B$ independent multiplier draws and compute $T_{n}^{(1)},\ldots,T_{n}^{(B)}$. Then, we let $\widehat{c}_{n, 1-\alpha}$ denote the empirical conditional $1 - \alpha$ quantile of $T_n^{(\mathrm{b})}$. With these definitions in place, in the following theorem we establish the conditions required for simultaneous inference across $K_n$ $L$-moments.

\begin{theorem}[Simultaneous Inference]
\label{simultaneous-band}
Suppose the assumptions of \cref{gaussian_process_limit} hold. Namely, suppose that the quantile function $Q$ is Lipschitz on $(0, 1)$. Let $K_n < n$ be a deterministic sequence where $K_n\to\infty$. Then, suppose that $\sigma_k \geq c_\sigma / k$ for $1 \leq k \leq K_n$ for some fixed $c_\sigma > 0$. Further suppose that $ K_n^5( \log(enK_n))^4 / n \longrightarrow 0.$ For either estimator, let
\begin{align*}
     T_n^{\circ}=\max_{k\leq K_n} \left\{ \frac{\sqrt{n}|\lambda_{k,n}^{\circ}-\lambda_k^*(F)|} {\widehat{\sigma}_{k,n}} \right\}.
\end{align*}
Then, the multiplier approximation satisfies
\begin{align*}
   \sup_{t\in\mathbb{R}} \left|\mathbb{P}(T_n^{\circ}\leq t)-\mathbb{P}(T_n^{(\mathrm{b})}\leq t \mid \mathcal{D}_n)\right| \overset{P}{\longrightarrow} 0. 
\end{align*}
Consequently, for every fixed $\alpha\in(0,1)$, the intervals $I_{k,n}^{\circ} =[\lambda_{k,n}^{\circ} \ \pm\ c_{n,1-\alpha}^* \widehat{\sigma}_{k,n}/\sqrt{n}]$ satisfy $  \mathbb{P}(\lambda_k^*(F)\in I_{k,n}^{\circ}\text{ for all } k\leq K_n) \to1-\alpha $. In addition, for $u \in [0, 1]$, the band
\begin{align*}
    B_n^{\circ}(u) =\left[Q_{K_n,n}^{\circ}(u) \ \pm\ \frac{c_{n,1-\alpha}^*}{\sqrt{n}} \sum_{k=1}^{K_n}\widehat{\sigma}_{k,n}|\phi_k(u)|\right],
\end{align*}
obeys $\liminf_{n\to\infty} \mathbb{P}((\Pi_{K_n}Q)(u)\in B_n^{\circ}(u) \text{ for every } u\in[0,1])\geq 1-\alpha$. These conclusions remain valid when the conditional critical value is estimated from $B_n\to\infty$ independent multiplier draws.
\end{theorem}

\cref{simultaneous-band} permits joint inference on an increasing
number of $L$-moments under either the unbiased or empirical estimators. Unlike the preceding $L^2$ limits, this result controls the largest standardized coordinate error, which is why we require a more restrictive growth condition on $K_n$. The coordinate intervals have asymptotically exact simultaneous coverage, whereas the reconstructed band can be conservative because it bounds the curve's estimation error by summing the absolute contributions of the coordinate errors. Importantly, this band covers the projection $\Pi_{K_n}Q$, rather than $Q$ itself: it accounts for sampling uncertainty in the retained coordinates but not the approximation error from truncation. Coverage of the full quantile function would additionally require controlling that remainder uniformly over ranks. 

Here, we have established estimation and inference in a one-sample setting. In \cref{did_estimation}, we extend the estimation and inference results developed above to several difference-in-differences settings under additive $L$-moment parallel trends. We consider both repeated cross-sectional and panel data, allowing for dependence across periods within individuals in the latter setting. We also extend the approach to staggered treatment adoption, following the cohort-time framework of \citet{CALLAWAY2021200}, where $L$-moment treatment effects are identified for individual cohorts and post-treatment periods and can be aggregated using fixed weights. In each setting, the estimators admit an asymptotic linear representation obtained by combining the influence contributions of the relevant $L$-moment estimators. These representations allow us to apply the simultaneous inference procedure of \cref{simultaneous-band} directly to the difference-in-differences estimators. Specifically, we construct jackknife estimates of the influence contributions for each $L$-moment and combine them using the same differences as in the corresponding DiD estimator. The resulting influence contributions can then be used to estimate the covariance matrix and implement the Gaussian multiplier bootstrap of \cref{simultaneous-band}, with a maximum statistic standardized by the estimated standard errors.\footnote{The same multiplier is assigned to each independent sampling unit across all $L$-moments and comparisons, preserving the dependence induced by repeated observations in panel data and shared comparison groups under staggered adoption.} \cref{did_estimation} provides the corresponding influence contributions, covariance estimators, and bootstrap implementation for each setting. We also establish estimation under the changes-in-changes framework, although we do not cover inference as this extension is less straightforward than in the additive model.

\section{Choosing $K$} \label{choosing_k_section}

Now we consider a data-dependent method to choose $K$. When practitioners are interested in particular distributional components then $K$ can be chosen by scientific interest; however, for distributional reconstruction $K$ determines both the approximation error and the amount of sampling variation. We'll focus on estimating a single quantile function; the extension to differences of quantile functions follows easily.

Let $Y_1,\ldots, Y_n \overset{iid}{\sim} F$ where $F \in \mathcal{P}_2(\mathbb{R})$, with quantile function $Q\, \mathord{:} \, [0,1]\to\mathbb R$. Throughout, we write $\lambda^*_k = \lambda^*_k(F)$ and we retain the notation $Q^\circ_{K, n} = \sum^K_{k=1} \lambda^\circ_{k, n} \phi_k$ for either estimator, as outlined in \cref{jackknife-bootstrap}. For each fixed $K$, orthonormality gives the risk decomposition
\begin{align*}
    \mathcal{R}^\circ_n(K; Q) = \mathbb{E} \! \left[ \| Q^\circ_{K, n} - Q \|^2_2 \right] = \sum^K_{k=1} \mathbb{E} \! \left[(\lambda^\circ_{k, n} - \lambda^*_k)^2 \right] + R^2_K(Q).
\end{align*}
When the unbiased $U$-statistic estimator is used, the first term in the summation is exactly the coordinatewise variances of each $L$-moment. The second term, $R^2_K(Q)$, is the squared projection remainder explored in \cref{decay_section}. Next, we consider selection rules for $K$ assuming a finite Legendre expansion.

\subsection{Selection Under a Finite Expansion} \label{finite_basis_section}

To begin, we consider a strong sparsity assumption. That is, we suppose that $Q$ has a finite Legendre expansion such that there exists some last nonzero coordinate $K_0$. Equivalently, we assume $\lambda_k = 0$ for all $k > K_0$. Now, let $p_n=\kappa\log(n)/n$ denote a penalty, where $n \geq 2$ and $\kappa > 0$ is fixed. Define
\begin{align} \label{eq:selector}
    \widehat{K}_s =  \min\argmin_{1 \leq K \leq K_n} \left\{-\sum_{k=1}^K (\widetilde{\lambda}^*_{k, n})^2 + K p_n\right\}.
\end{align}
Since $\|Q_n - \widetilde{Q}_{K,n} \|^2_2 = \|Q_n\|^2_2 - \sum^K_{k=1} (\widetilde{\lambda}^*_{k, n})^2$ this criterion trades off the fit to $Q_n$ against the number of retained coordinates; this follows from classical penalized projection approaches for model selection \citep{Barron1999}. Notably, the minimizer exists: $\sum^K_{k = 1}(\widetilde{\lambda}^*_{k, n})^2= \| \widetilde{Q}_{K, n} \|_2^2<\infty$, while $Kp_n\to\infty$ as $K\to\infty$ for each fixed $n$.

\begin{theorem}[Selection Under a Finite Expansion] \label{thm::sparse}
Suppose  $Q \in C^1([0, 1])$ and that
$0<c\le Q^\prime(u)\le L<\infty$ for $u \in [0, 1]$.
Assume also that $\lambda_k = 0$ for all $k > K_0$ where
$\lambda_{K_0}\ne 0$, and that $K_n \geq K_0$ for all sufficiently large $n$ (i.e.\ assume $K_n \to \infty$). Then, for every $\kappa > 0$, 
\begin{align*}
    \Pp(\widehat{K}_s \ne K_0) &= O(n^{-2\kappa/L^2}), \\
    \mathbb{E}[\|\widetilde{Q}_{\widehat{K}_s, n} - Q \|^2_2] &= \frac{V_{K_0}(F)}{n} + o(n^{-1})
\end{align*}
where $V_K(F) = \sum^K_{k=1} \sigma^2_{k}$. If $\kappa > L^2 / 2$ then we can further say that $\Pp(\widehat{K}_s \ne K_0) = o(n^{-1})$. Moreover, the untruncated empirical quantile estimator has risk
\begin{align*}
    \mathbb{E}[ \| Q_n-Q \|_2^2 ] =\frac{V_\infty(F)}{n}+o(n^{-1}),
\end{align*}
where $V_\infty(F) = \int_0^1 u(1-u)Q^\prime(u)^2du > V_{K_0}(F)$, so the truncated estimator strictly improves the first-order risk constant.
\end{theorem}

By \cref{thm::sparse} we can see that this selected estimator achieves the same first-order risk as the projection that knows $K_0$, but a strictly smaller first-order risk constant than $Q_n$. Notably, this holds under a strict sparsity assumption. Next, we explore settings in which the coefficients approach zero with $n$.

\subsection{Selection by Estimated Risk}

In the previous section our results depended on the strong sparsity assumption.
The reader may wonder whether we can relax
this to the case where the $\lambda_k$'s decrease but are not exactly zero.
A natural idea in this case is to
choose $K$ using risk estimation, following the principles underlying Mallows' $C_p$ \citep{mallows73}. For this approach, we use the unbiased estimator $\widehat{\lambda}^*_{k, n}$ and we let $v_{k, n} = \mathbb{V}(\widehat{\lambda}^*_{k, n})$ denote its variance. Then,
\begin{align*}
     \mathcal{R}_n(K; Q) =  \mathbb{E} \! \left[ \| \widehat{Q}_{K, n} - Q \|^2_2 \right]= \sum^K_{k=1} v_{k,n} + R^2_K(Q) = \|Q\|^2_2 + \sum^K_{k=1} \left\{ v_{k, n} - (\lambda^*_k)^2 \right\}.
\end{align*}
Since $ \|Q\|^2_2$ does not depend on $K$
it suffices to minimize
$R(K) =  \sum^K_{k=1} \{ v_{k, n} - (\lambda^*_k)^2\}$. In particular, if $M_{k,n}$ is an unbiased estimate of $(\lambda^*_k)^2$, then
\begin{align*}
    \widehat{R}(K) = \sum^K_{k=1} \left\{(\widehat{\lambda}^*_{k,n})^2 - 2M_{k, n} \right\}
\end{align*}
satisfies $\mathbb{E}[\widehat{R}(K)] = \mathcal{R}_n(K; Q) - \|Q\|^2_2$, and so $\widehat{R}(K)$ estimates risk up to a constant for every $K$. It remains to define $M_{k,n}$; for this, we employ repeated sample splitting. Fix $K_{\max}\leq\lfloor n/2\rfloor$ and, independently of the outcomes, draw $B$ uniform random partitions $(\mathcal{I}_b, \mathcal{J}_b)$ of $\{1,\ldots,n\}$ with $|\mathcal{I}_b|= \lfloor n/2\rfloor$ and $|\mathcal{J}_b|=n-\lfloor n/2\rfloor$. For $S_A=(Y_i:i\in A)$, let $\widehat{\lambda}_k^*(S_A)$ denote the unbiased estimator computed on that subsample. Then, we let
\begin{align*}
    M_{k,n} =\frac{1}{B}\sum_{b=1}^B \widehat{\lambda}_k^*(S_{\mathcal{I}_b}) \widehat{\lambda}_k^*(S_{\mathcal{J}_b}).
\end{align*}
Because two factors in each product use disjoint independent observations their product has expectation $(\lambda_k^*)^2$. Then, $\widehat{K}_R = \min\operatorname*{arg\,min}_{1\leq K\leq K_{\max}} \widehat{R}(K)$ and
\begin{align*}
    \widehat{Q}_{\widehat{K}_R,n} =\sum_{k=1}^{\widehat{K}_R}\widehat{\lambda}_{k,n}^*\phi_k.
\end{align*}
We explore the performance of this selection procedure via simulation study in \cref{risk_appendix}. In the following section, we explore the benefits of the $L$-moment framework through an empirical application.

\section{Empirical Application} \label{simulations_and_application}

We illustrate the usefulness of $L$-moment difference-in-differences using the introduction of Medicare Part D, following the empirical design of \citet{EngelhardtGruber2011}. Medicare Part D, implemented in 2006, substantially expanded public prescription-drug coverage for Medicare beneficiaries. A central objective of the program was not only to reduce average out-of-pocket prescription-drug expenditures, since the financial-protection value of insurance depends importantly on whether the program reduces unusually large out-of-pocket expenditures rather than merely lowering expenditure on average. This makes the setting particularly well suited to studying treatment effects on features of the outcome distribution beyond its mean. \citet{EngelhardtGruber2011} argue that because out-of-pocket prescription-drug spending is highly right-skewed, effects on the mean may provide an incomplete description of the program's financial-protection benefits and complement their mean estimates with difference-in-differences at selected quantiles of the spending distribution. The application therefore closely parallels the ``missing middle'' motivating our approach. A conventional difference-in-differences analysis estimates the effect of Part D on mean spending, imposing parallel trends only on the first $L$-moment. At the other extreme, comparing changes throughout the quantile function provides a much richer description of distributional effects, but implicitly requires the counterfactual evolution of the distribution to be correctly specified at a much finer level. $L$-moment DiD provides an intermediate approach. By estimating treatment effects on the first several $L$-moments, we can separately characterize economically meaningful changes in location, scale, and shape without requiring parallel trends for the entire quantile function. This distributional emphasis has continued in subsequent work on Medicare Part D. For example, \citet{CarvalhoEtAl2019} use MEPS data from 1997--2011 and apply a difference-in-differences design to concentration and generalized concentration indices of prescription-drug expenditures, explicitly studying how Part D affected the distribution of spending across the income distribution.

\begin{table}[htbp]
\centering \scalebox{.82}{
\begin{threeparttable}
\caption{Effect of Medicare Part D on Out-of-Pocket Prescription Drug Spending}
\label{tab:oop_lmoment_did}
\begin{tabular}{lrrcc}
\toprule
 & Estimate & SE & Pointwise 95\% CI & Simultaneous 95\% CI \\
\midrule
\multicolumn{5}{l}{\textit{$L$-moments (2007 dollars)}} \\
$\lambda_1$ & $-$279.0 & 46.6 & [$-$370.4, $-$187.6] & [$-$382.4, $-$175.6] \\
$\lambda_2$ & $-$170.1 & 36.2 & [$-$241.0, $-$99.1] & [$-$250.4, $-$89.8] \\
$\lambda_3$ & $-$69.1 & 28.0 & [$-$124.0, $-$14.2] & [$-$131.3, $-$6.9] \\
$\lambda_4$ & $-$24.7 & 23.1 & [$-$70.1, 20.7] & [$-$76.1, 26.6] \\
\addlinespace
\multicolumn{5}{l}{\textit{$L$-moment ratios}} \\
Gini coefficient ($\tau_2$) & 0.013 & 0.014 & [$-$0.015, 0.041] & --- \\
$L$-skewness ($\tau_3$) & 0.034 & 0.026 & [$-$0.016, 0.085] & --- \\
$L$-kurtosis ($\tau_4$) & 0.054 & 0.031 & [$-$0.007, 0.116] & --- \\
\addlinespace
\multicolumn{5}{l}{\textit{Effect heterogeneity (2007 dollars)}} \\
SD of individual effects, lower bound & 294.6 & 62.7 & $[171.7, \infty)$ & --- \\
\midrule
\multicolumn{5}{l}{Shared multiplier-bootstrap critical value: $\hat c = 2.218$} \\
\bottomrule
\end{tabular}
\begin{tablenotes}[flushleft]\footnotesize
\item \textit{Notes:} 10{,}000 Gaussian multiplier draws used in calculating $\hat{c}$.  The SD lower bound is $\sqrt{3}\,|\hat\Delta_{\lambda_2}|$; its confidence interval is one-sided, $\sqrt{3}\max\{0, |\hat\Delta_{\lambda_2}| - 1.96\,\mathrm{SE}\}$. MEPS data of \cite{EngelhardtGruber2011}, age 60+ sample, 2007 vs. 2005 comparison. Cell sizes: treated 3{,}592 / 3{,}470, control 1{,}154 / 1{,}237 (pre / post).
\end{tablenotes}
\end{threeparttable}}
\end{table}

We use the Medical Expenditure Panel Survey (MEPS) data underlying
\citet{EngelhardtGruber2011}, using waves of the MEPS Full-Year Consolidated Files, where 2006 is a transition year and 2007 is the only fully treated year. MEPS is a nationally representative survey containing detailed individual-level information on health insurance, utilization, and medical expenditures. Of particular importance here, prescription drug expenditures are separately recorded according to source of payment. Following Engelhardt and Gruber's DiD design, our baseline sample consists of individuals aged 60+, with those aged 65+ forming the treated group and those aged 60--64 serving as the comparison group. The treatment occurs with the implementation of Part D, so that the basic design has the standard block-treatment structure.\footnote{Engelhardt and Gruber also consider using only individuals age 65--70 as the treated group, and we find similar results as presented here with this restricted sample. They also use the average values of the years 2002--2005 as their pre-treatment for their estimate of the average effect and find a similar result as we do using only 2005 ($-293$ instead of $-279$).} 

For our analysis, we estimate the effects on the first four $L$-moments as these are the most interpretable, as well as the effect on the Gini coefficient $\lambda_2/\lambda_1$, $L$-skewness $\lambda_3/\lambda_2$, and $L$-kurtosis $\lambda_4/\lambda_2$. For the $L$-moment ratio effects, these are estimated using their observed versus their counterfactual values implied by the estimates for the $L$-moments and their standard errors are computed using the delta method.\footnote{For example, with the Gini coefficient, this is computed using the treated group's observed post-period $\widehat{\lambda}_2/\widehat{\lambda}_1$ minus its counterfactual, with counterfactual $L$-moments $\widehat{\lambda}_k(T,\mathrm{pre}) + \widehat{\lambda}_k(C,\mathrm{post}) - \widehat{\lambda}_k(C,\mathrm{pre})$.} As in \cite{EngelhardtGruber2011}, we find a substantial decrease in average out-of-pocket spending, but we also find a statistically significant decrease in the second and third $L$-moments, while the effect on the fourth $L$-moment is not statistically significant. However, it is interesting that there are (statistically insignificant) positive effects on the Gini coefficient, as well as on $L$-skewness and $L$-kurtosis. For the Gini coefficient, this is consistent with the idea that while the dispersion of spending was decreased, the proportional decrease in dispersion was similar to the proportional decrease in average, keeping the relative degree of inequality similar. We also use the lower bound on the variance of individual treatment effects in \cref{reconstruction_subsection}. That lower bound implies that the standard deviation of individual treatment effects among the treated group is lower bounded by $\sqrt{3}$ times the magnitude of the effect on the second $L$-moment, which in this case gives a lower bound of \$294.6 for the standard deviation of how much the elderly out-of-pocket spending was affected by Medicare Part D in 2007. Intuitively, because of how much the dispersion of out-of-pocket spending was reduced in absolute terms, we can be confident that there must have been significant heterogeneity around the average effect, with in this case the lower bound on the standard deviation of treatment effects even exceeding the magnitude of the average treatment effect.\footnote{This interpretation relies on the counterfactual population of the treated group and the actual population being the same, which is reasonable in this setting, given that treatment group membership is only based on age.}

\section{Discussion and Conclusion} \label{discussion_and_conclusion}

We have argued that an $L$-moment-based framework for parallel trends has several advantages. First, the implementation closely follows the conventional difference-in-differences approaches, where the sample $L$-moments replace sample means in the additive model. The jackknife and multiplier bootstrap provide standard errors and simultaneous inference without the need for density estimation, and these procedures apply to repeated cross-sectional, panel, and staggered-adoption settings. Additionally, researchers can use the framework to approximate the full counterfactual quantile function or to study only specific interpretable features of the distribution. The point is that distributional policy evaluation does not require identifying the entire counterfactual distribution. Imposing the parallel trends assumption on only selected $L$-moments identifies the corresponding effects \emph{without requiring untargeted components to be zero or negligible}. For example, the effects on dispersion can still be identified under the corresponding restriction even when the parallel trends assumption for averages fails. 

Furthermore, the identified higher-order $L$-moment effects tell us something about how differently treatment affects specific individuals. If everyone experienced the same effect, the treatment would only shift the outcome distribution, which would leave all higher-order $L$-moments unchanged. Nonzero higher-order effects therefore imply that individual effects must differ somehow. Under the finite-second-moment conditions of \cref{individual_heterogeneity_bound}, we can quantify this implication through a lower bound on the variance of individual treatment effects, without requiring rank invariance or even identification of the average effect.

Selecting a set $\mathcal{K}$ of features of interest is also distinct from choosing a truncation level for reconstruction: our approximation results do support the latter, but adding causally interpreted coordinates also adds identifying restrictions. Future work could adapt existing pre-trend methods to $L$-moments \citep{Roth2022,Dette2024,BorusyakJaravelSpiess2024} and develop sensitivity analyses allowing bounded violations of the targeted restrictions \citep{rambachanRoth2023,ManskiPepper2000}. Another direction is to use the framework to actually compare distributional methods. While we focus on the additive model, the changes-in-changes extension illustrates how the same decomposition yields $L$-moment estimates under alternative identifying assumptions. Comparing these estimates can show which conclusions depend on the method used and help assess which features of counterfactual evolution we can credibly compare between groups.


\section*{Acknowledgments}

GPT-6 Astra was used for drafting, editing, and proof assistance. Claude Opus 5.5 was used during the coding of the simulation study and application. \medskip

This paper is a product of the Iowa Agriculture and Home Economics Experiment Station, Ames, Iowa. Project No. IOW05806 is supported by USDA/NIFA and State of Iowa funds. Any opinions, findings, conclusions, or recommendations expressed in this publication are those of the authors and do not necessarily reflect the views of the U.S. Department of Agriculture. 

\section*{References}
\vspace{-1cm}
\bibliographystyle{abbrvnat}
\bibliography{references}

\setlength{\parindent}{0cm}

\appendix
\crefalias{section}{appendix}

\begin{center}
{\large\bf SUPPLEMENTARY MATERIAL}
\end{center}

\begin{description}

\item \cref{regularity_conditions}: Contains an extended discussion of regularity conditions for $L$-moments.

\item \cref{additive_vs_cic}: Contains an extended discussion on the differences between the extrapolations defined under the additive-quantile and changes-in-changes parallel trends assumptions.

\item \cref{did_estimation}: Contains methods for estimation and inference of the additive model and an extended discussion of estimation under the changes-in-changes identification framework.

\item \cref{risk_appendix}: Contains an extended discussion of selecting $K$ via risk estimation.

\item \cref{sec:proofs}: Contains all proofs from the main text and appendix, including:
\begin{description}
    \item \cref{parallel_trends_spectrum_corollary_proof}: Proof of \cref{parallel_trends_spectrum_corollary}.
    \item \cref{identification_theorem_proof}: Proof of \cref{identification_theorem}.
    \item \cref{identification_cic_proof}: Proof of \cref{identification_cic}.
    \item \cref{individual_heterogeneity_bound_proof}: Proof of \cref{individual_heterogeneity_bound}.
    \item \cref{moment_bound_proof}: Proof of \cref{moment_bound}.
    \item \cref{higher_order_energy_bound_proof}: Proof of \cref{higher_order_energy_bound}.
    \item \cref{rk_bound_general_proof}: Proof of \cref{rk_bound_general}.
    \item \cref{decay_smoothness_proof}: Proof of \cref{decay_smoothness}.
    \item \cref{counterfactual_cancellation_proof}: Proof of \cref{counterfactual_cancellation}.
    \item \cref{lmoment_eif_proof}: Proof of \cref{lmoment_eif}.
    \item \cref{lmoment_effiency_bound_proof}: Proof of \cref{lmoment_efficiency_bound}.
    \item \cref{gaussian_process_limit_proof}: Proof of \cref{gaussian_process_limit}.
    \item \cref{u_statistic_functional_limit_proof}: Proof of \cref{u_statistic_functional_limit}.
    \item \cref{simultaneous-band_proof}: Proof of \cref{simultaneous-band}.
    \item \cref{additive_cic_bound_proof}: Proof of \cref{additive_cic_bound}.
    \item \cref{lmoment_disagreement_lower_bound_proof}: Proof of \cref{lmoment_disagreement_lower_bound}.
    \item \cref{thm::sparse_proof}: Proof of \cref{thm::sparse}.
\end{description}
\end{description}

\section{Regularity of $L$-Moments and Quantile Functions} \label{regularity_conditions}

In this section we formalize the regularity conditions required for the three roles $L$-moments play in the identification and approximation of quantile functions. Namely, determination of a quantile function by its complete coefficient sequence, approximation by finite summation of coefficients, and pointwise convergence of that truncation. Recall that, for a random variable $X$ with quantile function $Q$,
\begin{align*}
     \int_0^1|Q(u)|^p du =\mathbb E[|X|^p]
\end{align*}
for $1\leq p<\infty$. In particular, $Q\in L^1(0,1)$ is equivalent to a finite absolute first moment, whereas $Q\in L^2(0,1)$ is equivalent to a finite second moment.

\subsection{Guarantees Under a Finite First Moment}
To begin, recall that we define $\phi_k(u) =\sqrt{2k-1} P_{k-1}^*(u)$ where $P_{k-1}^*$ is the shifted Legendre polynomial of degree $k-1$. For $f\in L^1(0,1)$, define its $k$th normalized coefficient by $c_k(f) =\int_0^1 f(u)\phi_k(u)du$, and define the associated $K$-term truncation by
\begin{align} \label{pi_approximation}
    \Pi_Kf(u) = \sum_{k=1}^K c_k(f)\phi_k(u).
\end{align}
Immediately, because each $\phi_k$ is continuous, and hence bounded, on $[0,1]$, every coefficient is finite whenever $f\in L^1(0,1)$, since
\begin{align*}
    |c_k(f)| \leq \|f\|_{L^1}\|\phi_k\|_{L^\infty} <\infty.
\end{align*}
Thus, $L^1$ integrability is sufficient for the complete coefficient sequence $\{c_k(f)\}^\infty_{k=1}$ to be well-defined. \medskip

As shown in the proof of \cref{parallel_trends_spectrum_corollary}, when $f \in L^1(0, 1)$, then the coefficient sequence uniquely defines $f$. That is, if $f,g\in L^1(0,1)$, then $c_k(f)=c_k(g)$ for every integer $k \geq 1$ if and only if $f = g$ Lebesgue-almost everywhere on $(0, 1)$. The key observation is recognizing that every polynomial $p$ can be written as a finite linear combination
\begin{align*}
    p(u)=\sum_{k=1}^{m+1}a_k\phi_k(u)
\end{align*}
for some $m$ and some coefficients $a_1,\ldots,a_{m+1}$. Thus, letting $h = f - g \in L^1(0, 1)$, for every polynomial $p$ it follows that
\begin{align*}
    \int_0^1h(u)\phi_k(u) du = 0 \qquad \Longleftrightarrow \qquad \int_0^1h(u)p(u)du = 0.
\end{align*}
The polynomials are uniformly dense in $C[0, 1]$. Thus, for every $\ell \in C[0,1]$, there is a sequence of polynomials $\{p_m\}$ satisfying $\|\ell -p_m\|_{L^\infty} \to 0$ \citep{rudin1976principles}. It then follows that
\begin{align*}
    \left|\int_0^1h(u)\ell(u)du\right| =\left|\int_0^1h(u)\{\ell(u)-p_m(u)\}du\right| \leq \|h\|_{L^1}\|\ell-p_m\|_{L^\infty} \longrightarrow 0.
\end{align*}
Then, by uniqueness in the Riesz--Markov representation theorem, the finite signed measure $h(u)du$ is therefore the zero measure, so $h=0$ Lebesgue-almost everywhere \citep{rudin1987real}. Importantly, this is a \textit{uniqueness} result. It says that the coefficient map $f \mapsto\{c_k(f):k\geq1\}$ is injective. That is, if an exact complete coefficient sequence comes from an $L^1$ function, there is at most one such function, up to almost-everywhere equality, that could have generated it. This is sufficient for identification, as shown in \cref{identification_theorem}. However, it does not guarantee anything about the convergence of the sequence of truncations $\{\Pi_K f \}_{K \geq 1}$. \medskip 

It may be tempting to infer the stronger statement, that for every $f \in L^1(0, 1)$
\begin{align} \label{false_convergence}
  \lim_{K \to \infty}  \|f- \Pi_Kf\|_{L^1} = 0.
\end{align}
However, this is not true in general. To understand why, let $\mathcal{V}_K =\operatorname{span}\{\phi_1,\ldots,\phi_K\}$ denote the space of polynomials of degree at most $K-1$. Uniform density of polynomials in $C[0,1]$, together with density of $C[0,1]$ in $L^1(0,1)$, implies that
\begin{align*}
    \lim_{K\to\infty} \left\{ \inf_{p\in\mathcal{V}_K} \|f-p\|_{L^1} \right\} =0.
\end{align*}
This guarantees the existence of increasingly accurate polynomial approximations, but does not say that the particular polynomial $\Pi_Kf$ is one of them. The distinction between these two approximation rules is visible even at $K=1$. Because $\phi_1 = 1$, then $\Pi_1f=\int_0^1f(u)du$, so the first Legendre truncation is the mean of $f(U)$ for $U\sim\operatorname{Uniform}(0,1)$. However, a best constant approximation under $L^1$ loss is a median of $f(U)$. For example, if $f(u) =\mathbb{I}(u>0.9)$, then $\Pi_1f=0.1$, whereas the unique $L^1$-best constant is zero. Thus, even at the first coordinate, $\Pi_K$ is not the solution to the best-$L^1$ approximation problem. \medskip

Notably, $L^1$ reconstruction of the form \cref{false_convergence} is possible, however, not under the approximation defined by the shifted Legendre polynomials. Instead, we can consider a Haar expansion \citep{Haar1910}. For $J\geq0$, partition $(0,1)$ into the dyadic intervals
\begin{align*}
    I_{J,m} =[m2^{-J},(m+1)2^{-J})\cap(0,1), \qquad m=0,\ldots,2^J-1,
\end{align*}
and define the approximation
\begin{align*}
    \mathsf{H}_J f(u) = \sum_{m=0}^{2^J-1} \left\{2^J\int_{I_{J,m}}f(v)dv\right\} \mathbb{I}(u\in I_{J,m}).
\end{align*}
Thus, $\mathsf{H}_Jf$ replaces $f$ on each interval by its average over that interval. It can be shown that, for every $f \in L^1(0, 1)$
\begin{align*}
    \lim_{J\to\infty}\|f-\mathsf{H}_Jf\|_{L^1}=0.
\end{align*}
Under this interval-based construction, instead of $L$-moment-based restrictions as in \cref{lpt_definition} and \cref{cic_pt} we could instead define Haar parallel trends 
\begin{align*}
    \int_{I_{J,m}}\mathcal D^0(u)du=0, \qquad m=0,\ldots,2^J-1.
\end{align*}
At $J=0$, this is mean parallel trends; at $J=1$, it imposes
parallel trends separately on the integrated quantiles in the lower
and upper halves of the distribution. Finer resolutions yield
restrictions on progressively smaller rank intervals. While this is a valid approach for defining an alternative spectrum of parallel trends restrictions, it loses the clean location, dispersion, skewness, and shape coordinates interpretation of $L$-moments described in \cref{pt_spectrum_section}. In the next section, we establish convergence results under the shifted Legendre polynomials when $f \in L^2(0, 1)$.

\subsection{Guarantees Under a Finite Second Moment}

A finite first moment guarantees that the complete $L$-moment sequence uniquely determines the distribution. A finite second moment additionally guarantees convergence of the truncation $\lim_{K \to \infty} \| f - \Pi_K f \|_{L^2}$ for $f \in L^2(0, 1)$. \medskip

\textbf{Orthogonal Projection and Reconstruction:} To begin, recall that the normalized shifted Legendre polynomials form a complete orthonormal basis of $L^2(0, 1)$ \citep{hosking1990}. Thus, it follows that the coefficient truncation $\Pi_K$ is the unique orthogonal projection onto $\mathcal{V}_K$:
\begin{align*}
    \Pi_Kf =\sum_{k=1}^Kc_k(f)\phi_k =\operatorname*{arg\,min}_{v\in\mathcal{V}_K}\|f-v\|_{L^2}.
\end{align*}
Thus, unlike the general $L^1$ case, each truncation is already the best approximation of its degree in integrated squared error. By completeness and Parseval's identity \citep{conway1985course}, it follows that $\lim_{K\to\infty}\|f-\Pi_Kf\|_{L^2} = 0$. Furthermore,
\begin{align*}
    \|f\|_{L^2}^2=\sum_{k=1}^{\infty}|c_k(f)|^2 \qquad \text{and} \qquad  \|f-\Pi_Kf\|_{L^2}^2=\sum_{k>K}|c_k(f)|^2.
\end{align*}
Thus, the squared approximation error cannot increase as coefficients are added and equals the sum of squared omitted coefficients. No smoothness conditions are required. Consequently, for a quantile function, $ \|Q\|^2_{L^2} = \mathbb{E}[X^2]<\infty$ and $c_k(Q) = \lambda^*_k(Q)$. Thus, a finite second moment implies
\begin{align*}
       \lim_{K \to \infty} \left\| Q- \sum_{k=1}^K\lambda_k^*(Q)\phi_k \right\|_{L^2} = 0.
    \end{align*}
This particular quantile approximation dates back to \cite{sillitto69}. Note that this also implies $L^1$ convergence, since $\|f\|_{L^1}\leq\|f\|_{L^2}$ on $(0,1)$. \medskip

\textbf{Pointwise Convergence:} Norm convergence alone does not control the error at a specified rank $u \in (0, 1)$. Quantile functions, however, are nondecreasing and therefore have bounded variation on every compact subinterval of $(0,1)$. Together with square integrability, this satisfies the local variation and endpoint-integrability conditions in Hobson's pointwise convergence theorem \citep{Hobson1931}. Consequently, for $u \in (0, 1)$ it follows that
\begin{align*}
    \lim_{K\to\infty}\Pi_K Q(u) =\frac{Q(u-)+Q(u+)}{2},
\end{align*}
where we define 
\begin{align*}
    Q(u-) &=\lim_{v\uparrow u}Q(v) = \sup_{0<v<u}Q(v),\\
    Q(u+) &=\lim_{v\downarrow u}Q(v) =\inf_{u<v<1}Q(v)
\end{align*}
for the left- and right-hand limits, respectively. At a continuity point, both limits equal $Q(u)$; at a discontinuity, they give the lower and upper endpoints of the jump.

\section{Additive-Quantile Parallel Trends vs. Changes-in-Changes} \label{additive_vs_cic}

In \cref{pt_spectrum_section} we showed that the parallel trends restrictions outlined in \cref{lpt_definition} and \cref{cic_pt} constitute two separate, non-nested spectra and two separate approaches to imputation of the missing counterfactual $Q^0_{11}$. Notably, the disagreement between these two extrapolation rules is completely determined by observed quantities. To formalize this difference, we define the control group's transport map as
\begin{align} \label{control_group_transport_map}
    S(x) = Q_{01}(F_{00}(x))
\end{align}
and $\delta(x) = S(x) - x$ as its displacement from the identity map. Thus, $S$ sends a baseline control outcome $x$ to the post-period control outcome at the same marginal rank, and $\delta$ is the rank-matched change between the two control distributions. Notably, both extrapolation rules begin with the same treated baseline outcome $Q_{10}(u)$; what differs between the approaches is how the control is constructed. 

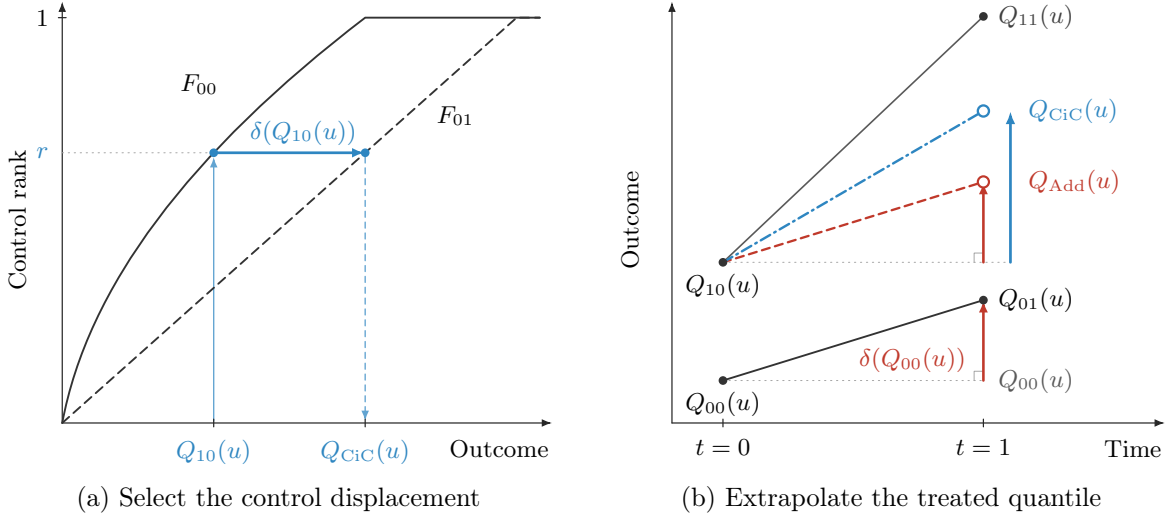
\begin{figure}[h]
\centering
\begingroup

\definecolor{didAdd}{HTML}{C0392B}
\definecolor{didCic}{HTML}{2E86C1}
\tikzset{did panel/.style={
  x=1cm,y=1cm,
  font=\fontsize{9}{10.8}\selectfont,
  line cap=round,line join=round,
  every node/.style={inner sep=1.3pt,outer sep=0pt},
  did axis/.style={black!85,line width=.45pt,-{Latex[length=1.35mm]}},
  did tick/.style={black!85,line width=.45pt},
  did add arrow/.style={didAdd,line width=.95pt,-{Latex[length=1.6mm]}},
  did cic arrow/.style={didCic,line width=1pt,-{Latex[length=1.7mm]}},
  did observed/.style={black!80,line width=.7pt},
  did additive/.style={didAdd,line width=.85pt,dash pattern=on 3pt off 2pt},
  did cic/.style={didCic,line width=.9pt,dash pattern=on 4pt off 1.7pt on .65pt off 1.7pt},
  did guide/.style={black!35,line width=.35pt,dash pattern=on .5pt off 1.7pt},
  did quiet/.style={font=\fontsize{8}{9.6}\selectfont,text=black!65}
}}
\def\didOutcomeScale{.49}
\pgfmathsetmacro{\didU}{1/4}
\pgfmathsetmacro{\didR}{5/9+4*\didU/9}
\pgfmathsetmacro{\didQZero}{2*\didU+6*\didU^2}
\pgfmathsetmacro{\didQOne}{12*\didU}
\pgfmathsetmacro{\didQTreat}{2*\didR+6*\didR^2}
\pgfmathsetmacro{\didQCic}{12*\didR}
\pgfmathsetmacro{\didDAdd}{\didQOne-\didQZero}
\pgfmathsetmacro{\didDCic}{\didQCic-\didQTreat}
\pgfmathsetmacro{\didQAdd}{\didQTreat+\didDAdd}
\pgfmathsetmacro{\didQObs}{\didQCic+5/2}

\captionsetup[subfigure]{justification=centering,singlelinecheck=false,skip=5pt}
\begin{subfigure}[t]{.48\linewidth}
  \centering
  \resizebox{\linewidth}{!}{%
    \begin{tikzpicture}[did panel]
      \path[use as bounding box] (-.20,-.60) rectangle (7.12,5.52);
\begin{scope}
\def\didOx{.66}
\let\didSx\didOutcomeScale
\def\didSy{5.25}
\pgfmathsetmacro{\didYr}{\didSy*\didR}
\pgfmathsetmacro{\didXTreat}{\didOx+\didSx*\didQTreat}
\pgfmathsetmacro{\didXCic}{\didOx+\didSx*\didQCic}
\coordinate (did-ar0) at (\didXTreat,\didYr);
\coordinate (did-ar1) at (\didXCic,\didYr);

\draw[did axis] (\didOx,0) -- (7.00,0);
\draw[did axis] (\didOx,0) -- (\didOx,5.46);
\node[rotate=90] at (.08,2.63) {Control rank};
\node[anchor=north east] at (7.00,-.18) {Outcome};
\node[anchor=east] at (.53,\didSy) {$1$};
\node[anchor=east,text=didCic] at (.53,\didYr) {$r$};
\draw[did tick] (.61,\didSy) -- (.71,\didSy);

\draw[did guide] (\didOx,\didYr) -- (did-ar0);
\draw[didCic!75,line width=.5pt,-{Latex[length=1.35mm]}]
  (\didXTreat,.025) -- (\didXTreat,\didYr-.055);
\draw[didCic!75,line width=.5pt,dash pattern=on 2.2pt off 1.8pt,
      -{Latex[length=1.35mm]}]
  (\didXCic,\didYr-.055) -- (\didXCic,.025);

\draw[did observed]
  (\didOx,0)
  .. controls ({\didOx+\didSx*2/3},{\didSy/3})
          and ({\didOx+\didSx*10/3},{2*\didSy/3})
  .. ({\didOx+\didSx*8},\didSy) -- (6.84,\didSy);
\draw[did observed,dash pattern=on 3.4pt off 2.1pt]
  (\didOx,0) -- ({\didOx+\didSx*12},\didSy) -- (6.84,\didSy);
\node[anchor=east] at (2.72,4.39) {$F_{00}$};
\node[anchor=west] at (5.46,3.98) {$F_{01}$};

\draw[did cic arrow]
  (did-ar0) -- (did-ar1);
\node[anchor=south,text=didCic]
  at ({(\didXTreat+\didXCic)/2+.18},\didYr+.06) {$\delta(Q_{10}(u))$};
\foreach \p in {did-ar0,did-ar1} \fill[didCic] (\p) circle (1.6pt);
\foreach \xx in {\didXTreat,\didXCic}
  \draw[did tick] (\xx,-.05) -- (\xx,.05);
\node[anchor=north,text=didCic] at (\didXTreat,-.17) {$Q_{10}(u)$};
\node[anchor=north,text=didCic] at (\didXCic,-.17) {$Q_{\mathrm{CiC}}(u)$};

\end{scope}
    \end{tikzpicture}%
  }
  \caption{Select the control displacement}
  \label{leftside_plot}
\end{subfigure}\hfill%
\begin{subfigure}[t]{.48\linewidth}
  \centering
  \resizebox{\linewidth}{!}{%
    \begin{tikzpicture}[did panel]
      \path[use as bounding box] (-.20,-.60) rectangle (7.12,5.52);
\begin{scope}
\def\didTZero{1.28}
\def\didTOne{4.65}
\pgfmathsetmacro{\didBlueX}{\didTOne+.35}
\def\didYBase{.55}
\pgfmathsetmacro{\didYZero}{\didYBase}
\pgfmathsetmacro{\didYOne}{\didYBase+\didOutcomeScale*(\didQOne-\didQZero)}
\pgfmathsetmacro{\didYTreat}{\didYBase+\didOutcomeScale*(\didQTreat-\didQZero)}
\pgfmathsetmacro{\didYAdd}{\didYTreat+\didOutcomeScale*\didDAdd}
\pgfmathsetmacro{\didYCic}{\didYTreat+\didOutcomeScale*\didDCic}
\pgfmathsetmacro{\didYObs}{\didYBase+\didOutcomeScale*(\didQObs-\didQZero)}
\coordinate (did-b00) at (\didTZero,\didYZero);
\coordinate (did-b01) at (\didTOne,\didYOne);
\coordinate (did-b10) at (\didTZero,\didYTreat);
\coordinate (did-b11) at (\didTOne,\didYObs);
\coordinate (did-bAdd) at (\didTOne,\didYAdd);
\coordinate (did-bCic) at (\didTOne,\didYCic);

\draw[did axis] (.62,0) -- (7.00,0);
\draw[did axis] (.62,0) -- (.62,5.46);
\node[rotate=90] at (.08,2.63) {Outcome};
\foreach \xx in {\didTZero,\didTOne}
  \draw[did tick] (\xx,-.05) -- (\xx,.05);
\node[anchor=north] at (\didTZero,-.17) {$t=0$};
\node[anchor=north] at (\didTOne,-.17) {$t=1$};
\node[anchor=north east] at (7.00,-.18) {Time};

\draw[did guide] (did-b00) -- (\didTOne,\didYZero);
\draw[did guide] (did-b10) -- (\didBlueX,\didYTreat);

\foreach \yy in {\didYZero,\didYTreat}
  \draw[black!35,line width=.35pt]
    (\didTOne-.12,\yy) -- (\didTOne-.12,\yy+.12) -- (\didTOne,\yy+.12);
\draw[did add arrow] (\didTOne,\didYZero) -- (did-b01);
\draw[did add arrow] (\didTOne,\didYTreat) -- (did-bAdd);
\draw[did cic arrow] (\didBlueX,\didYTreat) -- (\didBlueX,\didYCic);

\draw[did observed] (did-b00) -- (did-b01);
\draw[did observed,draw=black!65,line width=.6pt] (did-b10) -- (did-b11);
\draw[did additive] (did-b10) -- (did-bAdd);
\draw[did cic] (did-b10) -- (did-bCic);
\foreach \p in {did-b00,did-b01,did-b10,did-b11}
  \fill[black!80] (\p) circle (1.6pt);
\draw[didAdd,fill=white,line width=.8pt] (did-bAdd) circle (2.05pt);
\draw[didCic,fill=white,line width=.85pt] (did-bCic) circle (2.05pt);

\node[anchor=north] at (\didTZero,\didYZero-.10) {$Q_{00}(u)$};
\node[anchor=north] at (\didTZero,\didYTreat-.10) {$Q_{10}(u)$};

\node[anchor=east,text=didAdd] at (\didTOne-.18,\didYZero+.27)
  {$\delta(Q_{00}(u))$};
\node[anchor=west,text=black!65] at (4.8,\didYZero) {$Q_{00}(u)$};
\node[anchor=west] at (4.8,\didYOne) {$Q_{01}(u)$};
\node[anchor=west,text=didAdd] at (5.19,\didYAdd) {$Q_{\mathrm{Add}}(u)$};
\node[anchor=west,text=didCic] at (5.19,\didYCic) {$Q_{\mathrm{CiC}}(u)$};
\node[anchor=west,text=black!75] at (4.8,\didYObs) {$Q_{11}(u)$};
\end{scope}
    \end{tikzpicture}%
  }
  \caption{Extrapolate the treated quantile}
  \label{rightside_plot}
\end{subfigure}
\endgroup

\caption{Extrapolated values by model.}
\label{additive_vs_cic_plot}
\end{figure}

Under additive quantile parallel trends, the extrapolation rule is simply the observed control change at the same rank $u$, such that
\begin{align} \label{additive_delta}
     Q_{\mathrm{Add}}(u) &= Q_{10}(u) + \Big( Q_{01}(u) - Q_{00}(u) \Big) = Q_{10}(u)+\delta( Q_{00}(u)).
\end{align}
This is neatly visualized as the right-triangle in the control group of \cref{rightside_plot}. Under changes-in-changes, the control is \textit{not} constructed at the same rank $u$. Instead, under CiC we start with $Q_{10}(u)$ and find the rank occupied by this value in the baseline control distribution, $r(u) = F_{00}(Q_{10}(u))$, as visualized in \cref{leftside_plot}. Then, because $Q_{00}(r(u)) = Q_{10}(u)$, it follows that 
\begin{align*}
    Q_{\mathrm{CiC}}(u) &= Q_{01}\!\left(r(u)\right) = Q_{00}\!\left(r(u)\right) + \underbrace{Q_{01}\!\left(r(u)\right) - Q_{00}\!\left(r(u)\right)}_{S(Q_{10}(u))-Q_{10}(u)}  = Q_{10}(u)+\delta(Q_{10}(u)).
\end{align*}
Thus, the difference between these two extrapolations is simply $\Gamma(u) = \delta(Q_{10}(u))-\delta(Q_{00}(u))$, which is observable. From here, we write the difference between these two extrapolations in $L$-moment coordinates. Let
\begin{align*}
    \gamma_k = \langle \Gamma, \phi_k \rangle =  \int^1_0 Q_{\mathrm{CiC}}(u) \phi_k(u) du - \Big(\lambda_k^*(Q_{10}) +\lambda_k^*(Q_{01}) -\lambda_k^*(Q_{00}) \Big)
\end{align*}
denote the $k$th discrepancy coefficient. Henceforth, we refer to the sequence $\{\gamma_k:k\geq1\}$ as the \textit{discrepancy spectrum}. Notably, we may write each component as the difference in counterfactuals, $\gamma_k = \theta_{k, \mathrm{Add}} - \theta_{k, \mathrm{CiC}}$, as defined in \cref{additive_counterfactual} and \cref{cic_counterfactual}, respectively. Thus, $\gamma_k$ can be interpreted as a measure of model dependence. It is the amount by which the $k$th component changes when an analyst replaces the changes-in-changes extrapolation with the additive quantile extrapolation. In the following theorem, we establish an upper bound on the discrepancy between these two models to better understand when they agree and disagree.

\begin{theorem}[Discrepancy Bound] \label{additive_cic_bound}
Let $\mathcal{I} \subseteq \mathbb{R}$ be an interval containing $\operatorname{supp}(F_{00})\cup\operatorname{supp}(F_{10})$ and let $S:\mathcal{I} \to \mathbb{R}$ be the control-group transport map defined in \cref{control_group_transport_map}. Assume that $S\!\left(Q_{00}(u)\right)=Q_{01}(u)$ for Lebesgue-almost every $u \in (0, 1)$; this holds, for example, when $F_{00}$ is continuous. Finally, suppose that $S$ is absolutely continuous on every compact subinterval of $\mathcal{I}$. Then, $\Gamma \in L^2(0, 1)$ and
\begin{align*}
    \Gamma(u) = \int^{Q_{10}(u)}_{Q_{00}(u)} \big(S^\prime(x) - 1 \big) dx
\end{align*}
for Lebesgue-almost every $u \in (0, 1)$. Furthermore, 
\begin{align*}
   \| \Gamma \|_{L^2} = \left( \sum^{\infty}_{k=1} \gamma^2_k \right)^{1/2} \leq \operatorname*{ess\,sup}_{x\in\mathcal{I}}|S^\prime(x)-1| W_2(F_{10}, F_{00}).
\end{align*}
\end{theorem}

By \cref{additive_cic_bound} we can see that the disagreement between the additive-quantile and CiC extrapolations is bounded above by two pieces. The first piece concerns the control group's transport map, $S(x) = Q_{01}(F_{00}(x))$. In particular, $S^\prime(x) = 1$ almost everywhere on $\mathcal{I}$ implies that $S$ is simply a location-based shift, i.e. $S(x) = x + c$ for some constant $c$. In this setting, the control distribution could change by a large amount; however, every baseline outcome experiences the same additive change, so the additive-quantile and CiC extrapolations coincide. More generally,
\begin{align*}
    S^\prime(x) = \frac{f_{00}(x)}{f_{01}(S(x))},
\end{align*}
and so $S^\prime(x) > 1$ suggests that nearby outcome levels are more spread apart, and $0 < S^\prime(x) < 1$ suggests compression of outcome levels. The second piece of the bound $W_2(F_{10}, F_{00})$ measures the distributional distance between the baseline treatment and control groups. It is zero if and only if $F_{10}=F_{00}$ (or equivalently, $Q_{10}=Q_{00}$ almost everywhere). Thus, in this setting, it is trivially true that matching by rank and matching by baseline outcome level are the same.

\subsection{$L$-Moment Disagreement} \label{l_moment_disagreement}

In addition to considering when the methods will agree on the whole quantile function, with additional restrictions on the transport map, we can also show when they will disagree for specific $L$-moments. Recall that
\begin{align*}
    \gamma_k  &=\int_0^1 ( Q_{\mathrm{CiC}}(u)-Q_{\mathrm{Add}}(u)) \phi_k(u) du =\theta_{k,\mathrm{Add}}-\theta_{k,\mathrm{CiC}}.
\end{align*}
Thus, $\gamma_k$ measures the difference between the predicted counterfactual coefficients, or equivalently the difference between the additive and CiC treatment-effect predictions. \cref{lmoment_disagreement_lower_bound} shows that when the transport map is close to being affine, we can establish a lower bound on how similar the models' predicted $L$-moments are.

\begin{proposition}[$L$-Moment-Specific Lower Bound]
\label{lmoment_disagreement_lower_bound}
Suppose $Q_{00},Q_{10}\in L^1(0,1)$, and let $S:\mathcal{I}\to\mathbb{R}$ be the control-group transport map $S=Q_{01}\circ F_{00}$ on an interval $\mathcal{I}$ containing $\operatorname{supp}(F_{00})\cup\operatorname{supp}(F_{10})$. Assume that $S(Q_{00}(u))=Q_{01}(u)$ for Lebesgue-almost every $u\in(0,1)$. If, for some $a,b\in\mathbb{R}$ and
$\varepsilon\in[0,\infty)$,
\begin{align*}
    \sup_{x\in\mathcal{I}}|S(x)-(a+bx)| \leq\varepsilon,
\end{align*}
then, for every integer $k\geq1$,
\begin{align}
    |\gamma_k| \geq \left[ |b-1|\, \big|\lambda_k^*(Q_{10})-\lambda_k^*(Q_{00})\big| -2\varepsilon \right]_+,
    \label{eq:lmoment_disagreement_lower_bound}
\end{align}
where $[x]_+=\max\{x,0\}$. Consequently, if $ |b-1|\, |\lambda_k^*(Q_{10})-\lambda_k^*(Q_{00}) | >2\varepsilon$ then the two extrapolation rules cannot agree on the $k$th $L$-moment.
\end{proposition}

The lower bound separates two sources of disagreement between the models. The term
$|b-1|$ measures the extent to which the affine component of the control transport departs from a common location shift (as noted above when the mapping is only a common shift the methods predict the same shift in quantile functions and therefore the same shift in all $L$-moments), while $|    \lambda_k^*(Q_{10})-\lambda_k^*(Q_{00})|$ measures baseline imbalance in the particular distributional feature represented by the $k$th $L$-moment. The approximation error $\varepsilon$ captures departures of the observed control transport from an affine location--scale transformation. Thus, when the control transport is close to affine, substantial baseline imbalance in the $k$th $L$-moment and a nontrivial scale component $b\neq1$ force the CiC and additive-quantile models to make different predictions for that $L$-moment. Exact agreement requires either that the affine component be sufficiently close to a pure location shift, that the baseline $k$th $L$-moments be sufficiently similar, or that the nonlinear remainder be large enough to offset the affine discrepancy. In the exact affine case, $\varepsilon=0$, so
\begin{align*}
    |\gamma_k| =|b-1|\, \big|\lambda_k^*(Q_{10})-\lambda_k^*(Q_{00})\big|.
\end{align*}
If $b=1$, the map is a common location shift and the two
extrapolations coincide regardless of the baseline distributions.
If $b\neq1$, they agree on coordinate $k$ if and only if
$\lambda_k^*(Q_{10})=\lambda_k^*(Q_{00})$.
This feature-specific condition is substantially weaker than equality of the entire baseline distributions, which is required for full quantile agreement under an affine transport with $b\neq1$.

\section{Difference-in-Differences Estimation and Inference}\label{did_estimation}

In this section we extend our one-sample estimation procedure from \cref{estimation_inference} to several difference-in-differences settings. We first operate under the additive quantile parallel trends assumption, then later extend our setup to include the changes-in-changes framework.

\subsection{Additive Model Estimation and Inference}\label{add_estimation}

 Now, we extend these results to several canonical difference-in-differences settings. Specifically, repeated cross sections, panel data, and staggered adoption. Here, we work under the additive $L$-moment parallel trends assumption given in \cref{parallel_trends_spectrum_corollary}. Each of these settings follows from a common setup. Suppose $K$, the number of periods, and the collection of reported comparisons are fixed. Assume that every observed marginal distribution used below has a finite second moment and that each relevant cell or cohort has a positive limiting share of the independent sampling units. Then, for some set $\mathcal{I}$, denote a sample $S=(z_i : i\in\mathcal{I})$ of size $m>K$, with order statistics $z_{(1)},\ldots,z_{(m)}$, and let $\lambda_k^\circ(S)$ denote either of the estimators from \cref{estimator_section}. That is,
\begin{align*}
    \widehat{\lambda}_k^*(S) &=\frac{\sqrt{2k-1}}{m} \sum_{q=0}^{k-1}\xi_{kq} \sum_{r=q+1}^{m} \frac{\binom{r-1}{q}}{\binom{m-1}{q}}z_{(r)}, \\
    \widetilde{\lambda}_k^*(S) &=\sum_{r=1}^{m}z_{(r)} \int_{(r-1)/m}^{r/m}\phi_k(u)du.
\end{align*}
Next, to define the jackknife contribution, for $i\in\mathcal{I}$ define $\overline\lambda_k^{\circ,(-)}(S) =\frac{1}{m}\sum_{r\in\mathcal{I}} \lambda_k^\circ(S^{(-r)})$ and 
\begin{align} \label{did_jackknife}
    \widehat{\ell}_{i,k}(S) =(m-1)\left\{ \overline{\lambda}_k^{\circ,(-)}(S) -\lambda_k^\circ(S^{(-i)})\right\}.
\end{align}
Thus $\sum_{i\in\mathcal{I}} \widehat{\ell}_{i,k}(S)=0$ exactly. For the $U$-statistic, the average deleted estimate equals the full-sample estimate, so \cref{did_jackknife} reduces to $(m-1)\{\widehat{\lambda}_k^*(S) -\widehat{\lambda}_k^*(S^{(-i)})\}$. Following \cref{jackknife-bootstrap}, we estimate the covariance of the marginal influence contributions by
\begin{align*}
     \widehat{\Sigma}_{jk}(S) =\frac{1}{m}\sum_{i\in\mathcal{I}} \widehat{\ell}_{i,j}(S)\widehat{\ell}_{i,k}(S),
\end{align*}
for $j, k \leq K$. With this machinery in place, in what follows we formalize estimation and inference in each difference-in-differences setting.

\paragraph{Repeated Cross Sections.}

Let $Z_i=(G_i,T_i,Y_i)$ be independent and identically distributed, where $G_i\in\{0,1\}$ identifies the comparison and treated groups and $T_i\in\{0,1\}$ identifies the pre- and post-treatment periods. As discussed in the repeated-cross-section formulation of \cite{abadie2005semiparametric}, eventual treatment-group membership must be observed at both dates, and the date-specific samples must represent the same underlying group populations. Define $\mathcal{I}_{gt}=\{i:G_i=g,T_i=t\}$, $n_{gt}=|\mathcal{I}_{gt}|$, and the set $S_{gt}=(Y_i:i\in\mathcal{I}_{gt})$. In this setting we define  $\lambda_{k,gt}^\circ=\lambda_k^\circ(S_{gt})$ as the $k$th normalized $L$-moment in group $(g, t)$. Under $\mathcal{L}(\{k\})$, the $k$th coefficient is estimated by
\begin{align} \label{did_target}
   \widehat{\theta}_k^\circ = \lambda_{k,11}^\circ-\lambda_{k,10}^\circ -\lambda_{k,01}^\circ+\lambda_{k,00}^\circ.
\end{align}
With $s_{gt}=(2g-1)(2t-1)$ and $p_{gt}=\mathbb{P}(G=g,T=t)$, \cref{lmoment_linearization} gives the influence contribution $\sum_{g,t}s_{gt}\mathbb{I}(G_i=g,T_i=t) \ell_k(Y_i;F_{gt})/p_{gt}$. This yields the estimated counterpart,
\begin{align*}
    \widehat{\Psi}^{\mathrm{RC}}_{i,k} =\sum_{g,t\in\{0,1\}}s_{gt} \left(\frac{n \mathbb{I}(i\in\mathcal{I}_{gt})}{n_{gt}} \right) \widehat{\ell}^{\,gt}_{i,k}.
\end{align*}
The estimated covariance is then $\widehat{V}^{\mathrm{RC}}_{jk} =\sum_{g,t\in\{0,1\}} n^{-1}_{gt}\widehat{\Sigma}_{jk,gt}.$ The same formulas apply to four independent cell samples with fixed sizes and $n=\sum_{g,t}n_{gt}$. For $k=1$, the point estimator simply reduces to the usual unadjusted mean difference-in-differences estimator.

\paragraph{Panel Data.}
Now, suppose that $Z_i=(G_i,Y_{i0},Y_{i1})$ are independent and identically distributed but outcomes may be arbitrarily dependent within a person. Define $\mathcal{I}_g=\{i:G_i=g\}$, $n_g=|\mathcal I_g|$, and $S_{gt}=(Y_{it}:i\in\mathcal{I}_g)$. Here, we compute $\lambda_{k,gt}^\circ=\lambda_k^\circ(S_{gt})$ by sorting outcomes separately at each date. The original person labels, rather than the within-date ranks, identify deletions, so that $S_{gt}^{(-i)} =(Y_{rt}:r\in\mathcal{I}_g\setminus\{i\})$. Next, define the group-specific change and its delete-person version by $\widehat{\Delta}_{k,g}^\circ =\lambda_{k,g1}^\circ-\lambda_{k,g0}^\circ$ and
\begin{align*}
    \widehat{\Delta}_{k,g}^{\circ,(-i)} =\lambda_k^\circ(S_{g1}^{(-i)}) -\lambda_k^\circ(S_{g0}^{(-i)}).
\end{align*}
The estimator in this setting remains \cref{did_target} and can be written as $\widehat{\theta}_k^\circ =\widehat{\Delta}_{k,1}^\circ-\widehat{\Delta}_{k,0}^\circ$. Letting $\widehat{\ell}^{\,gt}_{i,k} =\widehat{\ell}_{i,k}(S_{gt})$, the estimated within-person influence difference is
\begin{align} \label{did_panel_difference}
    \widehat{D}_{i,k,g} &=\widehat{\ell}^{\,g1}_{i,k} -\widehat{\ell}^{\,g0}_{i,k} =(n_g-1)\left\{ \frac{1}{n_g}\sum_{r\in\mathcal{I}_g} \widehat{\Delta}_{k,g}^{\circ,(-r)} -\widehat{\Delta}_{k,g}^{\circ,(-i)}\right\}.
\end{align}
In particular, both terms delete the same person. The estimated covariance of these differences and the covariance of the difference-in-differences estimates are then given by
\begin{align} \label{did_panel_covariance}
    \widehat{\Gamma}_{jk,g} =\frac{1}{n_g}\sum_{i\in\mathcal{I}_g} \widehat{D}_{i,j,g}\widehat{D}_{i,k,g},
    \qquad
    \widehat{V}^{\mathrm{P}}_{jk} =\frac{\widehat{\Gamma}_{jk,1}}{n_1} +\frac{\widehat{\Gamma}_{jk,0}}{n_0},
\end{align}
with standard error $\widehat{\operatorname{se}}^{\mathrm{P}}_k =\sqrt{\widehat{V}^{\mathrm{P}}_{kk}}$. Finally, for joint inference, the full-sample contribution of person $i$ is
\begin{align*}
     \widehat{\Psi}^{\mathrm{P}}_{i,k} =\frac{n\,\mathbb{I}(i\in\mathcal{I}_1)}{n_1} \widehat{D}_{i,k,1} -\frac{n\,\mathbb{I}(i\in\mathcal{I}_0)}{n_0} \widehat{D}_{i,k,0}.
\end{align*}
Note that the covariance in \cref{did_panel_covariance} retains the within-person dependence across periods, as required for panel DiD inference \citep{SANTANNA2020101}.

\paragraph{Staggered Adoption.}
When treatment begins at different dates, we construct a separate two-group, two-period comparison for each adoption cohort and outcome date.  Following the cohort-time formulation of \citet{CALLAWAY2021200}, let $A_i\in\{2,\ldots,T,\infty\}$ denote the first treatment date, where we use $A_i=\infty$ for units untreated throughout the observation window. We assume treatment is absorbing, such that once a unit adopts, it remains treated. We say a cohort is the population with a common value $A_i=a$. Write $F_{a,s}$ for the observed outcome distribution in cohort $a$ at date $s$, and $F^0_{a,s}$ for its distribution when remaining untreated. For $t\geq a$, define
\begin{align*}
    \theta_k(a,t) =\lambda_k^*(F_{a,t})-\lambda_k^*(F^0_{a,t}).
\end{align*}
Thus $\theta_k(a,t)$ concerns the period-$t$ outcome distribution of cohort $a$, such that $t-a$ records the elapsed time since adoption. For $k=1$ this is just the usual cohort-time average treatment effect. Now, choose a baseline $b=a-1$ as the cohort's last pretreatment date, and initially use never-treated units as controls. Assume no anticipation before each cohort's adoption date, so that $F_{a,b}=F^0_{a,b}$ for $b=a-1$. We also assume the coordinate-specific staggered parallel trends
\begin{align} \label{did_staggered_pt}
    \lambda_k^*(F^0_{a,t})-\lambda_k^*(F_{a,b})
    =\lambda_k^*(F_{\infty,t})-\lambda_k^*(F_{\infty,b}),
\end{align}
for each reported pair $(a,t)$. This identifies the missing counterfactual $L$-moment as 
\begin{align*}
     \lambda_k^*(F^0_{a,t}) =\lambda_k^*(F_{a,b}) +\{\lambda_k^*(F_{\infty,t})-\lambda_k^*(F_{\infty,b})\},
\end{align*}
and so $\theta_k(a,t) =\lambda_k^*(F_{a,t})-\lambda_k^*(F_{a,b}) -\lambda_k^*(F_{\infty,t})+\lambda_k^*(F_{\infty,b})$. The estimator is then given by
\begin{align} \label{did_staggered_estimator}
     \widehat{\theta}_k^\circ(a,t) =\lambda_{k,a,t}^\circ-\lambda_{k,a,b}^\circ -\lambda_{k,\infty,t}^\circ+\lambda_{k,\infty,b}^\circ.
\end{align}
For repeated cross sections, suppose we observe i.i.d. records $Z_i=(A_i,T_i,Y_i)$, with $T_i\in\{1,\ldots,T\}$. Note that cohort membership $A_i$ must be known even for units observed before adoption, and the date-specific samples must represent the same underlying cohort populations \citep{CALLAWAY2021200}. Now, define $\mathcal{I}_{c,s}=\{i:A_i=c,T_i=s\}$, $n_{c,s}=|\mathcal{I}_{c,s}|$, and $S_{c,s}=(Y_i:i\in\mathcal{I}_{c,s})$. Set $\lambda_{k,c,s}^\circ=\lambda_k^\circ(S_{c,s})$ and $\widehat{\ell}^{\,c,s}_{i,k} =\widehat{\ell}_{i,k}(S_{c,s})$. For each $(a,t)$, only the four cells $(a,t),(a,b),(\infty,t),(\infty,b)$ enter \cref{did_staggered_estimator}. Consequently, the estimated influence contribution for the cohort-time effect is given by
\begin{align*}
    \widehat{\Psi}^{\mathrm{RC}}_{i,k}(a,t) =n\bigg\{\frac{\mathbb{I}(i\in\mathcal{I}_{a,t})}{n_{a,t}} \widehat{\ell}^{\,a,t}_{i,k} -\frac{\mathbb{I}(i\in\mathcal{I}_{a,b})}{n_{a,b}} \widehat{\ell}^{\,a,b}_{i,k} -\frac{\mathbb{I}(i\in\mathcal{I}_{\infty,t})} {n_{\infty,t}} \widehat{\ell}^{\,\infty,t}_{i,k} +\frac{\mathbb{I}(i\in\mathcal{I}_{\infty,b})}{n_{\infty,b}} \widehat{\ell}^{\,\infty,b}_{i,k}\bigg\}.
\end{align*}
For a complete panel, observe i.i.d. histories $Z_i=(A_i,Y_{i1},\ldots,Y_{iT})$. Define $\mathcal I_c=\{i:A_i=c\}$, $n_c=|\mathcal I_c|$, and $S_{c,s}=(Y_{is}:i\in\mathcal I_c)$, and again set $\lambda_{k,c,s}^\circ=\lambda_k^\circ(S_{c,s})$ and $\widehat\ell^{\,c,s}_{i,k}=\widehat\ell_{i,k}(S_{c,s})$. Delete the same person from both dates when computing the marginal jackknife contributions. The contribution for a within-cohort change, and then for its DiD contrast, are
\begin{align*}
    \widehat{D}^{\,c}_{i,k}(b,t) &=\widehat{\ell}^{\,c,t}_{i,k} -\widehat{\ell}^{\,c,b}_{i,k},
    \nonumber\\
    \widehat{\Psi}^{\mathrm{P}}_{i,k}(a,t)
    &=n\left\{ \frac{\mathbb{I}(i\in\mathcal{I}_a)}{n_a} \widehat{D}^{\,a}_{i,k}(b,t) -\frac{\mathbb{I}(i\in\mathcal{I}_\infty)}{n_\infty} \widehat{D}^{\,\infty}_{i,k}(b,t)\right\}.
\end{align*}
As in \cref{did_panel_difference}, taking the difference within each person retains the cross-period covariance. Cohort-time effects can also be summarized by fixed weighted averages \citep{CALLAWAY2021200}. For a fixed finite collection $\mathcal J$ of admissible pairs and fixed, nonrandom weights $w_{a,t}$, define $\theta_{k,w} =\sum_{(a,t)\in\mathcal J}w_{a,t}\theta_k(a,t)$. Its estimator and estimated influence contribution are
\begin{align*}
    \widehat\theta_{k,w}^{\circ} =\sum_{(a,t)\in\mathcal{J}} w_{a,t}\widehat\theta_k^{\circ}(a,t),
    \qquad
    \widehat{\Psi}_{i,k,w} =\sum_{(a,t)\in\mathcal{J}} w_{a,t}\widehat\Psi_{i,k}(a,t).
\end{align*}
An average at exposure duration $e\geq0$ is obtained by restricting $\mathcal J$ to pairs satisfying $t-a=e$ and choosing nonnegative weights summing to one.

\paragraph{Standard Errors and Joint Inference.} Inference in any of the aforementioned settings follows directly from the results established in \cref{jackknife-bootstrap}. Let $\widehat{\boldsymbol\theta}=(\widehat\theta_r)_{r=1}^d$ collect any fixed selection of the preceding effects or aggregates, and let $\widehat\Psi_i=(\widehat\Psi_{i,r})_{r=1}^d$ stack their contributions. The index $r$ now labels reported coordinates, which may differ in moment order, cohort, or date. Define the covariance
\begin{align*}
     \widehat{V} =\frac{1}{n^2}\sum_{i=1}^n
      \widehat{\Psi}_i \widehat{\Psi}_i^\top
\end{align*}
and standard error $\widehat{\operatorname{se}}_r= \sqrt{\widehat{V}_{rr}}$. Joint asymptotic normality follows from \cref{lmoment_linearization} and following the discussion in \cref{jackknife-bootstrap}, $n\widehat{V}$ consistently estimates the covariance in that limit. For each bootstrap draw, generate $\xi_1,\ldots,\xi_n\overset{iid}{\sim}\mathcal N(0,1)$ and calculate $B_r^{(\mathrm{b})}=\frac{1}{n}\sum_{i=1}^n\xi_i\widehat\Psi_{i,r}$ and
\begin{align*}
    T_{\max}^{(\mathrm{b})}=\max_{1\leq r\leq d} \left\{
      \frac{|B_r^{(\mathrm{b})}|}{\widehat{\operatorname{se}}_r} \right\}.
\end{align*}
For example, in the $2\times2$ repeated-cross-section design, $B_k^{(\mathrm{b})} =\sum_{g,t\in\{0,1\}}\frac{s_{gt}}{n_{gt}} \sum_{i\in\mathcal{I}_{gt}}\xi_i\widehat{\ell}^{\,gt}_{i,k}.$ Then, for $r = 1, \ldots, d$, the intervals
\begin{align*}
     \left[\widehat{\theta}_r-c_{1-\alpha}^{(\mathrm{b})}\widehat{\operatorname{se}}_r, \, \widehat{\theta}_r+c_{1-\alpha}^{(\mathrm{b})}\widehat{\operatorname{se}}_r\right]
\end{align*}
have asymptotic simultaneous coverage; $c_{1-\alpha}^{(\mathrm{b})}$ is the conditional $(1-\alpha)$ quantile of $T_{\max}^{(\mathrm{b})}$, approximated by repeated draws. Within a draw, use the same multiplier for each original sampling unit across all moments, dates, and comparisons. Thus, with panel data, a person receives a single multiplier for their entire history, and with staggered adoption, an observation reused across several cohort-time comparisons retains its multiplier. This preserves the dependence induced by pairing and shared comparison samples \citep{CALLAWAY2021200}, using the multiplier construction of \cref{jackknife-bootstrap}.

\subsection{Changes-in-Changes Estimation} \label{cic_estimation}

In this section, we establish an estimation procedure under the changes-in-changes framework. Suppose the conditions of \cref{identification_cic} hold. Specifically, we work under the restriction $\mathcal{L}_{\mathrm{CiC}}(1\mathord{:}K)$ for some fixed $K$. Then, let $h(y)=Q_{01}(F_{00}(y))$, and let $F_{\mathrm C}$ denote the CiC-imputed counterfactual distribution that is induced by $h$, whose $k$th normalized $L$-moment is given by $C_k=\lambda_k^*(F_{\mathrm{C}})$. Then $\theta_k=\lambda_k^*(F_{11})-C_k$ for $k\leq K$. Notably, the map $h$ is unknown and must be estimated from the control samples. Consequently, estimating $C_k$ requires the three observed distributions $F_{10}$, $F_{00}$, and $F_{01}$, even when only one $L$-moment is targeted.

\paragraph{Estimating the Map and the Imputed Moments.}
Let $\widehat{F}_{gt}$ denote the empirical distribution in population $(g,t)$, formed according to some sampling design, and let $\widehat{Q}_{gt}$ be its empirical quantile function. We estimate the map $h$ by $\widehat{h}(y) = \widehat{Q}_{01}(\widehat{F}_{00}(y)).$ Next, for treated baseline observations $X_1,\ldots,X_m$, let $\widehat{Z}_i=\widehat{h}(X_i)$ and
\begin{align*}
    \widehat{F}_{\mathrm{C}}(z) =\frac{1}{m}\sum_{i=1}^m \mathbb{I}(\widehat{Z}_i\leq z).
\end{align*}
Applying the empirical estimator of \cref{estimator_section}
to these transformed observations gives
\begin{align*}
    \widehat{C}_k &=\lambda_k^*(\widehat{F}_{\mathrm{C}}) =\sum_{j=1}^m\widehat{Z}_{(j)} \int_{(j-1)/m}^{j/m}\phi_k(u)du =\int_0^1 \widehat{Q}_{01}\!\left( \widehat{F}_{00}\!\left(\widehat{Q}_{10}(u)\right) \right)\phi_k(u) du,
\end{align*}
where $\widehat Z_{(1)}\leq\cdots\leq\widehat Z_{(m)}$. The last equality follows because $\widehat{h}$ is nondecreasing, so it preserves the ordering of the baseline sample, allowing for ties. Then, one can estimate the treatment effects by
\begin{align} \label{cic_estimator}
     \widehat{\theta}_k^{\mathrm{C}} =\lambda_k^*(\widehat{F}_{11})-\widehat{C}_k.
\end{align}
Thus, the changes-in-changes estimator uses the same one-sample $L$-moment calculation as before, after estimating and applying a mapping. However, all transformed observations share the estimated map, $\widehat{h}$. Sampling uncertainty in $\widehat{C}_k$ therefore comes from the treated baseline sample and both control samples, rather than from the transformed sample alone. Bootstrap draws must recompute the map as well as the moments, which we describe below. Furthermore, while the $U$-statistic formula can be used instead of the empirical $L$-moment estimator, its finite-sample unbiasedness property does not generally survive estimation of $h$.

\paragraph{Repeated Cross Sections.}
Suppose $Z_i=(G_i,T_i,Y_i)$ are i.i.d., that every group-period cell has positive probability, and that the date-specific samples represent the same underlying group populations. Define $\mathcal{I}_{gt}=\{i:G_i=g,T_i=t\}$, $n_{gt}=|\mathcal{I}_{gt}|$, and
\begin{align*}
    \widehat{F}_{gt}(y) =\frac{1}{n_{gt}}\sum_{i\in\mathcal{I}_{gt}} \mathbb{I}(Y_i\leq y).
\end{align*}
Then, we can simply apply the preceding construction to define our estimator with $m=n_{10}$ and $(X_1,\ldots,X_m)=(Y_i:i\in\mathcal{I}_{10})$. The same estimator applies to four independently sampled cells of fixed sizes with positive limiting sample fractions.

\paragraph{Panel Data.}
For i.i.d. histories $Z_i=(G_i,Y_{i0},Y_{i1})$, with $\mathbb{P}(G=g)>0$ for $g\in\{0,1\}$, define $\mathcal{I}_g=\{i:G_i=g\}$, $n_g=|\mathcal{I}_g|$, and
\begin{align*}
   \widehat{F}_{gt}(y) =\frac{1}{n_g}\sum_{i\in\mathcal{I}_g} \mathbb{I}(Y_{it}\leq y). 
\end{align*}
Use the same marginal estimator in \cref{cic_estimator}, now transforming $Y_{i0}$ for $i\in\mathcal{I}_1$ and comparing its imputed distribution with the empirical distribution of $Y_{i1}$ in that group. Outcomes are sorted separately at each date, but person labels must be retained for resampling. Resampling complete histories preserves the within-person covariance, as in the panel CiC analysis of
\citet{athey_imbens_2006}.

\paragraph{Staggered Adoption.}
Use the cohort-time notation of \cref{add_estimation}, following \citet{CALLAWAY2021200}. For cohort $a$, date $t\geq a$, and baseline $b=a-1$, define
\begin{align*}
    C_k(a,t)=\int_0^1 Q_{\infty,t}\!\left(F_{\infty,b}\!\left(Q_{a,b}(u)\right)\right) \phi_k(u)du.
\end{align*}
Under consistency, no anticipation, the comparison-specific support and integrability conditions, and the restrictions $\lambda_k^*(F^0_{a,t})=C_k(a,t)$ for $k\leq K$, estimate $\theta_k(a,t)$ by
\begin{align*}
    \widehat{\theta}_k^{\, \mathrm{C}}(a,t) =\lambda_k^*(\widehat{F}_{a,t})-\widehat{C}_k(a,t),
\end{align*}
where $\widehat C_k(a,t)$ is obtained by replacing the three distributions in $C_k(a,t)$ by their empirical counterparts. Equivalently, compute the empirical $L$-moments after applying $\widehat{Q}_{\infty,t}\circ\widehat{F}_{\infty,b}$ to the cohort-$a$ baseline sample. Form the empirical distributions from the repeated cross sections or complete panels as above, and retain only comparisons with positive limiting sample shares. Not-yet-treated controls can replace never-treated controls by using the same population $\{A>t\}$ at both dates and imposing the CiC moment restrictions on this pooled comparison population. Repeated cross sections must represent the same cohort mixture at both dates. For a fixed finite collection $\mathcal{J}$ and fixed, nonrandom weights, aggregate using $\widehat{\theta}_{k,w}^{ \, \mathrm{C}} =\sum_{(a,t)\in\mathcal{J}}w_{a,t} \widehat{\theta}_k^{\, \mathrm{C}}(a,t)$.

\section{Selecting $K$ via Risk Estimation} \label{risk_appendix}
In this section we use simulations to assess how well the risk-minimizing choice of $K$ performs under four designs calibrated to the 2007 treated cell of the empirical application (out-of-pocket spending for the age 65+ group with $n = 3{,}470$). The designs differ in how quickly the $L$-moment coefficients of the quantile function decay. In design~A (exactly sparse), $X \sim \operatorname{Uniform}[0, 2\bar{x}]$, which matches the cell mean, so only the first two coefficients are nonzero. In design~B (bounded, fast decay), $Q(u) = U(e^{cu}-1)/(e^{c}-1)$, where $U$ is the cell's 90th percentile and $c$ is chosen to match the cell mean of spending below $U$. Its coefficients are all nonzero but decay geometrically. Design~C (heavy tail) is a zero-inflated generalized Pareto distribution whose zero share matches that of the cell (10.6\%), whose shape $\xi = 0.47$ is the Hill estimate from the top 5\% of positive spending, and whose scale matches the median of positive spending. Design~D draws with replacement from the cell itself, so that $Q$ is the cell's empirical quantile function.

\begin{table}[htbp]
\centering \scalebox{.9}{
\begin{threeparttable}
\caption{Choosing $K$ by the Unbiased Risk Estimate}
\label{tab:sim_choose_K}
\begin{tabular}{lrccrrrrr}
\toprule
 & & & & & \multicolumn{4}{c}{Risk ($\times 10^{3}$)} \\
\cmidrule(lr){6-9}
 & $K_{\max}$ & $K^{*}$: median [IQR] & $\widehat{K}_R$: median [IQR] & \% $\widehat{K}_R = K^{*}$ & $\widehat{K}_R$ & $K^{*}$ & $K = 4$ & $Q_n$ \\
\midrule
\multicolumn{9}{l}{\textit{A. Uniform (exactly sparse)}} \\
& 10 & 2 [2, 2] & 2 [2, 2] & 82.1 & 0.0454 & 0.0383 & 0.0515 & 0.0638 \\
& 50 & 2 [2, 2] & 2 [2, 2] & 82.1 & 0.0454 & 0.0383 & 0.0515 & 0.0638 \\
\addlinespace
\multicolumn{9}{l}{\textit{B. Bounded, fast decay}} \\
& 10 & 6 [5, 6] & 5 [5, 6] & 25.3 & 0.0936 & 0.0884 & 0.204 & 0.108 \\
& 50 & 6 [5, 6] & 5 [5, 6] & 24.7 & 0.0944 & 0.0884 & 0.204 & 0.108 \\
\addlinespace
\multicolumn{9}{l}{\textit{C. Heavy tail (zero-inflated GPD)}} \\
& 10 & 10 [10, 10] & 10 [10, 10] & 98.9 & 5{,}800 & 5{,}790 & 6{,}490 & 5{,}540 \\
& 50 & 50 [50, 50] & 48 [36, 50] & 44.0 & 5{,}080 & 4{,}970 & 6{,}490 & 5{,}540 \\
\addlinespace
\multicolumn{9}{l}{\textit{D. Resampled data}} \\
& 10 & 10 [10, 10] & 10 [10, 10] & 100.0 & 180 & 180 & 409 & 23.9 \\
& 50 & 46 [42, 50] & 41 [35, 49] & 5.0 & 22.2 & 20.1 & 409 & 23.9 \\
\bottomrule
\end{tabular}
\begin{tablenotes}[flushleft]\footnotesize
\item \textit{Notes:} 1{,}000 replications of samples of size $n = 3{,}470$. Risk is the mean over replications of $\|\hat Q - Q\|_{2}^{2}$ (2007 dollars squared) for the $K$-term $L$-moment estimate at $\widehat{K}_R$, at the sample-specific oracle $K^{*}$ (the $K \le K_{\max}$ minimizing the realized loss in each replication), and at $K = 4$, and for the empirical quantile function $Q_n$. $\widehat{K}_R$ minimizes $\widehat{R}(K)$ over $K \le K_{\max}$ with $M_{k,n}$ from 200 random half-partitions. Designs calibrated to out-of-pocket prescription drug spending of the age 65+ group in 2007 (\cite{EngelhardtGruber2011} data, age 60+ sample): A uniform on $[0, 2\bar{x}]$; B $Q(u) = U(e^{cu}-1)/(e^{c}-1)$ with $U$ the 90th percentile; C zero-inflated generalized Pareto with Hill tail index; D resampling the observed data.
\end{tablenotes}
\end{threeparttable}}
\end{table}

The selection rule performs well when the $L$-moment coefficients decay quickly. In design~A, the sample-specific oracle is $K^{*} = 2$ in every replication, because each coefficient beyond the second adds only estimation error, and $\widehat{K}_R$ matches it in 82\% of replications. In design~B, $K^{*}$ is concentrated on 5--6 and $\widehat{K}_R$ on 5. Although $\widehat{K}_R$ is often one term short of $K^{*}$ (they agree in about 25\% of replications), the cost is small: the risk at $\widehat{K}_R$ is 0.0944, compared with 0.0884 at $K^{*}$, 0.204 at $K = 4$, and 0.108 for $Q_n$, all in thousands of squared dollars. In both designs, the results do not depend on $K_{\max}$, so a large $K_{\max}$ costs essentially nothing when the coefficients are sparse. When the coefficients decay slowly, as in designs~C and~D, a small $K_{\max}$ binds, so $\widehat{K}_R = K^{*} = K_{\max}$ in nearly every replication for $K_{\max} = 10$, and the truncated estimator has higher risk than $Q_n$ (in design~D 180 vs.\ 23.9). In design~C, $K^{*}$ remains at the cap of 50 in most replications, whereas $\widehat{K}_R$ stops earlier more often (median 48, interquartile range 36--50), typically in samples whose extreme draws make the high-order coefficients unreliable. Its risk of 5{,}080 lies between the oracle's 4{,}970 and the 5{,}540 of $Q_n$. \medskip

With $K_{\max} = 50$, design~D shows that the rule can find an interior optimum when the coefficients decay slowly. $K^{*}$ has a median of 46 (interquartile range 42--50) and $\widehat{K}_R$ a median of 41 (35--49). The two rarely coincide exactly, but the risk at $\widehat{K}_R$ (22.2) is close to that at $K^{*}$ (20.1) and below that of $Q_n$ (23.9). The design also informs the choice of $K$ in the application. Applied to the 2007 treated cell itself with $K_{\max} = 50$, the rule selects $\widehat{K}_R = 44$ for most choices of the random half-partitions, and between 43 and 46 for all of them, since $\widehat{R}(K)$ is nearly flat between 40 and 50. This value lies near the center of the design~D distributions of both $\widehat{K}_R$ and $K^{*}$, consistent with the idea that the selected $K$ is close to the risk-minimizing truncation for these data. The results support using a moderately large $K_{\max}$, since it incurs little cost when the coefficients are sparse and prevents a binding cap from forcing excessive truncation when they decay slowly. Overall, while the selected $K$ is sensitive in heavy-tailed designs, the truncated quantile using $\widehat{K}_R$ still consistently has lower risk than the empirical quantile function.

\section{Proofs} \label{sec:proofs}

\subsection{Proof of \cref{parallel_trends_spectrum_corollary}} \label{parallel_trends_spectrum_corollary_proof}

We split the proof of \cref{parallel_trends_spectrum_corollary} into three pieces, corresponding to each result.
\begin{enumerate}
    \item $\mathcal{L}(\{1\})$ is equivalent to M-PT.

    \begin{proof}[\textbf{Proof:}]
        To begin, recall that $\phi_1=P_0^* = 1$. Thus, the first defect coefficient is given by
        \begin{align*}
            d^0_1 = \int_0^1\mathcal D^0(u) du = \int^1_0  \left[ \left\{Q_{11}^0(u)-Q_{10}^0(u)\right\} -\left\{Q_{01}^0(u)-Q_{00}^0(u)\right\} \right] du.
        \end{align*}
For any integrable random variable, the integral of its quantile function is its expectation. Hence, we immediately have that
\begin{align*}
    d_1^0 &= \mathbb{E}[Y_1(0)\mid G=1] - \mathbb{E}[Y_0(0)\mid G=1] - \mathbb{E}[Y_1(0)\mid G=0] + \mathbb{E}[Y_0(0)\mid G=0]\\
    &= \mathbb{E}[Y_1(0)-Y_0(0)\mid G=1] - \mathbb{E}[Y_1(0)-Y_0(0)\mid G=0].
\end{align*}
Consequently, $d_1^0=0$ if and only if mean parallel trends, as defined in \cref{mean_pt}, holds. Since $\mathcal{L}(\{1\})$ is precisely the restriction $d_1^0=0$, this completes the proof.
    \end{proof}
    \item $\mathcal{L}(1\mathord{:}K+1)$ implies $\mathcal{L}(1\mathord{:}K)$ for every $K\geq1$.

    \begin{proof}[\textbf{Proof:}] This follows immediately, by definition. Recall that $\mathcal L(1\mathord{:}K+1)$ requires $d_k^0=0$ for every $k\in\{1,\ldots,K+1\}$. Therefore, in particular, $d_k^0=0$ for every $k\in\{1,\ldots,K\}$, which is equivalent to $\mathcal{L}(1\mathord{:}K)$.
    \end{proof}
    \item $\mathcal{L}(\mathbb{N})$ is equivalent to Q-PT.
    \begin{proof}[\textbf{Proof:}] Throughout the proof, let $\lambda$ denote the Lebesgue measure on $[0,1]$. Because $\mathcal{D}^0\in L^1(0,1)$ and each $\phi_k$ is a polynomial (and therefore bounded on $[0,1]$) it follows that every defect coefficient $d_k^0$ is well-defined.  First, assume that additive quantile parallel trends holds. That is, assume that $\mathcal{D}^0(u)=0$ for Lebesgue-almost every $u\in(0,1)$. Then, for every $k \in \mathbb{N}$ it follows that
    \begin{align} \label{qpt_k_equals_zero}
         d^0_k = \int_0^1\mathcal{D}^0(u)\phi_k(u)du = 0
    \end{align}
     and so $\mathcal{L}(\mathbb{N})$ holds. Conversely, suppose that $\mathcal{L}(\mathbb{N})$ holds, so that \cref{qpt_k_equals_zero} holds for all $k \in \mathbb{N}$. We will show that these equalities force $\mathcal{D}^0$ to vanish almost everywhere. \medskip 

Let $\mathscr{P}_m$ denote the vector space of real polynomials of degree at most $m$. Since $\phi_k=\sqrt{2k-1}P^*_{k-1}$ is a polynomial of degree $k-1$ with a nonzero leading coefficient, the collection $\{\phi_1,\ldots,\phi_{m+1}\}$ is a basis of the $(m+1)$-dimensional space $\mathscr{P}_m$. Consequently, every polynomial $p$ can be written as a finite linear combination
\begin{align*}
    p(u)=\sum_{k=1}^{m+1}a_k\phi_k(u)
\end{align*}
for some $m$ and some coefficients $a_1,\ldots,a_{m+1}$. Thus, for every polynomial $p$ it follows that
\begin{align} \label{polynomial_annihilation}
    \int_0^1\mathcal{D}^0(u)p(u)du = 0.
\end{align}
In fact, we can extend this result from all polynomials to all continuous functions. The Stone--Weierstrass theorem implies that the polynomials are uniformly dense in $C[0,1]$ \citep{rudin1976principles}: for every $h\in C[0,1]$, there exists a sequence of polynomials $\{p_m\}$ such that
\begin{align} \label{uniform_polynomial_approximation}
    \|h-p_m\|_{\infty} = \sup_{u\in[0,1]}|h(u)-p_m(u)| \longrightarrow 0.
\end{align}
For any $h\in C[0,1]$, choose a sequence satisfying \cref{uniform_polynomial_approximation}. Then, since \cref{polynomial_annihilation} gives that $\int_0^1\mathcal D^0p_m=0$ for every $m$, we have
\begin{align*}
    \left| \int_0^1\mathcal D^0(u)h(u)du \right| &= \left| \int_0^1 \mathcal{D}^0(u)[h(u)-p_m(u)]du \right|\\
    &\leq \int_0^1 |\mathcal D^0(u)|\, |h(u)-p_m(u)|du\\
    &\leq \|\mathcal D^0\|_{L^1} \|h-p_m\|_{\infty}.
\end{align*}
Then, it follows that $\|\mathcal D^0\|_{L^1} \|h-p_m\|_{\infty} \to 0$ as $m \to \infty$. Hence, $\int_0^1\mathcal D^0(u)h(u)du = 0$ for every $h\in C[0,1]$. From here, our goal is to translate our annihilation result for continuous functions to an almost-everywhere statement about $\mathcal{D}^0$. Define the finite signed Borel measure $\nu$ on $[0,1]$ by
\begin{align}
    \nu(A) = \int_A\mathcal{D}^0(u) du,
\end{align}
for $A\in\mathcal{B}([0,1])$. Note that this measure is finite since $ |\nu|([0,1]) = \int_0^1|\mathcal D^0(u)|du <\infty$. Then, since $\int_0^1\mathcal D^0(u)h(u)du = 0$ for every $h\in C[0,1]$, it follows that
\begin{align*}
      \int_{[0,1]}h\,d\nu=0
\end{align*}
for every $h \in C[0, 1]$. From here, recall that the Riesz--Markov representation theorem implies that a finite signed regular Borel measure
on a compact interval is uniquely determined by its integrals against continuous functions \citep{rudin1987real}. Since the zero measure and $\nu$ have the same integral against every continuous function, this uniqueness result implies that $\nu(A)=0$ for every Borel set $A\subseteq[0,1]$. Importantly, this implies that $\mathcal{D}^0$ vanishes almost everywhere. To see this, for each integer $n \in \mathbb{N}$ define
\begin{align*}
    A_n^+ = \left\{u\in(0,1): \mathcal{D}^0(u)>\frac{1}{n} \right\} \qquad \text{and} \qquad A_n^- = \left\{ u\in(0,1): \mathcal{D}^0(u)<-\frac{1}{n} \right\}.
\end{align*}
Then, because $\nu = 0$, it follows that
\begin{align*}
     0 &= \nu(A_n^+) = \int_{A_n^+}\mathcal{D}^0(u)du \geq \frac{1}{n}\lambda(A_n^+), \\
     0 &= -\nu(A_n^-) = \int_{A_n^-}-\mathcal{D}^0(u)du \geq \frac{1}{n}\lambda(A_n^-),
\end{align*}
and so $\lambda(A_n^+)=0$ and $\lambda(A_n^-)=0$. Finally, since $\{u:\mathcal{D}^0(u)\neq 0 \} = \bigcup_{n=1}^{\infty} \left(A_n^+\cup A_n^-\right)$, it follows that $\mathcal{D}^0(u)=0$ for Lebesgue-almost every $u \in (0, 1)$, which is exactly the additive quantile parallel trends assumption. \medskip

Note: this proves something stronger than $\mathcal{L}(\mathbb{N}) \Rightarrow \mathcal{D}^0(u)=0$ for Lebesgue-almost every $u\in(0,1)$. In fact, the same argument applies to any function $f \in L^1(0, 1)$, not just the additive quantile parallel trends defect $\mathcal{D}^0(u)$. For posterity, we record this as a technical lemma in what follows, which we will later use for identification in the changes-in-changes framework.
\begin{lemma}[$L^1$ Determination by Shifted-Legendre Coefficients]
\label{legendre_determination}
Let $f\in L^1(0,1)$. If
\begin{align*}
    \int_0^1f(u)\phi_k(u)du=0
\end{align*}
for every $k \in \mathbb{N}$, then $f=0$ Lebesgue-almost everywhere.
\end{lemma}

Finally, note that if we instead assume that $\mathcal{D}^0(u) \in L^2(0, 1)$ then the proof greatly simplifies. If $\mathcal{L}(\mathbb{N})$ holds, then $\langle\mathcal{D}^0,\phi_k\rangle_{L^2}=0$ for every $k\geq1$. Since $\{\phi_k\}_{k\geq1}$ is a complete orthonormal basis of $L^2(0,1)$, by Parseval's identity \citep{conway1985course}, it follows that
\begin{align*}
    \|\mathcal{D}^0\|_{L^2}^2 =\sum_{k=1}^{\infty} \left|\langle\mathcal{D}^0,\phi_k\rangle_{L^2}\right|^2 =0.
\end{align*}
Thus, it follows that $\mathcal{D}^0=0$ Lebesgue-almost everywhere, which is exactly Q-PT.
    \end{proof}
\end{enumerate}

\subsection{Proof of \cref{identification_theorem}} \label{identification_theorem_proof}

\begin{proof}[\textbf{Proof:}] 

To begin, observe that in the treated group, consistency implies that $Q_{11}=Q_{11}^1$. Furthermore, consistency, together with the no anticipation assumption, implies that $Q_{10}^0=Q_{10}^1=Q_{10}$. Finally, in the control group, under consistency we have that $Q_{00}=Q_{00}^0$ and $Q_{01}=Q_{01}^0$. Thus, the untreated quantile-trend defect can be written as
\begin{align*}
    \mathcal{D}^0(u) = \{Q_{11}^0(u)-Q_{10}(u)\} - \{Q_{01}(u)-Q_{00}(u)\},
\end{align*}
and consequently, the only unobserved object is the treated group's post-period untreated counterfactual quantile function $Q_{11}^0$. Now, fix $k \in \mathcal{K}$. The restriction $d_k^0=0$ is given by
\begin{align*}
     0 = d_k^0 &= \int_0^1 \mathcal{D}^0(u)\phi_k(u)du = \left\{ \lambda_k^*(Q_{11}^0) - \lambda_k^*(Q_{10}^0) \right\} - \left\{ \lambda_k^*(Q_{01}^0) - \lambda_k^*(Q_{00}^0) \right\}.
\end{align*}
Thus, after rearranging, the $k$th restriction identifies exactly the $k$th coefficient of the missing counterfactual quantile function:
\begin{align*}
    \lambda_k^*(Q_{11}^0) =\lambda_k^*(Q_{10})+\lambda_k^*(Q_{01}) -\lambda_k^*(Q_{00}).
\end{align*}
Substituting this equality back into the definition of $\theta_k$, using $Q_{11}^1=Q_{11}$, we obtain
\begin{align*}
     \theta_k &= \int_0^1 \left( Q_{11}^1(u)-Q_{11}^0(u)
    \right)\phi_k(u)du \\
    &= \lambda_k^*(Q_{11}^1) - \lambda_k^*(Q_{11}^0) \\[0.025in]
    &= \lambda_k^*(Q_{11})-\lambda_k^*(Q_{10}) -\lambda_k^*(Q_{01})+\lambda_k^*(Q_{00}),
\end{align*}
which is a functional of the four observed cell distributions. In particular, since $\phi_1 = 1$ the first normalized $L$-moment is the mean, and consequently,
\begin{align*}
    \theta_1 &=  \int_0^1 \left(Q_{11}^1(u)-Q_{11}^0(u) \right)du \\
    &=  \mathbb{E}\!\left[Y_1(1)\mid G=1\right] - \mathbb{E}\!\left[Y_1(0)\mid G=1\right] \\[0.025in]
    &= \mathbb{E} [Y_1(1)-Y_1(0)\mid G=1].
\end{align*}
Again, note that we do not require rank invariance or any other assumption about the joint distribution of $(Y_1(1),Y_1(0))$ within the treated group. Thus, $\theta_1$ is the usual average treatment effect on the treated. Finally, suppose that $\mathcal{L}(\mathbb{N})$ holds. The preceding argument identifies
\begin{align*}
    \theta_k = \int_0^1 \tau_{\mathrm{QTT}}(u)\phi_k(u)du
\end{align*}
for every integer $ k \geq 1$, where $\tau_{\mathrm{QTT}}(u)=Q^1_{11}(u)-Q^0_{11}(u)$. Because $\tau_{\mathrm{QTT}} \in L^1(0,1)$ under our finite absolute first moment assumption, by \cref{parallel_trends_spectrum_corollary} it follows that the coefficient sequence $\{\theta_1, \theta_2, \ldots \}$ uniquely determines $\tau_{\mathrm{QTT}}$ up to Lebesgue-almost-everywhere equality. Since $Q_{11}=Q_{11}^1$ is observed, $Q_{11}^0 = Q_{11}-\tau_{\mathrm{QTT}}$ is therefore also identified up to almost-everywhere equality.

\end{proof}

\subsection{Proof of \cref{identification_cic}} \label{identification_cic_proof}

The proof of \cref{identification_cic} follows similarly to the proofs of \cref{parallel_trends_spectrum_corollary} and \cref{identification_theorem}. However, under the changes-in-changes framework we now require new regularity conditions in order to ensure the composition $Q_{\mathrm{CiC}}(u) = Q_{01}\!\left( F_{00}\!\left(Q_{10}(u)\right) \right)$ is well-defined. Notably, Assumption~3.4 of \cite{athey_imbens_2006} requires the latent-support condition $\mathbb{U}_1 \subset \mathbb{U}_0$ in order to ensure the treated baseline support is contained in the control baseline support and that the treated counterfactual post-period support is contained in the control post-period support. In the $L$-moment framework, there is no latent variable $U$ or structural model $h_t$. Instead, we impose the observed-outcome support condition
\begin{align*}
    \operatorname{supp}(Y_0\mid G=1) \subseteq \operatorname{supp}(Y_0\mid G=0)
\end{align*}
which ensures that $Q_{01}(F_{00}(\cdot))$ is evaluated only at baseline outcome levels represented in the control group. Without this support condition, we can only identify the portion of the counterfactual distribution corresponding to the common support, e.g.\ see Corollary~3.1 of \cite{athey_imbens_2006}. Additionally, we assume that $0 < F_{00}(Q_{10}(u)) < 1$ for Lebesgue-almost every $u \in (0, 1)$ so that the composition $Q_{01}(F_{00}(Q_{10}(u)))$ is evaluated at a rank in the domain of $Q_{01}$. Finally, we must also assume $Q_{\mathrm{CiC}} \in L^1(0, 1)$ to ensure all normalized $L$-moments are finite. Because we assume finite absolute first moments of all potential outcome distributions, i.e., $Q_{11}^0,Q_{11}^1 \in L^1(0,1)$, it then follows that
\begin{align} \label{cic_in_l1}
      \mathcal{D}_{\mathrm{CiC}}^0 = Q_{11}^0-Q_{\mathrm{CiC}} \in L^1(0,1).
\end{align}
From here, we again split the proof into pieces.
\begin{enumerate}
    \item[1.] If $\mathcal{L}_{\mathrm{CiC}}(\mathcal{K})$ holds for some $\mathcal{K} \subseteq \mathbb{N}$, then, for every $k\in\mathcal{K}$,
    \begin{align*}
         \theta_k = \lambda_k^*(Q_{11}) - \int^1_0 Q_{\mathrm{CiC}}(u) \phi_k(u) du.
    \end{align*}
    In particular, if $1\in\mathcal{K}$, then
    \begin{align*}
        \theta_1 = \mathbb{E}[Y_1\mid G=1] - \int_0^1 Q_{\mathrm{CiC}}(u)du = \mathbb{E}\!\left[ Y_1(1)-Y_1(0) \mid G=1 \right].
    \end{align*}
    \begin{proof}[\textbf{Proof:}] To begin, note that, as shown in the proof of \cref{identification_theorem}, under consistency and no anticipation it follows that in the treated group $Q_{11}=Q_{11}^1$ and $Q_{10}^0=Q_{10}^1=Q_{10}$, and in the untreated group $Q_{00}=Q_{00}^0$ and $Q_{01}=Q_{01}^0$. Thus, the only quantile function that remains unobserved is $Q_{11}^{0}$. Next, suppose that $\mathcal{L}_{\mathrm{CiC}}(\mathcal{K})$ holds, and fix $k\in\mathcal{K}$. By definition,
\begin{align*}
     0 = d_{k,\mathrm{CiC}}^0 &= \int_0^1 \mathcal{D}_{\mathrm{CiC}}^0(u)\phi_k(u)du\\
     &= \int_0^1 \left( Q_{11}^0(u)-Q_{\mathrm{CiC}}(u) \right)\phi_k(u)du\\
    &= \lambda_k^*(Q_{11}^0) - \int_0^1 Q_{\mathrm{CiC}}(u)\phi_k(u)du.
\end{align*}
Therefore,
\begin{align*}
    \lambda_k^*(Q_{11}^0) = \int_0^1 Q_{\mathrm{CiC}}(u)\phi_k(u)du.
\end{align*}
Thus, we have identified exactly the $k$th normalized $L$-moment of the missing counterfactual distribution \textit{without} imposing equality between $Q_{11}^0$ and $Q_{\mathrm{CiC}}$ pointwise and furthermore, without placing any restrictions on coefficients indexed by $\ell\notin\mathcal{K}$. From here, because 
\begin{align*}
     \theta_k = \int_0^1 \left( Q_{11}^1(u)-Q_{11}^0(u)
    \right)\phi_k(u)du = \lambda_k^*(Q_{11}^1) - \lambda_k^*(Q_{11}^0)
\end{align*}
after rearranging we immediately obtain our final identified result,
\begin{align*}
     \theta_k = \lambda_k^*(Q_{11}) - \int_0^1 Q_{\mathrm{CiC}}(u)\phi_k(u)du.
\end{align*}
In the case of $\mathcal{L}_{\mathrm{CiC}}(\{1\})$ note that because $\phi_1 = 1$, then $\lambda_1^*(Q_{11}^0) = \int_0^1Q_{\mathrm{CiC}}(u)du$. Consequently,
\begin{align*}
     \theta_1 = \mathbb{E}[Y_1(1)\mid G=1] - \mathbb{E}[Y_1(0)\mid G=1] = \mathbb{E}[Y_1\mid G=1] -\int_0^1 Q_{\mathrm{CiC}}(u)du
\end{align*}
is the average treatment effect on the treated.
\end{proof}
    \item[2.] For every $K \geq 1$, $\mathcal{L}_{\mathrm{CiC}}(1\mathord{:}K+1) \implies \mathcal{L}_{\mathrm{CiC}}(1\mathord{:}K)$.
    \begin{proof}[\textbf{Proof:}] Again, this follows immediately, by construction. Since $\mathcal{L}_{\mathrm{CiC}}(1\mathord{:}K+1)$ requires $d_{k, \mathrm{CiC}}^0=0$ for every $k\in\{1,\ldots,K+1\}$, then necessarily we have $d_{k, \mathrm{CiC}}^0=0$ for every $k\in\{1,\ldots,K\}$, which is equivalent to $\mathcal{L}_{\mathrm{CiC}}(1\mathord{:}K)$. \end{proof}
    \item[3.] $\mathcal{L}_{\mathrm{CiC}}(\mathbb{N})$ holds if and only if $\mathcal{D}^0_{\mathrm{CiC}}(u) = 0$ for Lebesgue-almost every $u\in(0,1)$.

    \begin{proof}[\textbf{Proof:}] First suppose that $\mathcal{D}_{\mathrm{CiC}}^0(u)=0$ for Lebesgue-almost every $u \in (0, 1)$. Then, for every integer $k\geq1$,
\begin{align*}
    d_{k,\mathrm{CiC}}^0 = \int_0^1 \mathcal{D}_{\mathrm{CiC}}^0(u)\phi_k(u)du = 0,
\end{align*}
and therefore $\mathcal{L}_{\mathrm{CiC}}(\mathbb{N})$ holds. Conversely, suppose $\mathcal{L}_{\mathrm{CiC}}(\mathbb{N})$ holds. By \cref{cic_in_l1} we know that $\mathcal{D}_{\mathrm{CiC}}^0 \in L^1(0, 1)$; thus, we may apply \cref{legendre_determination}, which implies that $\mathcal{D}_{\mathrm{CiC}}^0(u)=0$ for Lebesgue-almost every $u \in (0, 1)$. Finally, since by definition $\mathcal{D}_{\mathrm{CiC}}^0 = Q_{11}^0-Q_{\mathrm{CiC}}$, it follows that $Q_{11}^0(u) = Q_{\mathrm{CiC}}(u)$ for Lebesgue-almost every $u \in (0, 1)$. Combining this with the fact that $Q_{11}^1=Q_{11}$, we then have
\begin{align*}
    \tau_{\mathrm{QTT}}(u) = Q_{11}(u)-Q_{\mathrm{CiC}}(u)
\end{align*}
almost everywhere, thereby completing the proof.
    \end{proof}
       
\end{enumerate}

\subsection{Proof of \cref{individual_heterogeneity_bound}}
\label{individual_heterogeneity_bound_proof}

\begin{proof}[\textbf{Proof:}]
For notational simplicity, we define $T=Y_1(1)-Y_1(0)$. Under the assumption that the conditional second moments are finite, all quantities below are finite. The conditional joint distribution of the two potential outcomes given $G=1$ is a coupling of $F_{11}^1$ and $F_{11}^0$. The optimal-coupling characterization of the Wasserstein distance therefore gives
\begin{align*}
    \mathbb{E}[T^2\mid G=1]
    \geq W_2^2(F_{11}^1,F_{11}^0).
\end{align*}
Since $\mathbb{E}[T\mid G=1]=\theta_1$, subtracting
$\theta_1^2$ and applying
\cref{wasserstein_treatment_decomposition} yields
\begin{align*}
    \mathbb{V}(T\mid G=1) \geq W_2^2(F_{11}^1,F_{11}^0)-\theta_1^2  =\sum_{k=2}^{\infty}\theta_k^2 \geq\sum_{k\in\mathcal{K}}\theta_k^2 =H_{\mathcal{K}}^2.
\end{align*}
Each selected coordinate is identified by
\cref{identification_theorem} or \cref{identification_cic}
under the corresponding restrictions.  For fixed marginals, let $U\sim\operatorname{Uniform}(0,1)$ and set $Y_1(d)=Q_{11}^d(U)$ for $d\in\{0,1\}$. This coupling satisfies
\begin{align*}
    \mathbb{E}[T^2\mid G=1] =\int_0^1
      \{Q_{11}^1(u)-Q_{11}^0(u)\}^2du =W_2^2(F_{11}^1,F_{11}^0),
\end{align*}
and hence attains the variance bound $W_2^2(F_{11}^1,F_{11}^0)-\theta_1^2$.
This establishes sharpness of the full marginal bound over the couplings.
\end{proof}

\subsection{Proof of \cref{moment_bound}} \label{moment_bound_proof}

\begin{proof}[\textbf{Proof:}] To begin, for $k \geq 2$ define the integrated Legendre weight
\begin{align} \label{ak_bound}
    A_k(t) = \int^t_0 \phi_k(u)du = \frac{P_k(2t - 1) - P_{k-2}(2t-1)}{2 \sqrt{2k-1}}
\end{align}
and note that $A_k(0) = A_k(1) = 0$. By \cite{szego1975orthogonal, Wang2023}, there exists some universal constant $C_0$ such that, for all $0 \leq t \leq 1$,
\begin{align*}
    |A_k(t)| \leq \frac{C_0}{\sqrt{k(k-1)}} \{t(1-t)\}^{1/4}.
\end{align*}
Next, let $F(z) = \mathbb{P}(X \leq z)$ and let $Q(u)=\inf\{x \in\mathbb{R}:F(x)\geq u\}$ denote the quantile function. Observe that for every $u \in (0, 1)$ and $a \in \mathbb{R}$ we may write
\begin{align} \label{quantile_diff}
    Q(u) - a = \int^\infty_a \mathbb{I}(Q(u) > z) dz - \int^a_{-\infty} \mathbb{I}(Q(u) \leq z) dz.
\end{align}
Note that because $\int^1_0 \phi_k(u) du = 0$ subtracting $a$ does not change the $L$-moment coefficient,
\begin{align*}
      \lambda_k^*(Q) =\int_0^1 Q(u)\phi_k(u)du =\int_0^1(Q(u)-a)\phi_k(u)du.
\end{align*}
Consequently, multiplying \cref{quantile_diff} by $\phi_k(u)$ and integrating across $(0, 1)$ yields
\begin{align*}
   \lambda_k^*(Q) &= \int_0^1\int_a^\infty \phi_k(u)\mathbb{I}(Q(u)>z)dz \, du - \int_0^1\int_{-\infty}^a \phi_k(u)\mathbb{I}(Q(u)\leq z)dz \,du \\
    &\overset{(i)}{=} \int_a^\infty \left\{\int_0^1 \phi_k(u)\mathbb{I}(Q(u)>z) du\right\}dz - \int_{-\infty}^a \left\{\int_0^1 \phi_k(u)\mathbb{I}(Q(u)\leq z) du\right\}dz \\
    &\overset{(ii)}{=}  \int_a^\infty \left\{\int_{F(z)}^1\phi_k(u) du\right\}dz - \int_{-\infty}^a \left\{\int_0^{F(z)}\phi_k(u)du\right\}dz \\
    &\overset{(iii)}{=} -\int_{\mathbb{R}} A_k(F(z))dz,
\end{align*}
where $(i)$ follows by Fubini's theorem, $(ii)$ follows since
\begin{align*}
    \int_0^1\phi_k(u)\mathbb{I}(Q(u)\leq z)du &=\int_0^{F(z)}\phi_k(u)du, \\
    \int_0^1 \phi_k(u)\mathbb{I}(Q(u)>z)du &= \int_{F(z)}^1 \phi_k(u)du,
\end{align*}
and $(iii)$ follows because $\int_{F(z)}^1 \phi_k(u)du = A_k(1)-A_k(F(z)) = -A_k(F(z))$ since $A_k(1) = 0$, and likewise,
\begin{align*}
    \int_0^{F(z)}\phi_k(u)du=A_k(F(z)).
\end{align*}
Consequently, applying \cref{ak_bound} we can see that
\begin{align*}
    |\lambda^*_k(Q)| \leq \frac{C_0}{\sqrt{k(k-1)}} \int_{\mathbb{R}} \big( F(z)(1 - F(z)) \big)^{1/4} dz.
\end{align*}
Now, fix some $a \in \mathbb{R}$ and let $m = \mathbb{E}[|X - a|^p]^{1/p}$. By Markov's inequality,
\begin{align*}
    \mathbb{P}(|X-a|\geq t) = \mathbb{P}(|X-a|^p\geq t^p) \leq\frac{\mathbb{E}[|X-a|^p]}{t^p} =\left(\frac{m}{t}\right)^p.
\end{align*}
Consequently, it follows that
\begin{align*}
    F(a - t) = \mathbb{P}(X \leq a - t) \leq \mathbb{P}(|X - a| \geq t) \leq  \min \left\{ 1, \left( \frac{m}{t} \right)^{p} \right\}
\end{align*}
and similarly, $1 - F(a + t) \leq \min \left\{ 1, \left( \frac{m}{t} \right)^{p} \right\}$. Now, our plan is to split $\int_{\mathbb{R}} \big( F(z)(1 - F(z)) \big)^{1/4} dz$ at $a$. First, note that because $0 \leq F(z) \leq 1$ it follows that for all $z \in \mathbb{R}$,
\begin{align*}
     (F(z)(1-F(z)))^{1/4}&\leq F(z)^{1/4}, \\
    (F(z)(1-F(z)))^{1/4}&\leq(1-F(z))^{1/4}.
\end{align*}
We'll use the first bound for values below $a$ and the second for values above $a$ to obtain a one-sided tail probability on both sides. That is,
\begin{align*}
     \int_{\mathbb{R}} (F(z)(1-F(z)))^{1/4}dz &= \int_{-\infty}^a (F(z)(1-F(z)))^{1/4}dz + \int_a^\infty (F(z)(1-F(z)))^{1/4}dz\\
    &\leq \int_{-\infty}^a F(z)^{1/4}dz +\int_a^\infty (1-F(z))^{1/4}dz.
\end{align*}
From here, in the lower-tail integral, substitute $t=a-z$ and in the upper-tail integral, substitute $t=z-a$. Then, it follows that
\begin{align*}
    \int_{\mathbb{R}} \big( F(z)(1 - F(z)) \big)^{1/4} dz &\leq \int^\infty_{0} \left( F(a - t)^{1/4} + (1 - F(a + t))^{1/4} \right)dt \\
    &\leq 2 \int^\infty_0 \min \left\{ 1, \left( \frac{m}{t} \right)^{p / 4} \right\} dt \\
    &= 2 \left(m + m^{p/4} \int^\infty_m t^{-p/4} dt \right) \\
    &= \left( \frac{2p}{p-4} \right) m.
\end{align*}
Then, taking the infimum over $a$ establishes for $k \geq 2$,
\begin{align*}
     |\lambda^*_k(Q)| \leq \frac{C_p M_p(Q)}{\sqrt{k(k-1)}}
\end{align*}
where $C_p = 2 C_0 p / (p-4)$. Finally, to complete the desired inequality, recall that
\begin{align*}
    R^2_K(Q) = \sum_{k > K} \left(\lambda^*_k(Q) \right)^2.
\end{align*}
Consequently, it is easy to see that
\begin{align*}
     R^2_K(Q) \leq C^2_p M^2_p(Q) \sum_{k > K} \frac{1}{k(k-1)} = \frac{C^2_p M^2_p(Q)}{K}
\end{align*}
because the remaining series telescopes. Observe that
\begin{align*}
    \frac{1}{k(k-1)}=\frac{1}{k-1}-\frac{1}{k}.
\end{align*}
Then, for every integer $N > K$,
\begin{align*}
    \sum_{k=K+1}^{N}\frac{1}{k(k-1)} =\sum_{k=K+1}^{N} \left(\frac{1}{k-1}-\frac{1}{k}\right) =\frac{1}{K}-\frac{1}{N}.
\end{align*}
Letting $N\to\infty$ therefore gives
$\sum_{k>K}(k(k-1))^{-1}=1/K$. Finally, taking square roots completes the proof.

\end{proof}

\subsection{Proof of \cref{higher_order_energy_bound}} \label{higher_order_energy_bound_proof}

\begin{proof}[\textbf{Proof:}] To begin, let $w(u)=u(1-u)$ and let $\ell_n(u)=\phi_{n+1}(u)=\sqrt{2n+1}P_n(2u-1)$ denote the $(n+1)$st shifted Legendre polynomial, so that $\ell_n$ has degree $n$. Then, for $n\geq s$, set
\begin{align*}
     \Lambda_{n,s}=\frac{(n+s)!}{(n-s)!}
\end{align*}
and $U_{n,s}(u)=w(u)^s\ell_n^{(s)}(u)$. First, our goal is to prove the higher-order derivative identity
\begin{align}  \label{deriv_identity}
     U_{n,s}^{(s)}(u) =(-1)^s\Lambda_{n,s}\ell_n(u).
\end{align}
To do so, we leverage classical Jacobi derivative identities. Let $P_j^{(\alpha,\beta)}$ denote the $j$th Jacobi polynomial and recall that, when $\alpha = \beta = 0$, then 
\begin{align*}
    P^{(0, 0)}_j(x) = P_j(x)
\end{align*}
is simply the $j$th Legendre polynomial, and after substituting $x = 2u - 1$, the $j$th shifted Legendre polynomial. However, differentiation changes these parameters. For example,
\begin{align*}
    \frac{d}{dx} P_j(x) = \frac{j + 1}{2} P^{(1, 1)}_{j-1}(x)
\end{align*}
so now $\alpha = \beta = 1$. More generally, by equation~18.9.15 of the Jacobi derivative identities \citep{NISTDLMF},
\begin{align*}
    \frac{d}{dx} P_n^{(\alpha,\beta)}(x) = \frac{1}{2}(n+\alpha+\beta+1) P_{n-1}^{(\alpha+1,\beta+1)}(x),
\end{align*}
and by equation~18.9.16,
\begin{align*}
     \frac{d}{dx} \left((1-x)^\alpha (1+x)^\beta P_n^{(\alpha,\beta)}(x) \right) = -2(n+1)(1-x)^{\alpha-1}(1+x)^{\beta-1} P_{n+1}^{(\alpha-1,\beta-1)}(x).
\end{align*}
Thus, applying the change of variables $x=2u-1$, it follows that
\begin{align}
     \ell_n^{(s)}(u) &=\sqrt{2n+1}\frac{(n+s)!}{n!} P_{n-s}^{(s,s)}(2u-1), \label{legendre-derivative}  \\
    \frac{d^s}{du^s} \left\{w(u)^s P_{n-s}^{(s,s)}(2u-1)\right\} &=(-1)^s\frac{n!}{(n-s)!}P_n(2u-1). \label{weighted-jacobi-derivative}
\end{align}
Substituting \cref{legendre-derivative} into $U_{n,s}(u)$ yields
\begin{align*}
    U_{n,s}(u) &= w(u)^s\ell_n^{(s)}(u) = \sqrt{2n+1}\frac{(n+s)!}{n!} w(u)^s P_{n-s}^{(s,s)}(2u-1).
\end{align*}
Consequently, taking $s$ derivatives, 
\begin{align*}
    U_{n,s}^{(s)}(u) &= \sqrt{2n+1}\frac{(n+s)!}{n!} \frac{d^s}{du^s} \left\{w(u)^s P_{n-s}^{(s,s)}(2u-1)\right\}\\
    &\overset{(i)}{=} \sqrt{2n+1}\frac{(n+s)!}{n!} (-1)^s\frac{n!}{(n-s)!}P_n(2u-1)\\
    &= (-1)^s\frac{(n+s)!}{(n-s)!}\ell_n(u) \\
    &=(-1)^s\Lambda_{n,s}\ell_n(u),
\end{align*}
where $(i)$ follows by \cref{weighted-jacobi-derivative}, thereby establishing \cref{deriv_identity}. From here, observe that we can write $U_{n,s}(u)=u^s A(u)$, where $A(u)=(1-u)^s\ell_n^{(s)}(u)$ is a polynomial. Thus, for any integer $0 \leq r < s$, by the product rule,
\begin{align*}
     U_{n,s}^{(r)}(0) = \sum_{j=0}^{r}\binom{r}{j} \left.\frac{d^j}{du^j}u^s\right|_{u=0} A^{(r-j)}(0) = 0.
\end{align*}
Note that we can simply factor out $(1-u)^s$ to give the identical argument at $u=1$. Therefore, $U_{n,s}^{(r)}(0)=U_{n,s}^{(r)}(1)=0$ for $r = 0, \ldots, s - 1$. Consequently, it follows that after repeatedly integrating by parts, for some $g\in C^{s-1}([0,1])$,
\begin{align*}
    \int_0^1 g(u)U_{n,s}^{(s)}(u)du &= \sum_{j=0}^{s-1}(-1)^j \big[g^{(j)}(u)U_{n,s}^{(s-1-j)}(u)\big]_{u=0}^{u=1} +(-1)^s\int_0^1g^{(s)}(u)U_{n,s}(u)du \\
    &= (-1)^s\int_0^1g^{(s)}(u)U_{n,s}(u)du
\end{align*}
since the boundary terms drop out. Thus, after rearranging and substituting in \cref{deriv_identity},
\begin{align*}
    \int_0^1 g^{(s)}(u)U_{n,s}(u)du &= (-1)^s \int_0^1 g(u)U_{n,s}^{(s)}(u)du \\
    &=  (-1)^s \int_0^1 g(u) \big((-1)^s\Lambda_{n,s}\ell_n(u) \big)du \\
    &= \Lambda_{n,s} \int_0^1 g(u) \ell_n(u) du \\
    &= \Lambda_{n,s} c_{n+1}(g)
\end{align*}
where we say $c_{n+1}(g) = \int_0^1 g(u)\ell_n(u)du$. Furthermore, note that this same calculation with $g= \ell_m$ for $m \neq n$ and $n, m \geq s$, followed by orthonormality of the Legendre polynomials, gives
\begin{align} \label{deriv_orthogonal}
    \int_0^1 w(u)^s \ell_n^{(s)}(u)\ell_m^{(s)}(u)du = \Lambda_{n,s} \int_0^1 \ell_n(u)\ell_m(u) du = \Lambda_{n,s} \mathbb{I}(n=m).
\end{align}

\medskip 

We now turn these identities into an upper bound on $R_K(g)$. First, define the coefficients 
\begin{align*}
     e_{n,s}(u)= \frac{w(u)^{s/2}\ell_n^{(s)}(u)}{\sqrt{\Lambda_{n,s}}},
\end{align*}
for $n \geq s$. By \cref{deriv_orthogonal}, this is an orthonormal family in $L^2(0, 1)$, and furthermore,
\begin{align*}
    \langle w^{s/2} g^{(s)}, e_{n, s} \rangle = \sqrt{\Lambda_{n,s}} c_{n+1}(g).
\end{align*}
Consequently, applying Bessel's inequality \citep{conway1985course} first to a finite summation in $K$, then extending to its limit, yields
\begin{align*}
    \sum_{n=s}^{\infty}\Lambda_{n,s}c_{n+1}(g)^2 \leq \int_0^1 w(u)^s|g^{(s)}(u)|^2du = \mathcal{E}_s(g).
\end{align*}
From here, recall that $R^2_K(g) = \sum^\infty_{n=K} c^2_{n+1}(g)$ and note that $\Lambda_{n, s}$ is increasing for $n \geq s$. Therefore, for $K \geq s$ it follows that
\begin{align} \label{rkg_bound}
     R^2_K(g) &\leq \frac{1}{\Lambda_{K,s}} \sum_{n=K}^{\infty}\Lambda_{n,s}c^2_{n+1}(g) \leq \frac{(K-s)!}{(K+s)!}\mathcal{E}_s(g).
\end{align}
From here, let $T_K= I -\Pi_K$. Because $\Pi_K$ is an orthogonal projection, for every $v\in L^2(0,1)$,
\begin{align*}
    \|T_Kv\|_2^2 =\|v\|_2^2-\|\Pi_Kv\|_2^2 \leq\|v\|_2^2.
\end{align*}
Thus, taking a projection remainder cannot increase the $L^2$ norm. Next, using linearity and the decomposition $f=(f-g)+g$, we obtain
\begin{align*}
    R_K(f) &=\|T_K(f-g)+T_Kg\|_2 \leq\|T_K(f-g)\|_2+\|T_Kg\|_2 \leq\|f-g\|_2 + R_K(g),
\end{align*}
thereby completing the proof, after applying \cref{rkg_bound}.
\end{proof}

\subsection{Proof of \cref{rk_bound_general}} \label{rk_bound_general_proof}

\begin{proof}[\textbf{Proof:}] To begin, fix $0 < \delta < 1/2$ and recall that applying \cref{higher_order_energy_bound} to the auxiliary function $g_\delta$ yields
\begin{align*}
      R_K(f) \leq \inf_{\delta \in (0, 1/2)} \left\{ E_s(f;\delta) + \sqrt{\frac{(K-s)!}{(K+s)!}}  I_s(f;\delta) \right\}.
\end{align*}
We proceed by bounding $ E^2_s(f;\delta) = \int_0^\delta [f(u)-p^-_{\delta}(u)]^2du +\int_{1-\delta}^1 [f(u)-p^+_{\delta}(u)]^2du$. For each fixed $0<u<\delta$, the function $f^{(s-1)}$ is absolutely continuous on $[u,\delta]$, since it is Lipschitz there. Furthermore, each lower derivative is continuously differentiable on that compact interval and is therefore absolutely continuous as well. Taylor's formula with an integral remainder therefore yields
\begin{align*}
     f(u)-p^-_\delta(u) =\frac{1}{(s-1)!}\int_\delta^u (u-t)^{s-1}f^{(s)}(t) dt,
\end{align*}
and so, after taking absolute values and reversing the integration limits,
\begin{align*}
        |f(u)-p^-_\delta(u)| \leq\frac{1}{(s-1)!}\int_u^\delta (t-u)^{s-1}|f^{(s)}(t)| dt.
\end{align*}
Note that for $t\in[u,\delta]$, requiring $\delta<1/2$ ensures $d(t)=t$. Applying the derivative envelope $|f^{(s)}(u)| \leq C d(u)^{-s-\alpha}$, then using $(t-u)^{s-1}\leq t^{s-1}$ and $t^{-\alpha}\leq u^{-\alpha}$,
we obtain
\begin{align*}
    |f(u)-p^-_\delta(u)| &\leq\frac{C}{(s-1)!} \int_u^\delta(t-u)^{s-1}t^{-s-\alpha} dt\\
    &\leq\frac{C}{(s-1)!}\int_u^\delta t^{-1-\alpha}dt\\
    &\leq\frac{C}{(s-1)!}u^{-\alpha} \int_u^\delta \left(\frac{1}{t} \right) dt\\
    &=\frac{C}{(s-1)!}u^{-\alpha}\log(\delta/u).
\end{align*}
Next, squaring and integrating and letting $u=\delta v$ gives
\begin{align*}
    \int_0^\delta|f(u)-p^-_\delta(u)|^2 du
    &\leq\frac{C^2}{((s-1)!)^2} \int_0^\delta u^{-2\alpha}(\log(\delta/u))^2du\\
    &=\frac{C^2\delta^{1-2\alpha}}{((s-1)!)^2} \int_0^1v^{-2\alpha}(\log(1/v))^2dv.
\end{align*}
Note that this integral is finite for $\alpha<1/2$. Next, substituting $x=-\log(v)$ and $y=(1-2\alpha)x$ yields
\begin{align*}
    \int_0^1v^{-2\alpha}(\log(1/v))^2dv &=\int_0^\infty x^2e^{-(1-2\alpha)x}dx\\
    &=\frac{1}{(1-2\alpha)^3} \int_0^\infty y^2e^{-y}dy =\frac{2}{(1-2\alpha)^3}.
\end{align*}
Consequently, we obtain the bound
\begin{align} \label{lower_endpoint_bound}
     \int_0^\delta|f(u)-p^-_\delta(u)|^2du \leq \frac{2C^2}{((s-1)!)^2(1-2\alpha)^3} \delta^{1-2\alpha}.
\end{align}
Following analogous steps, we obtain the exact same bound in \cref{lower_endpoint_bound} for the upper interval $\int_{1-\delta}^1 [f(u)-p^+_{\delta}(u)]^2du$. Thus, combining bounds and taking a square root yields the inequality
\begin{align*}
      E_s(f;\delta) \leq\frac{2C}{(s-1)!(1-2\alpha)^{3/2}} \delta^{1/2-\alpha}.
\end{align*}
We now bound the interior interval $I_s(f;\delta)$. For $0<u<1$, note that $w(u)=u(1-u)\leq d(u)$. Furthermore, if $u\leq1/2$ then $w(u)\leq u$,
and if $u\geq1/2$ then $w(u)\leq1-u$. Thus,
\begin{align*}
    I_s(f;\delta)^2
    &=\int_\delta^{1-\delta}w(u)^s|f^{(s)}(u)|^2du\\
    &\leq C^2\int_\delta^{1-\delta} w(u)^s d(u)^{-2s-2\alpha}du\\
    &\leq C^2\int_\delta^{1-\delta}d(u)^{-s-2\alpha}du\\
    &=C^2\left\{ \int_\delta^{1/2}u^{-s-2\alpha} du +\int_{1/2}^{1-\delta}(1-u)^{-s-2\alpha}du \right\}\\
    &=2C^2\int_\delta^{1/2}t^{-s-2\alpha}dt
\end{align*}
where in the last equality we use the substitution $t = 1 - u$. Because $s\geq2$ and $\alpha\geq0$, $s+2\alpha-1>0$. Direct integration then gives
\begin{align*}
  \int_\delta^{1/2}t^{-s-2\alpha}dt =\frac{\delta^{1-s-2\alpha}-(1/2)^{1-s-2\alpha}}{s+2\alpha-1} \leq\frac{\delta^{1-s-2\alpha}}{s+2\alpha-1},
\end{align*}
and so
\begin{align*}
     I_s(f;\delta) \leq\frac{\sqrt{2}C}{\sqrt{s+2\alpha-1}}  \delta^{(1-s-2\alpha)/2}.
\end{align*}
Finally, note that after canceling common factors,
\begin{align*}
      \frac{(K+s)!}{(K-s)!} =\left\{\prod_{j=-s+1}^{0}(K+j)\right\} \left\{\prod_{j=1}^{s}(K+j)\right\}.
\end{align*}
Each product contains $s$ factors and the smallest factor in the first product is $K-s+1$. Furthermore,
\begin{align*}
    K-s+1-\frac{K}{s}=\frac{(s-1)(K-s)} s \geq 0.
\end{align*}
Every factor in the first product is therefore at least $K/s$ and every factor in the second product is at least $K$. Thus,
\begin{align*}
    \frac{(K+s)!}{(K-s)!}\geq(K/s)^sK^s =\frac{K^{2s}}{s^s}
\end{align*}
and therefore, $\sqrt{\frac{(K-s)!}{(K+s)!}} \leq s^{s/2}K^{-s}$. Then, putting everything together we now have 
\begin{align*}
     R_K(f) \leq C\left\{\frac{2}{(s-1)!(1-2\alpha)^{3/2}} \delta^{1/2-\alpha} +\frac{\sqrt{2}\,s^{s/2}}{\sqrt{s+2\alpha-1}} K^{-s}\delta^{(1-s-2\alpha)/2} \right\}
\end{align*}
for $0 < \delta < 1/2$. To complete the proof, let $\delta=K^{-2}$. This choice of $\delta$ is admissible for every $K \geq s \geq 2$, since $0<K^{-2}\leq 1/4 < 1/2$. Then,
\begin{align*}
    \delta^{1/2-\alpha} &= K^{-1+2\alpha}, \\
    K^{-s}\delta^{(1-s-2\alpha)/2} &=K^{-s}K^{s+2\alpha-1}=K^{-1+2\alpha}.
\end{align*}
Therefore, if we define the universal constant depending only on $s$, $\alpha$, and $C$,
\begin{align*}
     C_{s,\alpha} = C\left\{ \frac{2}{(s-1)!(1-2\alpha)^{3/2}} +\frac{\sqrt{2}\,s^{s/2}}{\sqrt{s+2\alpha-1}} \right\},
\end{align*}
we obtain the final inequality of
\begin{align*}
    R_K(f)\leq C_{s,\alpha}K^{-1+2\alpha} =\frac{C_{s,\alpha}}{K^{1-2\alpha}},
\end{align*}
thereby completing the proof.
\end{proof}

\subsection{Proof of \cref{decay_smoothness}} \label{decay_smoothness_proof}

\begin{proof}[\textbf{Proof:}] Following the same steps as in the proof of \cref{higher_order_energy_bound}, it immediately follows that
\begin{align*}
     R^2_K(f) &\leq \frac{1}{\Lambda_{K,s}} \sum_{n=K}^{\infty}\Lambda_{n,s}c^2_{n+1}(f) \\
       &\leq \frac{(K-s)!}{(K+s)!} \int_0^1 (u(1-u))^s|f^{(s)}(u)|^2du\\
    &\leq \frac{(K-s)!}{(K+s)!} \frac{(s!)^2}{(2s+1)!} L^2_s
\end{align*}
where in the last inequality we use the Lipschitzness assumption $|f^{(s)}|\leq L_s$ almost everywhere and that
\begin{align*}
    \int_0^1 u^s(1-u)^s du= \frac{(s!)^2}{(2s+1)!}.
\end{align*}
Thus, taking square roots yields our desired inequality on $R_K(f)$. Finally, note that for a fixed integer $s \geq 1$ and $K\geq s$, canceling the common factorial terms gives
\begin{align*}
    \frac{(K-s)!}{(K+s)!} &=\frac{1}{(K-s+1)(K-s+2)\cdots(K+s)} =\frac{1}{K^{2s}}\prod_{j=-s+1}^{s} \left(1+\frac{j}{K}\right)^{-1}.
\end{align*}
Since $s$ is fixed, the number of factors in the product does not grow with $K$, and each factor $(1+j/K)^{-1}$ tends to one.
Consequently,
\begin{align*}
    \frac{(K-s)!/(K+s)!}{K^{-2s}} =\prod_{j=-s+1}^{s}\left(1+\frac{j}{K}\right)^{-1} \longrightarrow 1
\end{align*}
and therefore, $ \frac{(K-s)!}{(K+s)!} \sim K^{-2s}$ which gives us that $R_K(f)=O(K^{-s})$.

\end{proof}

\subsection{Proof of \cref{counterfactual_cancellation}} \label{counterfactual_cancellation_proof}

\begin{proof}[\textbf{Proof:}] To begin, observe that for $0<u<v<1$, the assumed lower bound on rank spacings along with the mean value theorem imply
\begin{align*}
    Q^1(v)-Q^1(u) =Q^0(v)-Q^0(u)+h(v)-h(u) \geq\bigl(m-\|h'\|_\infty\bigr)(v-u)>0.
\end{align*}
Thus $Q^1$ is strictly increasing. Since $h$ is continuous and bounded on $[0,1]$, adding it preserves the left-continuity of the usual quantile representative of $Q^0$ and its square integrability. Consequently, $Q^1$ is the quantile function of $Q^1(U)$ for $U\sim\operatorname{Uniform}(0,1)$ and belongs to $L^2(0,1)$. Next, let $T_K=I-\Pi_K$. Linearity gives us that $T_KQ^1=T_KQ^0+T_Kh$, so the reverse triangle inequality yields
\begin{align*}
    \big|R_K(Q^1)-R_K(Q^0)\big| =\big|\|T_KQ^0+T_Kh\|_2-\|T_KQ^0\|_2\big| \leq\|T_Kh\|_2=R_K(h).
\end{align*}
From here, observe that $R_K(h)=o(r_K)$ implies
\begin{align*}
    \left|\frac{R_K(Q^1)}{R_K(Q^0)}-1\right| \leq\frac{R_K(h)}{R_K(Q^0)}\longrightarrow 0
\end{align*}
where $R_K(Q^0) \asymp r_K>0$ ensures the denominator is positive and bounded below by a positive multiple of $r_K$ for all sufficiently large $K$. Finally, $Q^1-Q^0=h$ gives
\begin{align*}
    R_K(Q^1-Q^0)=R_K(h)=o(r_K),
\end{align*}
thereby completing the proof.
    
\end{proof}

\subsection{Proof of \cref{lmoment_eif}} \label{lmoment_eif_proof}

\begin{proof}[\textbf{Proof:}] To begin, fix some integer $j \geq 0$, let $h_j(y_0,\ldots,y_j) = \max\{y_0,\ldots,y_j\}$, and define $H_j(F)=\mathbb{E}_F\!\left[h_j(Y_0,\ldots,Y_j)\right]$ to be the expected maximum draw. Then, recall that the $k$th normalized $L$-moment can be written as
\begin{align*}
    \lambda^*_k(F) = \sqrt{2k-1} \sum_{j=0}^{k-1} \xi_{kj} \beta_j(F)
\end{align*}
where $H_j(F) =(j+1)\int_0^1Q_F(u)u^j du =(j+1)\beta_j(F)$ and $\xi_{kj} = (-1)^{k-1+j} \binom{k-1}{j}\binom{k-1+j}{j}$. Note that 
\begin{align*}
     \sum^{k-1}_{j=0} \xi_{kj} u^j = P^*_{k-1}(u)
\end{align*}
is simply the shifted Legendre polynomial of degree $k-1$. Now, to derive the efficient influence function for $\lambda^*_k(F)$ we may focus on $H_j(F)$. Let $\{F_\varepsilon:\varepsilon\in(-\delta,\delta)\}$ be an arbitrary regular parametric submodel through $F=F_0$, with score $s\in L_0^2(F)$. Because $F\in\mathcal{P}_2(\mathbb{R})$, $h_j\in L^2(F^{\otimes(j+1)})$. Differentiating the product expectation at $\varepsilon=0$ therefore yields
\begin{align} \label{h_deriv}
     \left.\frac{d}{d\varepsilon}H_j(F_\varepsilon) \right|_{\varepsilon=0} = \sum_{\ell=0}^j \mathbb{E}_F\!\left[h_j(Y_0,\ldots,Y_j)s(Y_\ell)\right].
\end{align}
Now, define $M_j(y;F) = \mathbb{E}_F\!\left[\max\{y,Y_1,\ldots,Y_j\}\right]$, where the expectation averages over $j$ independent draws from $F$, and set $M_0(y;F)=y$. Thus $M_j$ fixes one argument of the kernel and averages over the others. Independence of the draws and symmetry of $h_j$ imply, for each $\ell=0,\ldots,j$,
\begin{align*}
    \mathbb{E}_F\!\left[h_j(Y_0,\ldots,Y_j)\mid Y_\ell\right] = M_j(Y_\ell;F).
\end{align*}
Consequently, 
\begin{align*}
    \mathbb{E}_F\!\left[h_j(Y_0,\ldots,Y_j)s(Y_\ell)\right] &= \mathbb{E}_F\!\left[ s(Y_\ell) \mathbb{E}_F\!\left[h_j(Y_0,\ldots,Y_j)\mid Y_\ell\right] \right] \\
    &= \mathbb{E}_F\!\left[s(Y_\ell)M_j(Y_\ell;F)\right] \\
    &= \mathbb{E}_F\!\left[s(Y)M_j(Y;F)\right]
\end{align*}
where we say $Y \sim F$. Then, because all $j+1$ terms have the same expectation, it follows that \cref{h_deriv} simplifies to
\begin{align*}
     \left.\frac{d}{d\varepsilon}H_j(F_\varepsilon)
    \right|_{\varepsilon=0}
    =(j+1)\mathbb E_F\!\left[M_j(Y;F)s(Y)\right].
\end{align*}
From here, because $H_j(F) = (j+1)\beta_j(F)$, we can simply divide by $j + 1$ and use the fact that $\mathbb{E}_F[s(Y)] = 0$ to center our expression to arrive at
\begin{align*}
    \left.\frac{d}{d\varepsilon}\beta_j(F_\varepsilon) \right|_{\varepsilon=0} &=\mathbb{E}_F\!\left[\left(M_j(Y;F)-(j+1)\beta_j(F)\right)s(Y) \right].
\end{align*}
Thus, because $M_j(Y;F)-(j+1)\beta_j(F)$ is mean zero and belongs to $L^2(F)$ it follows that $y \mapsto M_j(y;F)-(j+1)\beta_j(F)$ is the pathwise gradient of $\beta_j(F)$. To complete the derivation of the influence function for the $k$th normalized $L$-moment, we can simply apply the linearity of the pathwise derivative to obtain
\begin{align*}
     \ell_k(y;F) = \sqrt{2k-1}\sum_{j=0}^{k-1}\xi_{kj} \left(M_j(y;F)-(j+1)\beta_j(F)\right),
\end{align*}
thereby completing the proof. Additionally, we establish an integral representation of the efficient influence function. Let $F(z) = \mathbb{P}(Y \leq z)$ denote the cumulative distribution function of $Y \sim F$. First, observe that the excess of the maximum draw above a fixed $y \in \mathbb{R}$ can be written as
\begin{align*}
    \max\{y, Y_1, \ldots, Y_j \} - y = \int^\infty_y \mathbb{I}(\max\{Y_1, \ldots, Y_j\} > z) dz.
\end{align*}
Thus, applying Tonelli's theorem gives us
\begin{align*}
    M_j(y; F) - y = \int^\infty_y \left(1 - \mathbb{P}(Y_1 \leq z, \ldots, Y_j \leq z) \right) dz = \int^\infty_y \left(1 - F(z)^j \right) dz.
\end{align*}
Next, we want to obtain an analogous representation of $(j+1)\beta_j(F) = H_j(F)$. Recall that $H_j(F) = \mathbb{E}[\max\{Y_0,\ldots,Y_j\}]$. Now, it follows that
\begin{align*}
    \max\{Y_0,\ldots,Y_j\}-y = \int_y^\infty \mathbb{I}\!\left(\max\{Y_0,\ldots,Y_j\}>z\right)dz - \int_{-\infty}^y \mathbb{I}\!\left(\max\{Y_0,\ldots,Y_j\}\leq z\right)dz.
\end{align*}
Note that since our draws are independent, $\mathbb{P}(\max\{Y_0, \ldots, Y_j \} \leq z) = F(z)^{j+1}$. Consequently, for every fixed $y \in \mathbb{R}$,
\begin{align*}
    H_j(F) = y + \int_y^\infty(1-F(z)^{j+1})dz -\int_{-\infty}^yF(z)^{j+1}dz.
\end{align*}
Now, combining expressions, we obtain
\begin{align*}
     M_j(y;F)-(j+1)\beta_j(F) &= \int_{-\infty}^yF(z)^{j+1} dz +\int_y^\infty (F(z)^{j+1}-F(z)^j) dz\\
    &= \int_{-\infty}^yF(z)^{j+1} dz -\int_y^\infty F(z)^j(1-F(z))dz\\
    &= \int_{\mathbb{R}}F(z)^j (F(z)-\mathbb{I}(y\leq z))dz.
\end{align*}
Finally, plugging this expression back into our influence function yields
\begin{align*}
       \ell_k(y;F) &= \sqrt{2k-1}\sum_{j=0}^{k-1}\xi_{kj} \int_{\mathbb{R}}F(z)^j (F(z)-\mathbb{I}(y\leq z)) dz\\
       &= \sqrt{2k-1}\int_{\mathbb{R}} \left\{\sum_{j=0}^{k-1}\xi_{kj}F(z)^j\right\} (F(z)-\mathbb I(y\leq z)) dz\\
       &= \int_{\mathbb{R}}\phi_k(F(z)) (F(z)-\mathbb I(y\leq z)) dz
\end{align*}
where we apply the polynomial identity $\sum_{j=0}^{k-1}\xi_{kj}t^j=P_{k-1}^*(t)$ at $t=F(z)$.
\end{proof}

\subsection{Proof of \cref{lmoment_efficiency_bound}} \label{lmoment_effiency_bound_proof}

\begin{proof}[\textbf{Proof:}] By \cref{lmoment_eif}, stacking the coordinatewise canonical gradients gives the efficient influence function for the vector of the first $K$ normalized $L$-moments. Its covariance matrix is therefore the nonparametric efficiency bound,
\begin{align*}
    \Sigma_K(F) = \mathbb{E}\!\left[\boldsymbol{\ell}_K(Y;F)\boldsymbol{\ell}_K(Y;F)^{\top} \right].
\end{align*}
Next, since $\mathbb{E}[M_a(Y;F)]=(a+1)\beta_a(F)$, expanding the order-statistic representation in \cref{lmoment_eif} immediately gives
\begin{align*}
    \Sigma_{jk}(F) = \sqrt{(2j-1)(2k-1)} \sum_{a=0}^{j-1}\sum_{b=0}^{k-1} \xi_{ja}\xi_{kb} \operatorname{Cov}\!\left( M_a(Y;F),M_b(Y;F) \right).
\end{align*}
Note that all terms are finite: the bound $|M_a(y;F)|\leq |y|+a\mathbb{E}_F|Y|$ implies that $M_a(Y;F)\in L^2(F)$. Next, to obtain the integral representation, define $A_y(z)=F(z)-\mathbb{I}(y\leq z)$ and let $Y^\prime$ be an independent copy of $Y$. For every $y\in\mathbb{R}$,
\begin{align*}
     \int_{\mathbb{R}}|A_y(z)|dz = \int_{-\infty}^{y}F(z)dz + \int_y^\infty(1-F(z))dz = \mathbb{E}[|Y^\prime-y|] \leq |y|+\mathbb{E}[|Y|].
\end{align*}
Because $Y$ has a finite second moment and $\|\phi_r\|_\infty=\sqrt{2r-1}$, this bound ensures absolute integrability of the product of the two influence-function integrands. Fubini's theorem therefore permits interchanging expectation and integration. Next, define
\begin{align*}
     \mathbb{E}[A_Y(z)A_Y(t)] = F(z\wedge t)-F(z)F(t) = C_F(z,t).
\end{align*}
Then, applying the integral representation in \cref{lmoment_eif} yields
\begin{align*}
    \Sigma_{jk}(F) = \int_{\mathbb{R}}\int_{\mathbb{R}} \phi_j(F(z))\phi_k(F(t))C_F(z,t)dz\,dt.
\end{align*}
Finally, $\ell_1(Y;F)= Y- \mathbb{E}[Y]$ and $\phi_1 = 1$, so the preceding identity gives
\begin{align*}
    \Sigma_{11}(F) = \mathbb{V}(Y) = \int_{\mathbb{R}}\int_{\mathbb{R}}C_F(z,t)dz\,dt.
\end{align*}
Since $C_F(z,t)=F(z\wedge t)(1-F(z\vee t)) \geq 0$, the uniform bounds on the Legendre weights imply
\begin{align*}
    |\Sigma_{jk}(F)| \leq \|\phi_j\|_\infty\|\phi_k\|_\infty \int_{\mathbb{R}} \int_{\mathbb{R}}C_F(z,t)dz\,dt =  \sqrt{(2j-1)(2k-1)} \mathbb{V}(Y),
\end{align*}
which completes the proof.
    
\end{proof}

\subsection{Proof of \cref{gaussian_process_limit}} \label{gaussian_process_limit_proof}

\begin{proof}[\textbf{Proof:}] To begin, let $Y_i=Q(U_i)$ denote the quantile coupling, where $U_i \overset{iid}{\sim} \operatorname{Uniform}(0, 1)$ and where $V_n(u)=U_{(\lceil nu\rceil)}$ for $0<u<1$ is the associated empirical quantile function. Note that monotonicity of $Q$ gives $Y_{(i)}=Q(U_{(i)})$ and hence $Q_n=Q \circ V_n$ (including when $Q$ has flat portions). By Corollary~21.5 and Example~21.6 of \cite{van2000asymptotic}, we immediately have that $\sqrt{n}(V_n-\mathrm{id})\rightsquigarrow\mathbb{B}$ in $\ell^\infty(0,1)$ where $\mathbb{B}$ is a standard Brownian bridge and $\mathrm{id}$ is the identity map. From here, the functional delta method yields
\begin{align*}
    \sqrt{n}(Q_n-Q) \rightsquigarrow Q^\prime \mathbb{B}
\end{align*}
in $L^2(0, 1)$. Because multiplication by $Q^\prime$ is a continuous linear map from $C[0,1]$ into $L^2(0,1)$, $Q^\prime\mathbb{B}$ is a centered Gaussian random element. Its covariance kernel is $Q^\prime(u)Q^\prime(v)(u\wedge v-uv)$, and
\begin{align*}
    \mathbb{E}\left[ \|Q^\prime \mathbb{B}\|_2^2 \right] =\int_0^1u(1-u)Q^\prime(u)^2 du\leq L^2/6
\end{align*}
under the assumption that $|Q(v)-Q(u)|\le L|v-u|$ for $u,v\in[0,1]$. Next, to establish the growing-$K_n$ projection, let $T_n=\Pi_{K_n}$. For every sequence $x_n\to x$ in $L^2(0,1)$,
\begin{align*}
    \|T_nx_n-x\|_2 \leq \|x_n-x\|_2+\|\Pi_{K_n}x-x\|_2 \longrightarrow 0.
\end{align*}
The first term vanishes by assumption, and the second by completeness of the orthonormal basis as $K_n\to\infty$; for more details, see the discussion in \cref{regularity_conditions}. The same implication holds along every subsequence. Thus, by the extended continuous mapping theorem \citep{van2000asymptotic},
\begin{align*}
    \sqrt{n} \left( \widetilde{Q}_{K_n,n}-\Pi_{K_n}Q \right) \rightsquigarrow \mathbb{G}_F
\end{align*}
in $L^2(0, 1)$. This completes the Gaussian process limit proof, without any restriction on $K_n$. Finally, observe that
\begin{align*}
    \sqrt{n}(\widetilde{Q}_{K_n,n}-Q) =\sqrt{n}( \widetilde{Q}_{K_n,n}-\Pi_{K_n}Q) +\sqrt{n}(\Pi_{K_n}Q-Q).
\end{align*}
Under the assumption that $\sqrt{n}\|Q-\Pi_{K_n}Q\|_2 \longrightarrow 0$, the last term converges to zero
in $L^2(0, 1)$, and so by Slutsky's theorem, the same Gaussian process limit holds.    
\end{proof}

\subsection{Proof of \cref{u_statistic_functional_limit}} \label{u_statistic_functional_limit_proof}

\begin{proof}[\textbf{Proof:}] In order to establish a Gaussian process limit under the collective unbiased $U$-statistic estimator, we need to bound $\mathbb{E}[R^2_{k,n}]$ as a function of $K_n$. Thus, we first derive the exact second moment of $R_{k,n}$ using only the assumption that $\mathbb{E}[|Y|]<\infty$. Later, we invoke the global Lipschitz condition to establish bounds. To begin, let $(n)_r=n(n-1)\cdots(n-r+1)$ and recall that for each $k$ the unbiased estimator is defined as
\begin{align*}
    \widehat{\lambda}_{k,n}^* = \sqrt{2k-1}\sum_{j=0}^{k-1}\xi_{kj}\widehat{\beta}_{j, n}
\end{align*}
where $ \widehat{\beta}_{j, n}=\frac{U_{n,j+1}(h_{j+1})}{j+1}$ for $j=0,\ldots,K-1$. Now, for $1\leq p\leq n$, define 
\begin{align} \label{bpn_definition}
    B_{p,n}(z)=\frac{1}{\binom{n}{p}} \sum_{1\leq i_1<\cdots<i_p\leq n} \prod_{s=1}^p\mathbb{I}(Y_{i_s}\leq z).
\end{align}
This is the empirical average of the cumulative distribution function indicators for $p$-sample maxima. Importantly, the tail-integral formula for the expectation of an integrable random variable gives
\begin{align*}
    p\widehat\beta_{p-1,n} &= \int_{\mathbb{R}} \left( \mathbb{I}(z \geq 0)-B_{p,n}(z) \right) dz, \\
    p\beta_{p-1}(F) &= \int_{\mathbb{R}}\left( \mathbb{I}(z\geq 0)-F(z)^p \right)dz
\end{align*}
and consequently, after subtracting and rearranging,
\begin{align*}
    \widehat{\lambda}_{k,n}^* - \lambda_k^*(F) =-\sqrt{2k-1}\sum_{p=1}^k\frac{\xi_{k,p-1}}p \int_{\mathbb{R}} \left( B_{p,n}(z)-F(z)^p\right) dz.
\end{align*}
Next, let $D_F(y,z)= \mathbb{I}(y \leq z)-F(z)$, and define
\begin{align*}
     W_{n,r}(z)=\frac{1}{\binom{n}{r}} \sum_{1\leq i_1<\cdots<i_r\leq n} \prod_{s=1}^r D_F(Y_{i_s},z),
\end{align*}
where $W_{n,0}(z)=1$. Then, expanding each indicator in \cref{bpn_definition} as $F(z)+D_F(Y_i,z)$ gives
\begin{align*}
     B_{p,n}(z)=\sum_{r=0}^p\binom pr F(z)^{p-r}W_{n,r}(z).
\end{align*}
The coefficient follows because a specified $r$-subset belongs to $\binom{n-r}{p-r}$ of the $p$-subsets and $\binom nr\binom{n-r}{p-r}/\binom np=\binom pr$. Next, since $\phi_k(v)=\sqrt{2k-1}\sum_{p=1}^k\xi_{k,p-1}v^{p-1}$, it follows that
\begin{align*}
     \widehat{\lambda}_{k,n}^* - \lambda_k^*(F) =-\sum_{r=1}^k\frac{1}{r!} \int_{\mathbb{R}}\phi_k^{(r-1)}(F(z))W_{n,r}(z)dz.
\end{align*}
From here, observe that the first-order term is simply the sample average of the cumulative distribution function expression for $\ell_k$ in \cref{lmoment_eif}. Consequently, the remainder is exactly
\begin{align} \label{rkn_definition}
    R_{k,n} =\widehat{\lambda}_{k,n}^*-\lambda_k^*(F) -\frac{1}{n}\sum_{i=1}^n\ell_k(Y_i;F) =-\sum_{r=2}^k\frac{1}{r!} \int_{\mathbb{R}}\phi_k^{(r-1)}(F(z))W_{n,r}(z)dz,
\end{align}
with $R_{1,n}=0$. Finally, let $C(z,t)=F(z\wedge t)-F(z)F(t)$ and $v(z)=F(z)(1-F(z))$ and note
\begin{align*}
    \mathbb{E}[W_{n,r}(z)W_{n,s}(t)] =\begin{cases}
   \binom nr^{-1}C(z,t)^r,&r=s,\\
   0,& r\neq s
  \end{cases}.
\end{align*}
If $r \neq s$, an index occurring in only one product can be integrated out, and so the expectation evaluates to zero. When the sets agree, they have the same size $r$, and independence gives $C(z,t)^r$. There are $\binom{n}{r}$ matching pairs. Thus, after squaring \cref{rkn_definition} and taking the expectation, all cross terms drop out and we arrive at
\begin{align*}
    \mathbb{E}[R^2_{k,n}] =\sum_{r=2}^k\frac{1}{r!(n)_r} \int_{\mathbb{R}^2} \phi_k^{(r-1)}(F(z))\phi_k^{(r-1)}(F(t)) C(z,t)^r dz\,dt,
\end{align*}
where we have used the fact that $\binom{n}{r}=(n)_r/r!$. Notably, this expression is finite and exact for every $k\leq n$ as long as $\mathbb{E}[|Y|] <\infty$, with no regularity assumption on $F$ or $Q$. Furthermore, for each
fixed $k$, we can also verify $\mathbb{E}[R^2_{k,n}]=O(n^{-2})$, as discussed in \cref{u_statistic_section}. \medskip

With this expression for $\mathbb{E}[R^2_{k,n}]$ we now turn to establishing a bound in terms of $K_n$. We define $a=Q(0)$ and $b=Q(1)$, so that $Y \in [a,b]$ almost surely. Now, let $w(u)=u(1-u)$ and $c(u,v)=u \wedge v-uv$. Applying the monotone change of variables $z=Q(u)$ and $t=Q(v)$ gives
\begin{align*}
     \mathbb{E}[R^2_{k,n}] =\sum_{r=2}^k\frac{1}{r!(n)_r} \int_0^1\!\int_0^1 \phi_k^{(r-1)}(u)\phi_k^{(r-1)}(v) Q^\prime(u)Q^\prime(v)c(u,v)^r du\,dv.
\end{align*}
Next, we use the polynomial derivative identities established in the proof of \cref{higher_order_energy_bound}. In particular, \cref{legendre-derivative}, with polynomial degree $k-1$ and derivative order $r-1$, gives
\begin{align*}
    \phi_k^{(r-1)}(u) = \sqrt{2k-1} \frac{(k+r-2)!}{(k-1)!} P_{k-r}^{(r-1,r-1)}(2u-1)
\end{align*}
for $1\leq r\leq k$ where $P_d^{(\alpha,\beta)}$ denotes the Jacobi polynomial of degree $d$. For $2\le r\le k$, taking both polynomial degree indices equal to $k-1$ and derivative order $s=r-1$ in \cref{deriv_orthogonal} therefore yields
\begin{align*}
 \int_0^1 w(u)^{r-1} (\phi_k^{(r-1)}(u))^2 du = \frac{(k+r-2)!}{(k-r)!} \int_0^1\phi_k(u)^2 du = \frac{(k+r-2)!}{(k-r)!}.
\end{align*}
The same identity holds for $r=1$ by the normalization
$\|\phi_k\|_2=1$. From here, because $0 \leq c(u,v) \leq \sqrt{w(u)w(v)}$, applying the Cauchy--Schwarz inequality gives
\begin{align*}
 &\int_0^1\!\int_0^1 |\phi_k^{(r-1)}(u)\phi_k^{(r-1)}(v)|
   Q^\prime(u)Q^\prime(v)c(u,v)^r du\,dv \leq  \left\{\int_0^1|\phi_k^{(r-1)}(u)|Q^\prime(u)w(u)^{r/2} du\right\}^2\\
 &\leq \left\{\int_0^1w(u)^{r-1}(\phi_k^{(r-1)}(u))^2 du\right\} \left\{\int_0^1w(u)Q^\prime(u)^2 du\right\}\\
 &\leq \frac{L^2}{6}\frac{(k+r-2)!}{(k-r)!}
\end{align*}
where the last inequality follows under the global Lipschitzness assumption, which implies that $0\leq Q^\prime(u) \le L$ for all $u \in (0, 1)$. Consequently,
\begin{align*}
   \mathbb{E}[R^2_{k,n}] \leq \frac{L^2}{6}\sum_{r=2}^k \frac{(k+r-2)!}{(k-r)! r!(n)_r}. 
\end{align*}
Then, for $2 \leq r \leq k \leq n/2$, we have $(n)_r\ge(n/2)^r$ and $(k+r-2)!/(k-r)! \leq (2k)^{2r-2}$. Setting $x=8k^2/n$ and applying the inequality
\begin{align*}
   \sum_{r\geq 2} \frac{x^r}{r!} \leq \frac{x^2e^x}{2} 
\end{align*}
we then obtain
\begin{align*}
    \mathbb{E}[R^2_{k,n}] \leq \frac{L^2}{24k^2} \sum_{r=2}^{\infty}\frac{x^r}{r!} \leq \frac{L^2x^2e^x}{48k^2} \leq \frac{4L^2}{3}\frac{k^2}{n^2}\exp \! \left(\frac{8k^2}{n}\right).
\end{align*}
Finally, orthonormality and $R_{1,n}=0$ now yield, for $2\leq K \leq n/2$,
\begin{align*}
    \mathbb{E} \left[ \left\|\sqrt{n}\sum_{k=1}^K R_{k,n}\phi_k\right\|_2^2 \right] =n\sum_{k=1}^K\mathbb{E}[R^2_{k,n}] \leq \frac{4L^2}{3}\frac{K^3}{n}\exp\! \left(\frac{8K^2}{n} \right).
\end{align*}
Under the assumption that $K_n^3/n \to 0$, eventually $K_n\leq n/2$ and $K_n^2/n \to 0$ and consequently,
\begin{align*}
    n\sum_{k=1}^{K_n}\mathbb{E}[R^2_{k,n}] \longrightarrow 0.
\end{align*}
Now, to complete the proof, we need to establish the Gaussian process limit. Under the coupling $Y_i=Q(U_i)$, let $G_n$ and $V_n$ be the empirical distribution and quantile functions of the uniform sample. For convenience, we define $\alpha_n= \sqrt{n}(G_n-\mathrm{id})$, $\beta_n= \sqrt{n}(V_n-\mathrm{id})$, and $X_n=\sqrt{n}(Q_n-Q)$. Finally, define
\begin{align*}
     S_n(u) = \frac{1}{\sqrt{n}}\sum_{i=1}^n Q^\prime(u)(u-\mathbb{I}(U_i\leq u)) = -Q^\prime(u)\alpha_n(u).
\end{align*}
Note that after a change of variables $z = Q(u)$ we can write the influence function of \cref{lmoment_eif} as $\ell_k(Y_i;F) =\int_0^1\phi_k(u)Q^\prime(u)(u-\mathbb{I}(U_i\leq u))du$ and so
\begin{align*}
    \langle S_n,\phi_k\rangle =\frac{1}{\sqrt{n}}\sum_{i=1}^n\ell_k(Y_i;F).
\end{align*}
Now, observe that both projection errors can be written as
\begin{align*}
    Z_n &= \sqrt{n}(\widehat{Q}_{K_n,n}-\Pi_{K_n}Q) = \Pi_{K_n}S_n + \sqrt{n} \sum_{k=1}^{K_n}R_{k,n} \phi_k, \\
    W_n &=\sqrt{n}(\widetilde{Q}_{K_n,n}-\Pi_{K_n}Q) =\Pi_{K_n}X_n =\Pi_{K_n}S_n+\Pi_{K_n}\rho_n 
\end{align*}
where $\rho_n(u)=X_n(u)-S_n(u) =\sqrt{n}(Q_n(u)-Q(u))+Q^\prime(u)\alpha_n(u)$ and $\|\rho_n\|_2=o_p(1)$. We can now see that
\begin{align*}
     \|Z_n-W_n\|_2 \leq \left\|\sqrt n\sum_{k=1}^{K_n}R_{k,n}\phi_k\right\|_2 +\|\rho_n\|_2 \to 0
\end{align*}
as $n \to \infty$ and $K_n^3/n \to 0$. Let $a_n=n\sum_{k=1}^{K_n}\mathbb{E}[R_{k,n}^2]$. Then, for every $\varepsilon>0$, Markov's inequality gives
\begin{align*}
     \mathbb{P}(\|Z_n-W_n\|_2>\varepsilon) \leq \frac{4 a_n}{\varepsilon^2} + \mathbb{P}(\|\rho_n\|_2>\varepsilon/2) \longrightarrow 0.
\end{align*}
Consequently, $\sqrt{n}\|\widehat{Q}_{K_n,n}-\widetilde{Q}_{K_n,n}\|_2 \overset{P}{\longrightarrow} 0$. Because $W_n \rightsquigarrow \mathbb{G}_F$ by \cref{gaussian_process_limit}, Slutsky's theorem then gives $Z_n \rightsquigarrow \mathbb{G}_F$ in $L^2(0,1)$, thereby completing the proof.
\end{proof}

\subsection{Proof of \cref{simultaneous-band}} \label{simultaneous-band_proof}

\begin{proof}[\textbf{Proof:}] The proof of \cref{simultaneous-band} proceeds in several steps. To keep things organized, we enumerate them here before proving each component one at a time.
\begin{enumerate}
    \item \textbf{Variance Bound:} We establish a bound on the jackknife variance estimator in terms of $k$ in order to facilitate the growing $K_n$ setting.
    \item \textbf{Remainder Bound:} We establish a bound on $\widetilde{R}_{k, n}$, the remainder under the empirical estimator. Notably, in the proof of \cref{u_statistic_functional_limit} we bounded the $U$-statistic remainder $R_{k, n}$. However, the proof of \cref{gaussian_process_limit} did not require an analogous bound on $\widetilde{R}_{k, n}$, which is why we establish it here.
    \item \textbf{Convergence Rates:} We establish the rates and conditions required for simultaneous inference. These will later be used to leverage \cite{Chernozhukov2015} and \cite{Chernozhukov2017} for bounds and limit theorems for the maxima of Gaussian random vectors.
    \item \textbf{Coverage:} We combine results from the previous three proofs to establish valid limiting coverage for a growing collection of $L$-moments.
\end{enumerate}

Throughout the proof, we use $C$ to denote a finite constant that may change between displays and may depend on the fixed constants $L$ and $c_\sigma$, but not on $n$, $k$, or the observation index. We write $K=K_n$ and $\ell_{i,k}=\ell_k(Y_i;F)$ for brevity. All order-dependent bounds are understood for sufficiently large $n$. \medskip 

\textbf{Variance Bound:} To begin, we establish bounds on the jackknife variance estimator in terms of $k$ in order to establish conditions that hold for $K_n \to \infty$. Note that, as discussed in \cref{jackknife-bootstrap}, we focus specifically on the jackknife variance construction under the unbiased $U$-statistic estimator. However, analogous bounds can be derived under the empirical estimator as well. Let
\begin{align*}
    a_{k,r} = \frac{1}{r!}\int_{\mathbb{R}^2} \phi_k^{(r-1)}(F(z))\phi_k^{(r-1)}(F(t)) \left(F(z\wedge t)-F(z)F(t)\right)^r dz\,dt.
\end{align*}
Then, as shown in the proof of \cref{u_statistic_functional_limit}, $v_{k,m} = \mathbb{E}[R^2_{k,m}] = \sum_{r=2}^k\frac{a_{k,r}}{(m)_r}$ and furthermore,
\begin{align*}
    0 \leq a_{k,r}\leq\frac{L^2}{6r!}\frac{(k+r-2)!}{(k-r)!}.
\end{align*}
The identity holds for $m\geq k$, and the coefficients $a_{k,r}$ depend on $F$, $k$, and $r$, but not $m$. This notation allows us to compare the remainder variances at sample sizes $n$ and $n-1$ by changing only the factorial denominators. In particular, uniformly for $k\leq K_n$ we know that
\begin{align} \label{variance_eq}
    \mathbb{V}(\widehat{\lambda}_{k,n}^*) = \frac{\sigma_k^2}{n}+v_{k,n}
\end{align}
and $v_{k,n} \leq  \frac{Ck^2}{n^2}$ for some absolute constant $C$. We can use these results to immediately bound the jackknife variance. For each $1\leq k<n$, subset counting gives
\begin{align*}
    \frac{1}{n}\sum_{j=1}^n\widehat{\lambda}_{k,n}^{*,(-j)} =\widehat{\lambda}_{k,n}^*.
\end{align*}
By exchangeability of the observations, $\mathbb{E}[ \widehat{\lambda}_{k,n}^*\widehat{\lambda}_{k,n}^{*,(-i)}]$
does not depend on the deleted index $i$. Consequently,
\begin{align*}
 \mathbb{E}\!\left[ \widehat{\lambda}_{k,n}^* \widehat{\lambda}_{k,n}^{*,(-i)} \right] = \frac{1}{n}\sum_{j=1}^n \mathbb{E}\!\left[ \widehat{\lambda}_{k,n}^* \widehat{\lambda}_{k,n}^{*,(-j)} \right] = \mathbb{E}\!\left[ \widehat{\lambda}_{k,n}^* \left\{\frac{1}{n}\sum_{j=1}^n \widehat{\lambda}_{k,n}^{*,(-j)}\right\} \right] = \mathbb{E}\!\left[ (\widehat{\lambda}_{k,n}^*)^2 \right].
\end{align*}
Expanding the square in the definition of the
jackknife contribution gives
\begin{align*}
 \mathbb{E}[\widehat{\ell}^2_{i,k}] &=
 (n-1)^2  \mathbb{E}\!\left[ \left(\widehat{\lambda}_{k,n}^* -\widehat{\lambda}_{k,n}^{*,(-i)}\right)^2 \right] = (n-1)^2\left\{ \mathbb{V}(\widehat{\lambda}_{k,n}^{*,(-i)}) - \mathbb{V}(\widehat{\lambda}_{k,n}^*) \right\}.
\end{align*}
The last equality uses the fact that the full and deleted
estimators are both unbiased for $\lambda_k^*(F)$. Next, as discussed in \cref{u_statistic_section},
\begin{align*}
   \mathbb{E}[\widehat{\lambda}_{k,n}^*\mid Y_i] = \lambda_k^*(F)+\frac{1}{n}\ell_k(Y_i;F). 
\end{align*}
Because the deleted estimator is independent of $Y_i$, $\mathbb{E}[\widehat{\lambda}_{k,n}^{*,(-i)}\mid Y_i] =\lambda_k^*(F)$. Thus, it follows that
\begin{align*}
 \mathbb{E}\!\left[ \widehat{\ell}_{i,k}\ell_k(Y_i;F)\right] = \mathbb{E}\!\left[ \ell_k(Y_i;F)\, \mathbb{E}[\widehat{\ell}_{i,k}\mid Y_i] \right] = \frac{n-1}{n} \mathbb{E}[\ell_k(Y_i;F)^2] = \frac{n-1}{n}\sigma_k^2.
\end{align*}
Combining this result with \cref{variance_eq} gives us
\begin{align*}
    \mathbb{E}\left[(\widehat{\ell}_{i,k}-\ell_{i,k})^2\right] =\frac{\sigma_k^2}{n}+(n-1)^2(v_{k,n-1}-v_{k,n}) =\frac{\sigma_k^2}{n} +(n-1)^2\sum_{r=2}^k\frac{r\,a_{k,r}}{(n-r)(n)_r}.
\end{align*}
Thus, applying the same inequalities to this summation as in the proof of \cref{u_statistic_functional_limit} it follows that for $k \leq K_n$,
\begin{align} \label{jackknife_var_bound}
     \mathbb{E}\! \left[(\widehat{\ell}_{i,k}-\ell_{i,k})^2\right] \leq \frac{\sigma_k^2+Ck^2}{n}
\end{align}
for some universal constant $C$ (not necessarily the same constant as before). \medskip

\textbf{Remainder Bound:} Let $\Delta_n=\|F_n-F\|_\infty$ and write $[a,b]=[Q(0),Q(1)]$. The integrated-CDF representation used in
\cref{u_statistic_functional_limit_proof} gives
\begin{align*}
    \widetilde{\lambda}_{k,n}^*-\lambda_k^*(F) = -\int_a^b\int_{F(z)}^{F_n(z)}\phi_k(t)dt\,dz.
\end{align*}
On the other hand, the CDF representation of the influence
function gives
\begin{align*}
    \frac{1}{n}\sum_{i=1}^n\ell_k(Y_i;F) = -\int_a^b\phi_k\{F(z)\}(F_n(z)-F(z))dz.
\end{align*}
Subtracting these identities yields the exact remainder
\begin{align}  \label{eq:ss-empirical-remainder-identity}
    \widetilde{R}_{k,n} = \widetilde{\lambda}_{k,n}^*-\lambda_k^*(F) -\frac{1}{n}\sum_{i=1}^n\ell_k(Y_i;F) = -\int_a^b\int_{F(z)}^{F_n(z)} \bigl[\phi_k(t)-\phi_k(F(z))\bigr]dt\,dz.
\end{align}
Now, for every $t$ between $F(z)$ and $F_n(z)$, the fundamental theorem of calculus along with the inequality $|F_n(z)-F(z)|\leq \Delta_n$ imply that
\begin{align*}
    \bigl|\phi_k(t)-\phi_k(F(z))\bigr| \leq \int_0^1|\phi_k^\prime(s)| \mathbb{I}(|s-F(z)|\leq \Delta_n) ds. 
\end{align*}
Note that the interval between $F(z)$ and $F_n(z)$ has length at most $\Delta_n$. Applying this bound directly in
\cref{eq:ss-empirical-remainder-identity} therefore gives
\begin{align*}
   |\widetilde{R}_{k,n}| &\leq \Delta_n\int_a^b\int_0^1|\phi^\prime_k(s)|
       \mathbb{I}(|s-F(z)| \leq\Delta_n)ds\,dz.
\end{align*}
We now apply the change of variables $z=Q(u)$. At almost every rank where $Q^\prime(u)>0$, $F(Q(u))=u$. Ranks where $Q^\prime(u)=0$ contribute nothing. Then, using $0\leq Q^\prime \leq L$ we obtain
\begin{align*}
 |\widetilde{R}_{k,n}| &\leq L\Delta_n\int_0^1\int_0^1 |\phi^\prime_k(s)|\mathbb{I}( |s-u|\leq \Delta_n) ds\,du\\
 &= L\Delta_n\int_0^1|\phi^\prime_k(s)| \left\{\int_0^1
          \mathbb{I}(|s-u|\leq \Delta_n)du\right\}ds\\
 &\leq 2L\Delta_n^2\int_0^1|\phi^\prime_k(s)|ds.
\end{align*}
The last inequality follows because, for each fixed $s$, the ranks $u$ satisfying $|s-u|\leq \Delta_n$ form an interval of length at most $2\Delta_n$. Finally, it remains to bound $\phi^\prime_k$. For $k\ge2$, $\phi^\prime_k$ has degree $k-2$, so $[0,1]$ can be partitioned into at most $k-1$ intervals on which $\phi_k$ is monotone. On each interval its variation is at most $2\|\phi_k\|_\infty=2\sqrt{2k-1}$. Hence
\begin{align*}
    \int_0^1|\phi^\prime_k(s)|ds \leq 2(k-1)\sqrt{2k-1}.
\end{align*}
Putting everything together, we have for $k \geq 2$
\begin{align} \label{r_tilde_bound}
     |\widetilde{R}_{k,n}| \leq 4L(k-1)\sqrt{2k-1}\Delta_n^2 \leq Ck^{3/2}\Delta_n^2,
\end{align}
where $C$ depends only on $L$, not on $n$ or $k$.
For $k=1$, $\phi_1$ is constant, so \cref{eq:ss-empirical-remainder-identity} gives $\widetilde R_{1,n}=0$ exactly. \medskip 

\textbf{Convergence Rates:} Now, we establish rates of convergence. Let $\Lambda_n=\log(enK)$ and $\tau_n=\sqrt{K^5_n/n}$ denote our rates. In particular, note that $O_p(\tau_n) = o_p(\Lambda^{-2}_n)$. We explain the significance of these rates further in the coverage section of the proof, when we apply results from \cite{Chernozhukov2015} and \cite{Chernozhukov2017}. Next, define $X_{i,k}=\ell_{i,k}/\sigma_k$, $\Gamma_{jk}= \mathbb{E}[(X_{i,j}X_{i,k})]$, and
\begin{align*}
    D_n=\max_{k\leq K_n} \left\{\frac{1}{n}\sum^n_{i=1} \frac{(\widehat{\ell}_{i,k}-\ell_{i,k})^2}{\sigma_k^2} \right\}^{1/2}.
\end{align*}
Note that $X_{i,k}$ is centered with variance one and $|X_{i,k}| \leq C K_n$. This follows, since under the rank representation of the influence function in \cref{lmoment_eif}, letting
$Y_i=Q(U_i)$,
\begin{align*}
    |\ell_k(Y_i;F)| \leq \int_0^1|\phi_k(u)|Q^\prime(u) |u-\mathbb{I}(U_i\leq u)|du \leq L\int_0^1|\phi_k(u)|du \leq L.
\end{align*}
The final inequality follows from Cauchy--Schwarz and
$\|\phi_k\|_2=1$. Hence
\begin{align*}
     |X_{i,k}| \leq \frac{L}{\sigma_k} \leq \frac{L}{c_\sigma}k \leq CK_n.
\end{align*}
Then, since the maximum is bounded by the sum, applying our variance bound \cref{jackknife_var_bound} (along with the assumption that $\sigma_k \geq c_\sigma/k$), it follows that
\begin{align*}
    \mathbb{E}[D^2_n] &\leq \sum_{k\leq K_n} \left( \frac{1}{n}\sum^n_{i=1} \mathbb{E}\! \left[\frac{(\widehat{\ell}_{i,k}-\ell_{i,k})^2}{\sigma_k^2} \right] \right) \leq \sum_{k\leq K_n} \left( \frac{1}{n}\sum^n_{i=1} \frac{1}{\sigma^2_k}\left(\frac{\sigma_k^2+Ck^2}{n} \right) \right) \\
    &\leq \sum_{k\leq K_n} \left( \frac{1}{n}\sum^n_{i=1} \left\{ \frac{1}{n} + \frac{C k^4}{c^2_\sigma n} \right\} \right) \leq \frac{1}{n} \sum_{k \leq K_n} \left( 1 + \frac{C k^4}{c^2_\sigma}\right) \leq \frac{K^5_n}{n}\left(1 + \frac{C}{c^2_\sigma} \right).
\end{align*}
Consequently, $D_n=O_p(\tau_n)$. Furthermore, following similar steps it follows that both
\begin{align*}
     \max_{k\leq K_n} \left\{ \frac{\sqrt{n}|R_{k,n}|}{\sigma_k} \right\} =O_p(\tau_n) \qquad \text{and} \qquad \max_{k\leq K_n} \left\{ \frac{\sqrt{n}|\widetilde{R}_{k,n}|}{\sigma_k} \right\}
       =O_p(\tau_n).
\end{align*}
For example, applying the remainder bound in \cref{r_tilde_bound}, it follows that
\begin{align*}
    \max_{k\leq K_n } \left\{ \frac{\sqrt{n}|\widetilde{R}_{k,n}|}{\sigma_k} \right\} &\leq \max_{k \leq K_n} \left\{ \frac{\sqrt{n}}{\sigma_k}\left(Ck^{3/2}\Delta_n^2 \right) \right\} = C \Delta^2_n \sqrt{n} \max_{k \leq K_n} \left\{ \frac{k^{3/2}}{\sigma_k} \right\} \\
    &\leq \frac{C \Delta^2_n \sqrt{n}}{c_\sigma} \max_{k \leq K_n} \left\{ k^{5/2} \right\} = \sqrt{\frac{K^5_n}{n}} \left( \frac{C}{c_\sigma} \left(\sqrt{n}\Delta_n \right)^2 \right),
\end{align*}
and so the $O_p(\tau_n)$ bound follows since  $\Delta_n = \|F_n-F\|_\infty =O_p(n^{-1/2})$ by the Dvoretzky--Kiefer--Wolfowitz inequality. Next, we establish the accuracy of the jackknife covariance matrix. Let
\begin{align*}
    E_n= \max_{j,k\le K_n} \left|\frac{1}{n}\sum_{i=1}^nX_{i,j}X_{i,k}-\Gamma_{jk}\right|
\end{align*}
denote the largest entrywise estimation error of the empirical second-moment matrix of the true standardized scores. This measures sampling error before accounting for estimation of the scores themselves. Independence across observations, $|X_{i,j}|\leq CK_n$, and $\mathbb{E}[X_{i,k}^2] = 1$ imply that
\begin{align*}
     \mathbb{E}[E^2_n] \leq \sum_{j,k\le K_n}
   \mathbb{E}\! \left[ \left\{\frac{1}{n}\sum_{i=1}^nX_{i,j}X_{i,k}-\Gamma_{jk}\right\}^2 \right] \leq \frac{1}{n}\sum_{j,k\le K_n} \mathbb E[X_{1,j}^2X_{1,k}^2] \leq \frac{CK_n^4}{n}.
\end{align*}
Hence, applying Markov's inequality yields $E_n=O_p(K^2/\sqrt{n})=O_p(\tau_n)$. Next, to account for estimation define
\begin{align*}
    e_{i,k} = \frac{\widehat{\ell}_{i,k}-\ell_{i,k}}{\sigma_k}, \qquad \frac{\widehat{\ell}_{i,k}}{\sigma_k}=X_{i,k}+e_{i,k},
\end{align*}
and recall that $\max_{k \leq K_n}\frac{1}{n}\sum_{i=1}^n e_{i,k}^2=D_n^2.$ Then, we can establish the decomposition
\begin{align*}
    \frac{\widehat{\Sigma}_{jk,n}-\Sigma_{jk}(F)}{\sigma_j\sigma_k} =\left\{\frac{1}{n}\sum_{i=1}^nX_{i,j}X_{i,k}-\Gamma_{jk}\right\} + \frac{1}{n}\sum_{i=1}^nX_{i,j}e_{i,k} +\frac{1}{n}\sum_{i=1}^ne_{i,j}X_{i,k} +\frac{1}{n}\sum_{i=1}^ne_{i,j}e_{i,k}.
\end{align*}
Bounding each term in this expansion one at a time, it follows that
\begin{align*}
    C_n = \max_{j,k\leq K_n} \left\{ \frac{|\widehat{\Sigma}_{jk,n}-\Sigma_{jk}(F)|}{\sigma_j\sigma_k} \right\} \leq E_n+2D_n\sqrt{1+E_n}+D_n^2 =O_p(\tau_n).
\end{align*}
We can use this same covariance bound $C_n$ to establish the relative accuracy of the estimated standard deviations as well. Since $\widehat{\Sigma}_{kk,n}=\widehat{\sigma}_{k,n}^2$ and $\Sigma_{kk}(F)=\sigma_k^2$,
\begin{align*}
     \max_{k\leq K_n} \left|\frac{\widehat{\sigma}_{k,n}^2}{\sigma_k^2}-1\right| \leq C_n.
\end{align*}
Let $\mathcal{A}_n=\{C_n\leq 1/2\}$. Then
$\mathbb{P}(\mathcal{A}_n)\to 1$, and on this event for $k \leq K_n$,
\begin{align*}
   (1-C_n)\sigma_k^2 \leq \widehat{\sigma}_{k,n}^2 \leq (1+C_n)\sigma_k^2.
\end{align*}
In particular, every estimated variance is positive on
$\mathcal{A}_n$. Therefore,
\begin{align*}
    \left|\frac{\widehat{\sigma}_{k,n}}{\sigma_k}-1\right| =\frac{ |\widehat{\sigma}_{k,n}^2/ \sigma_k^2-1| }{\widehat{\sigma}_{k,n}/\sigma_k+1} \leq\frac{C_n}{\sqrt{1-C_n}+1} \leq C_n,
\end{align*}
and furthermore, $\max_{k\leq K_n}|\sigma_k/\widehat{\sigma}_{k,n}-1| \leq \sqrt{2} C_n$ on $\mathcal{A}_n$. Finally, let $\widehat{\Gamma}_{jk} =\frac{\widehat{\Sigma}_{jk,n}} {\widehat{\sigma}_{j,n}\widehat{\sigma}_{k,n}}$. Then, it follows that
\begin{align*}
    \widehat{\Gamma}_{jk}-\Gamma_{jk} =\frac{\sigma_j\sigma_k}{\widehat{\sigma}_{j,n}\widehat{\sigma}_{k,n}}\frac{\widehat{\Sigma}_{jk,n}-\Sigma_{jk}(F)}{\sigma_j\sigma_k} + \Gamma_{jk} \left\{\frac{\sigma_j\sigma_k}{\widehat{\sigma}_{j,n}\widehat{\sigma}_{k,n}}-1\right\}.
\end{align*}
Since $|\Gamma_{jk}|\leq 1$, on $\mathcal{A}_n$ it follows that
\begin{align*}
    \delta_n = \max_{j,k\le K_n}|\widehat{\Gamma}_{jk}-\Gamma_{jk}| \leq \frac{2C_n}{1-C_n}\leq 4C_n.
\end{align*}
Then, because $\mathbb{P}(\mathcal{A}_n)\to1$, we conclude that
\begin{align*}
    \max_{k\leq K_n} \left|\frac{\widehat{\sigma}_{k,n}}{\sigma_k}-1\right| =O_p(\tau_n), \qquad \delta_n=O_p(\tau_n).
\end{align*}

 \medskip

\textbf{Coverage:} To complete the proof, we now establish $1 - \alpha$ coverage across a growing collection of $L$-moments. Let $G_K \sim \mathcal{N}_K(0,\Gamma)$ and define $H_n(t)=\mathbb{P}(\|G_K\|_\infty \leq t)$ and
\begin{align*}
     M_n= \max_{k\leq K_n }\left|n^{-1/2}\sum^n_{i=1} X_{i,k}\right|.
\end{align*}
The reference CDF $H_n$ is deterministic under the population distribution, although it may change with $n$ as the dimension increases. It serves as an intermediate comparison distribution and doesn't need to converge to a fixed CDF. By Proposition~2.1 of \cite{Chernozhukov2017} case~(E.1), it follows that for independent centered vectors with unit coordinate variances and coordinates bounded by $B$, the error of a Gaussian approximation over hyperrectangles is at most $C(B^2\log^7(enK_n)/n)^{1/6}$. Applying that proposition to $[-t,t]^{K_n}$ therefore gives
\begin{align*}
    \eta_n = \sup_{t \in \mathbb{R}}|\mathbb{P}(M_n \leq t)-H_n(t)| \leq C(K^2\Lambda_n^7/n)^{1/6} \longrightarrow 0
\end{align*}
since $K_n \le n^{1/5}$ and $\Lambda_n=O(\log{n})$ imply $K^2_n \Lambda_n^7/n\le n^{-3/5}\Lambda_n^7\to 0$. Note that since every coordinate of $G_K$ is a standard normal, the Gaussian tail bound and the union bound imply
\begin{align*}
    \mathbb{P}(\|G_K\|_\infty>x)\leq 2K_n e^{-x^2/2}
\end{align*}
for $x > 0$. Therefore, by the definition of $\eta_n$, $\mathbb{P}(M_n > x) \leq 2K_n e^{-x^2/2}+\eta_n$, so taking $x \propto \sqrt{\Lambda_n}$ with a sufficiently large constant gives $M_n=O_p(\sqrt{\Lambda_n})$. Now, for either estimator, the first-order expansion gives
\begin{align*}
 \frac{\sqrt{n}\{\lambda_{k,n}^{\circ}-\lambda_k^*(F)\}}
      {\widehat{\sigma}_{k,n}} =\frac{\sigma_k}{\widehat{\sigma}_{k,n}} \left\{\frac1{\sqrt n}\sum_{i=1}^nX_{i,k} +\frac{\sqrt{n} R_{k,n}^{\circ}}{\sigma_k}\right\}.
\end{align*}
Applying the elementary inequality $|\max_k|a_k|-\max_k|b_k||\leq \max_k|a_k-b_k|$, we obtain
\begin{align*}
     |T_n^{\circ}-M_n| \leq \max_{k\le K_n} \left\{\left| \frac{\sigma_k}{\widehat{\sigma}_{k,n}}-1\right|M_n \right\}
     +\max_{k\le K_n} \left\{ \frac{\sigma_k}{\widehat\sigma_{k,n}} \right\}
       \max_{k\le K_n} \left\{ \frac{\sqrt{n}|R_{k,n}^{\circ}|}{\sigma_k} \right\} =O_p(\tau_n\sqrt{\Lambda_n}),
\end{align*}
where $R_{k,n}^{\circ}$ is the remainder of the selected estimator. However, a small difference between statistics does not imply a uniformly small difference between their CDFs. To translate this approximation in probability into an approximation of CDFs, we apply Theorem~3(i) of \citet{Chernozhukov2015} to $(G_K,-G_K)$, which yields
\begin{align*}
   \omega_n(\varepsilon) = \sup_{t \in \mathbb{R}}\mathbb{P}\big( |\|G_K\|_\infty-t|\leq \varepsilon \big) \le 4\varepsilon( \sqrt{2\log(2K)}+1). 
\end{align*}
Now, take $\varepsilon_n=\sqrt{\tau_n}$ and set $p_n= \mathbb{P}(|T_n^{\circ}-M_n|>\varepsilon_n)$. Since $\tau_n\sqrt{\Lambda_n}/\varepsilon_n =\sqrt{\tau_n\Lambda_n}\to0$ and $\varepsilon_n\sqrt{\log(2K_n)}\to0$, note that both $p_n$ and $\omega_n(\varepsilon_n)$ vanish. Then, on the event $\{|T_n^{\circ}-M_n|\leq \varepsilon_n\}$ it follows that
\begin{align} \label{tn_inequality}
     \mathbb{P}(M_n \leq t-\varepsilon_n)-p_n \leq \mathbb{P}(T_n^{\circ}\leq t) \leq \mathbb{P}(M_n\leq t+\varepsilon_n)+p_n.
\end{align}
By the definition of $\eta_n$, the CDF of $M_n$ is within $\eta_n$ of $H_n$ at every threshold, including $t-\varepsilon_n$ and $t+\varepsilon_n$. Hence
\begin{align*}
    H_n(t-\varepsilon_n)- \eta_n -p_n \leq\mathbb{P}(T_n^\circ\leq t) \leq H_n(t+\varepsilon_n)+ \eta_n +p_n.
\end{align*}
Combining this inequality with the fact that
\begin{align*}
     \begin{split}
  0\leq H_n(t+\varepsilon_n)-H_n(t)
      &\leq \omega_n(\varepsilon_n),\\
  0\leq H_n(t)-H_n(t-\varepsilon_n)
      &\leq \omega_n(\varepsilon_n)
 \end{split}
\end{align*}
along with \cref{tn_inequality} gives
\begin{align*}
     -\eta_n -p_n-\omega_n(\varepsilon_n) \leq \mathbb{P}(T_n^\circ\leq t)-H_n(t) \leq \eta_n +p_n+\omega_n(\varepsilon_n).
\end{align*}
Since these inequalities are uniform in $t$, we immediately obtain
\begin{align} \label{uniform_result}
    \sup_{t\in\mathbb{R}} |\mathbb{P}(T_n^\circ\leq t)-H_n(t)| \leq \eta_n + p_n+ \omega_n(\varepsilon_n) \longrightarrow 0.
\end{align}
Next, conditional on the data, the vector with coordinates $Z_{k,n}^{(\mathrm{b})}/\widehat\sigma_{k,n}$ is exactly a centered Gaussian with covariance
\begin{align*}
    \operatorname{Cov}\!\left( \frac{Z_{j,n}^{(\mathrm{b})}}{\widehat{\sigma}_{j,n}}, \frac{Z_{k,n}^{(\mathrm{b})}}{\widehat{\sigma}_{k,n}} \mid \mathcal{D}_n\right) =\frac{\widehat{\Sigma}_{jk,n}} {\widehat{\sigma}_{j,n}\widehat{\sigma}_{k,n}} =\widehat{\Gamma}_{jk}. 
\end{align*}
Let $\widehat{H}_n(t)=\mathbb{P}(T_n^{(\mathrm{b})}\le t \mid \mathcal{D}_n)$. Theorem~2 of \citet{Chernozhukov2015}, applied to the doubled Gaussian vectors, gives
\begin{align} \label{delta_bound}
  \sup_{t \in \mathbb{R}}|\widehat{H}_n(t)-H_n(t)| \le C\delta_n^{1/3} \{1\vee\log(2K_n/\delta_n)\}^{2/3}=o_p(1).
\end{align}
This bound is interpreted as zero when $\delta_n=0$. Note that \cref{delta_bound} follows since $\delta_n=O_p(\tau_n)$ and $\tau_n\Lambda_n^2 \to 0$, so we have $\delta_n\Lambda_n^2=o_p(1)$. On $0<\delta_n\leq1$,
\begin{align*}
     \delta_n (1\vee\log(2K_n/\delta_n))^2 \leq C\left( \delta_n\Lambda_n^2 +\delta_n\log^2(1/\delta_n) \right) \overset{P}{\longrightarrow} 0.
\end{align*}
The second term vanishes because $x\log^2(1/x)\to 0$ as $x\downarrow0$. Taking cube roots establishes convergence. Then, applying the triangle inequality and our previous results yields
\begin{align*}
    \sup_{t \in \mathbb{R}}|\mathbb{P}(T_n^{\circ}\leq t) -\mathbb{P}(T_n^{(\mathrm{b})}\le t \mid \mathcal{D}_n)| \leq \sup_{t \in \mathbb{R}}|\widehat{H}_n(t)-H_n(t)| + \sup_{t\in\mathbb{R}} |\mathbb{P}(T_n^\circ\leq t)-H_n(t)| \overset{P}{\longrightarrow} 0.
\end{align*}
Finally, to obtain valid $1 - \alpha$ coverage, we need to account for the data dependence of the critical value. To do so, we bracket our critical value between nonrandom quantiles of $H_n$. Notably, the CDF $H_n$ is continuous because each absolute Gaussian coordinate has no atoms, and so $H_n\{H_n^{-1}(p)\}=p$ for $0<p<1$. Let $\gamma_n\downarrow0$ be a deterministic sequence such that
\begin{align*}
    \mathcal{E}_n =\mathcal{A}_n\cap\{\sup_{t \in \mathbb{R}}|\widehat{H}_n(t)-H_n(t)|\leq \gamma_n\}
\end{align*}
satisfies $\mathbb{P}(\mathcal{E}_n)\to 1$, and put $q_n^{\pm}=H_n^{-1}(1-\alpha\pm2\gamma_n)$. On $\mathcal{E}_n$, for all sufficiently large $n$, it follows that $\widehat{H}_n(q_n^-)\leq 1-\alpha-\gamma_n$ and $\widehat{H}_n(q_n^+)\geq 1-\alpha+\gamma_n$. Thus the conditional $(1-\alpha)$ quantile on $\mathcal{E}_n$ obeys
\begin{align*}
     q_n^-\le c_{n,1-\alpha}^*\le q_n^+.
\end{align*}
Applying \cref{uniform_result} at these two deterministic
thresholds gives
\begin{align*}
 1-\alpha-2\gamma_n-d_n^{\circ}-\mathbb{P}(\mathcal{E}_n^c) \leq \mathbb{P}(T_n^{\circ}\leq c_{n,1-\alpha}^*) \leq 1-\alpha+2\gamma_n+d_n^{\circ}+\mathbb{P}(\mathcal E_n^c)
\end{align*}
where we say $d_n^{\circ} = \sup_{t\in\mathbb{R}} |\mathbb{P}(T_n^\circ\leq t)-H_n(t)|$. Because all error terms vanish, it follows that $\mathbb{P}(T_n^{\circ}\leq c_{n,1-\alpha}^*) \longrightarrow 1-\alpha.$ On $\mathcal{A}_n$, this event is precisely simultaneous coverage by
\begin{align*}
    I_{k,n}^{\circ} =\left[\lambda_{k,n}^{\circ} \, \pm \, c_{n,1-\alpha}^* \frac{\widehat{\sigma}_{k,n}}{\sqrt{n}}\right],
\end{align*}
for $k \leq K_n$, and importantly, no fixed limiting CDF for the growing maximum is assumed. Since $\mathbb{P}(\mathcal{A}_n)\to1$, this proves asymptotically exact simultaneous coverage of the retained coefficients $k \leq K_n$. To obtain the projected-curve band, note that on the same event, for every $u\in[0,1]$,
\begin{align*}
 |Q_{K_n,n}^{\circ}(u)-(\Pi_{K_n}Q)(u)| \leq \sum_{k=1}^{K_n} |\lambda_{k,n}^{\circ}-\lambda_k^*(F)|\,|\phi_k(u)| \leq \frac{c_{n,1-\alpha}^*}{\sqrt{n}} \sum_{k=1}^{K_n}\widehat\sigma_{k,n}|\phi_k(u)|.
\end{align*}
Thus simultaneous coefficient coverage implies coverage at every rank by the band
\begin{align*}
    B_n^{\circ}(u)= \left[Q_{K_n,n}^{\circ}(u) \, \pm \, \frac{c_{n,1-\alpha}^*}{\sqrt{n}} \sum_{k=1}^{K_n}\widehat\sigma_{k,n}|\phi_k(u)|\right].
\end{align*}
Consequently, $\liminf_{n\to\infty} \mathbb{P}((\Pi_{K_n}Q)(u)\in B_n^{\circ}(u) \text{ for every }u\in[0,1])\geq 1-\alpha$. Finally, suppose the conditional critical value is computed from $B_n\to\infty$ independent multiplier draws. Let $\widehat{H}_{n,B_n}$ denote their empirical CDF. Conditional on the data, the Dvoretzky--Kiefer--Wolfowitz inequality gives, for every $\varepsilon>0$,
\begin{align*}
    \mathbb{P}\!\left(
    \sup_{t \in \mathbb{R}}|\widehat H_{n,B_n}(t)-\widehat H_n(t)|>\varepsilon
 \mid \mathcal{D}_n \right)\le2\exp(-2B_n\varepsilon^2).
\end{align*}
Hence the simulated CDF is uniformly $o_p(1)$ from $\widehat{H}_n$ and therefore from $H_n$. Replacing $\widehat{H}_n$ by $\widehat H_{n,B_n}$ in the same deterministic quantile brackets proves the coverage conclusions for the simulated critical value.
\end{proof}

\subsection{Proof of \cref{additive_cic_bound}} \label{additive_cic_bound_proof}

\begin{proof}[\textbf{Proof:}] To begin, suppose that $F_{00}$ is continuous. Then, $F_{00}(Q_{00}(u))=u$ for every $u\in(0,1)$ and hence
\begin{align*}
      S\!\left(Q_{00}(u)\right) =Q_{01}\!\left(F_{00}\!\left(Q_{00}(u)\right)\right) =Q_{01}(u),
\end{align*}
when $S$ is defined as in \cref{control_group_transport_map}. Next, let $\delta(x) = S(x) - x$. It then follows for almost every $u \in (0, 1)$ that
\begin{align*}
    Q_{\mathrm{Add}}(u) &=Q_{10}(u)+Q_{01}(u)-Q_{00}(u)\\ 
    &=Q_{10}(u) +S\!\left(Q_{00}(u)\right)-Q_{00}(u)\\
    &=Q_{10}(u)+\delta\!\left(Q_{00}(u)\right),
\end{align*}
and similarly, $Q_{\mathrm{CiC}}(u) =S\!\left(Q_{10}(u)\right) =Q_{10}(u)+\delta\!\left(Q_{10}(u)\right)$. Thus, the difference in extrapolations is given by
\begin{align*}
    \Gamma(u) =\delta\!\left(Q_{10}(u)\right) -\delta\!\left(Q_{00}(u)\right).
\end{align*}
Next, under the assumption that $S$ is absolutely continuous on every compact subinterval of $\mathcal{I}$, so too is $\delta$, such that $\delta^\prime(x)=S^\prime(x)-1$ for Lebesgue-almost every $x\in\mathcal{I}$. Furthermore, note that for almost every $u$, both $Q_{00}(u)$ and $Q_{10}(u)$ lie in $\mathcal{I}$. Thus, applying the fundamental theorem of calculus yields
\begin{align*}
    \delta\!\left(Q_{10}(u)\right) -\delta\!\left(Q_{00}(u)\right) =\int_{Q_{00}(u)}^{Q_{10}(u)}\delta^\prime(x)dx =\int_{Q_{00}(u)}^{Q_{10}(u)}\big( S^\prime(x)-1 \big) dx.
\end{align*}
From here, we can immediately obtain the pointwise bound
\begin{align*}
    |\Gamma(u)| &\leq \int_{\min\{Q_{00}(u),Q_{10}(u)\}}^{\max\{Q_{00}(u),Q_{10}(u)\}} |S^\prime(x)-1|dx \leq \operatorname*{ess\,sup}_{x\in\mathcal I}|S'(x)-1| \cdot |Q_{10}(u)-Q_{00}(u)|.
\end{align*}
Consequently, if we square and integrate both sides over $u$ we have 
\begin{align} \label{penultimate_eq}
       \|\Gamma\|_{L^2}^2 \leq \left(\operatorname*{ess\,sup}_{x\in\mathcal I}|S'(x)-1|\right)^2 \int_0^1|Q_{10}(u)-Q_{00}(u)|^2 du,
\end{align}
which completes the proof, after noting a few additional details. First, under the assumption that $F_{00},F_{10}\in\mathcal{P}_2(\mathbb{R})$, and because our outcomes are scalar valued, it follows by \cite{vallender1974} that
\begin{align*}
    \int_0^1|Q_{10}(u)-Q_{00}(u)|^2 du = W^2_2(F_{10}, F_{00}).
\end{align*}
Second, because $\{\phi_k\}_{k\geq1}$ is a complete orthonormal basis of
$L^2(0,1)$ and $\gamma_k=\langle\Gamma,\phi_k\rangle_{L^2}$, by Parseval's
identity \citep{conway1985course} it follows that
\begin{align*}
    \|\Gamma\|_{L^2(0,1)}^2  =\sum_{k=1}^{\infty}\gamma_k^2.
\end{align*}
Therefore, taking square roots to \cref{penultimate_eq} yields the final bound,
\begin{align*}
     \| \Gamma \|_{L^2} = \left( \sum^{\infty}_{k=1} \gamma^2_k \right)^{1/2} \leq \operatorname*{ess\,sup}_{x\in\mathcal I}|S'(x)-1| W_2(F_{10}, F_{00}).
\end{align*}
    
\end{proof}

\subsection{Proof of \cref{lmoment_disagreement_lower_bound}} \label{lmoment_disagreement_lower_bound_proof}

\begin{proof}[\textbf{Proof:}]
To begin, write $S(x)=a+bx+\rho(x)$, where we assume $\sup_{x\in\mathcal{I}}|\rho(x)| \leq \varepsilon$. Then, for $g\in\{0,1\}$, it follows that
\begin{align*}
     |S(Q_{g0}(u))| \leq |a|+|b| |Q_{g0}(u)|+\varepsilon.
\end{align*}
Next, for almost every $u\in(0,1)$, it follows that
\begin{align*}
    \Gamma(u) &=S(Q_{10}(u))-Q_{10}(u)-S(Q_{00}(u))+Q_{00}(u)\\
    &=(b-1)(Q_{10}(u)-Q_{00}(u)) +\rho(Q_{10}(u))-\rho(Q_{00}(u)).
\end{align*}
Multiplying by $\phi_k$ and integrating yields
\begin{align}
    \gamma_k &=(b-1)(\lambda_k^*(Q_{10})-\lambda_k^*(Q_{00}))
      +e_k,
    \label{eq:lmoment_affine_decomposition}
\end{align}
where
\begin{align*}
    e_k=\int_0^1 \big(\rho(Q_{10}(u))-\rho(Q_{00}(u))\big) \phi_k(u)du.
\end{align*}

Since the difference of the two remainder terms is bounded in
absolute value by $2\varepsilon$, Cauchy--Schwarz and
$\|\phi_k\|_2=1$ give
\[
    |e_k|
    \leq2\varepsilon\int_0^1|\phi_k(u)|\,du
    \leq2\varepsilon.
\]
The reverse triangle inequality applied to
\cref{eq:lmoment_affine_decomposition}, together with
$|\gamma_k|\geq0$, proves
\cref{eq:lmoment_disagreement_lower_bound}.
The strict inequality in $ |b-1|\, |\lambda_k^*(Q_{10})-\lambda_k^*(Q_{00})| >2\varepsilon$ makes
the lower bound positive. Finally, if $\varepsilon=0$, then $e_k=0$, which gives
\begin{align*}
     |\gamma_k| =|b-1|\, |\lambda_k^*(Q_{10})-\lambda_k^*(Q_{00})|,
\end{align*}
thereby completing the proof.
\end{proof}

\subsection{Proof of \cref{thm::sparse}} \label{thm::sparse_proof}

\begin{proof}[\textbf{Proof:}]
Throughout the proof we take $n$ to be sufficiently large such that $K_n \geq K_0$. Let $F_n$ denote the empirical distribution function and $D_n = \|F_n - F \|_\infty$. Then, the Dvoretzky--Kiefer--Wolfowitz inequality yields $\mathbb{P}( D_n \geq t) \leq 2 e^{-2n t^2}$ for some $t > 0$. For $u \in (0, 1)$ observe that
\begin{align*}
    Q( (u-D_n) \vee 0) \leq  Q_n(u) \leq Q( (u+D_n) \wedge 1).
\end{align*}
Since we assume $Q$ is $L$-Lipschitz, this implies that
$\|Q_n - Q \|_\infty \leq L D_n$. Hence, writing $Z_n = \|Q_n - Q\|_2$ we have that
\begin{align*}
    \mathbb{P}( Z_n > t) \leq 2 e^{-2 n t^2/L^2}.
\end{align*}
Parseval's identity then gives the simultaneous bound $\sum_{k \geq 1}(\widetilde{\lambda}^*_{k, n} - \lambda^*_k)^2 = \|Q_n - Q\|^2_2 = Z^2_n$. Now, let $C_n(K) = -\sum^K_{k=1} (\widetilde{\lambda}^*_{k, n})^2 + Kp_n$. For $K_0 < K \leq K_n$, the true coefficients vanish, so on $\{Z_n^2 \leq p_n\}$,
\begin{align*}
    C_n(K)-C_n(K_0) = (K-K_0)p_n - \sum_{k=K_0+1}^K(\widetilde{\lambda}^*_{k, n})^2 \geq  (K-K_0)p_n-Z_n^2 \ge 0.
\end{align*}
Even if equality holds, the smallest-minimizer convention rules out selection of $K>K_0$. Therefore,
\begin{align*}
    \Pp(\widehat{K}_s > K_0) \leq \Pp(Z_n^2>p_n) \le 2n^{-2\kappa/L^2}.
\end{align*}
Now, let $a_0=|\lambda^*_{K_0}|>0$. On $\{Z_n\le a_0/2\}$ we have $|\widetilde{\lambda}^*_{K_0, n}|\ge a_0/2$. For $K<K_0$,
\begin{align*}
    C_n(K)-C_n(K_0) =\sum_{k=K+1}^{K_0}(\widetilde{\lambda}^*_{k, n})^2 -(K_0-K)p_n \ge a_0^2/4-K_0p_n>0
\end{align*}
for all sufficiently large $n$. Thus
\begin{align*}
    \Pp(\widehat{K}_s < K_0)\le2\exp\left(-\frac{na_0^2}{2L^2}\right).
\end{align*}
When $K_0=1$, underselection is impossible. Combining the bounds gives, for all sufficiently large $n$,
\begin{align*}
    \mathbb{P}(\widehat{K}_s\neq K_0) \leq 2n^{-2\kappa/L^2} +2\exp\!\left(-\frac{na_0^2}{2L^2}\right) =O(n^{-2\kappa/L^2}).
\end{align*}
In particular, this probability is $o(n^{-1})$ when $\kappa>L^2/2$. \medskip

Now, we compare the selected and oracle risks. Let $B= \|Q \|_\infty$. Because the sample and $Q_n$ lie in the support interval, $\|Q_n \|_2\le B$. Since $\widetilde{Q}_{K,n}$ is an orthogonal projection, $\|\widetilde{Q}_{K,n}\|_2 \le B$ and $\|\widetilde{Q}_{K,n} - Q\|^2_2 \leq 4B^2$ for every $K$, including random $K$. On $\{\widehat{K}_s = K_0\}$, the adaptive and oracle estimators coincide. On $\{\widehat{K}_s>K_0\}$, the finite expansion $Q =\Pi_{K_0}Q$ and orthogonality imply
\begin{align*}
    0 \leq \|\widetilde{Q}_{\widehat{K}_s,n}-Q\|_2^2 -\|\widetilde{Q}_{K_0,n}-Q\|_2^2 =\sum_{k=K_0+1}^{\widehat{K}_s} (\widetilde{\lambda}_{k,n}^*)^2 \leq Z_n^2.
\end{align*}
Note that overselection requires $Z_n^2>p_n$. Integrating the
concentration bound then yields
\begin{align*}
    \mathbb E\!\left[Z_n^2\mathbb{I}(Z_n^2>p_n)\right]
    &=p_n\mathbb{P}(Z_n^2>p_n) +\int_{p_n}^\infty\mathbb{P}(Z_n^2>t) dt\\
    &\leq2\left(p_n+\frac{L^2}{2n}\right) \exp\!\left(-\frac{2np_n}{L^2}\right)\\
    &=\frac{2\kappa\log n+L^2}{n} n^{-2\kappa/L^2} =o(n^{-1})
\end{align*}
for every $\kappa>0$. Using the bounded loss only on the underselection event, we consequently obtain
\begin{align} \label{eq:compare}
    \left| \mathbb{E} \! \left[\|\widetilde{Q}_{\widehat K_s,n}-Q\|_2^2 \right] - \mathbb{E} \! \left[ \|\widetilde{Q}_{K_0,n}-Q\|_2^2 \right] \right| \leq \mathbb{E}\!\left[Z_n^2\mathbb{I}(Z_n^2>p_n)\right] +4B^2\mathbb{P}(\widehat{K}_s<K_0) =o(n^{-1}).
\end{align}
By \cref{gaussian_process_limit}, $\sqrt{n}(Q_n-Q)\rightsquigarrow\mathbb{G}_F$ in $L^2(0,1)$, where $\mathbb{G}_F$ has the same distribution as $Q^\prime \mathbb{B}$ for a standard Brownian bridge $\mathbb{B}$. The DKW bound also gives uniform integrability of $nZ^2_n$ (indeed, bounded moments of every fixed order). Thus
\begin{align*}
     n\mathbb{E} \! \left[\|Q_n-Q\|_2^2 \right] \longrightarrow \mathbb{E} \! \left[ \|\mathbb{G}_F\|_2^2 \right] =\int_0^1u(1-u)Q^\prime(u)^2 du =V_\infty(F).
\end{align*}
The same reasoning applied to each fixed coefficient, using $n(\widetilde\lambda_{k,n}^*-\lambda_k^*)^2\leq nZ_n^2$,
gives
\begin{align*}
     n\mathbb{E} \! \left[ (\widetilde\lambda_{k,n}^*-\lambda_k^*)^2\right] \longrightarrow \mathbb{E} \! \left[ \langle\mathbb{G}_F,\phi_k\rangle^2 \right] =\Sigma_{kk}(F).
\end{align*}
Since $\lambda_k^*=0$ for $k>K_0$, orthogonality now implies
\begin{align*}
    \mathbb{E} \! \left[ \|\widetilde{Q}_{K_0,n}-Q\|_2^2 \right] =\sum_{k=1}^{K_0} \mathbb{E} \! \left[(\widetilde{\lambda}_{k,n}^*-\lambda_k^*)^2\right] =\frac{V_{K_0}(F)}n+o(n^{-1}),
\end{align*}
where $V_{K_0}(F)=\sum_{k=1}^{K_0}\Sigma_{kk}(F)$, with $\Sigma_{kk}(F)$ as defined in \cref{lmoment_efficiency_bound}. Together with \cref{eq:compare}, this proves the risk expansion for every $\kappa>0$. Finally, Parseval applied to the Gaussian limit implies $V_\infty(F) =\sum_{k\ge1}\Sigma_{kk}(F)$. Since $Q^\prime$ is strictly positive, the Brownian-bridge covariance operator is strictly positive definite on nonzero $L^2$ functions, so $\Sigma_{kk}(F)>0$ for every $k$. Hence, $V_{K_0}(F) <V_\infty(F)$.
\end{proof}

\end{document}